%% file: manuscript.tex
\documentclass[a4paper,twocolumn,showpacs,superscriptaddress,floatfix,nofootinbib,prd]{revtex4-1}
\usepackage{graphicx}
\usepackage{epsf}
\usepackage{epsfig}
\usepackage{amssymb,amsmath}
\usepackage[usenames]{color}
\usepackage{amssymb}
\usepackage{times}
\usepackage{hyperref}
\hypersetup{
  colorlinks=true,        
  linkcolor=blue,         
  citecolor=cyan,         
}

\newcommand{\cf}{cf.,~}
\newcommand{\ie}{i.e.,~}
\newcommand{\eg}{e.g.,~}
\newcommand{\ms}{\,{\rm ms}}
\newcommand{\Msun}{M_{\odot}}
\newcommand{\mn}[1]{\texttt{#1}}

\newcommand{\Ibar}{I\hskip-0.22cm-}

\begin{document}

\title{Spectral properties of the post-merger gravitational-wave signal
  from binary neutron stars}
\author{Kentaro~Takami}
\affiliation{Institut f{\"u}r Theoretische Physik,
  Max-von-Laue-Stra{\ss}e 1, 60438 Frankfurt, Germany}
\author{Luciano~Rezzolla}
\affiliation{Institut f{\"u}r Theoretische Physik,
  Max-von-Laue-Stra{\ss}e 1, 60438 Frankfurt, Germany}
\author{Luca~Baiotti} 
\affiliation{Graduate School of Science, Osaka University, 560-0043 Toyonaka, Japan}

\begin{abstract}
Extending previous work by a number of authors, we have recently
presented a new approach in which the detection of gravitational waves
from merging neutron star binaries can be used to determine the equation
of state of matter at nuclear density and hence the structure of neutron
stars. In particular, after performing a large number of
numerical-relativity simulations of binaries with nuclear equations of
state, we have found that the post-merger emission is characterized by
two distinct and robust spectral features. While the high-frequency peak
was already shown to be associated with the oscillations of the
hypermassive neutron star produced by the merger and to depend on the
equation of state, we have highlighted that the low-frequency peak is
related to the merger process and to the total compactness of the stars
in the binary. This relation is essentially universal and provides a
powerful tool to set tight constraints on the equation of state. We here
provide additional information on the extensive analysis performed,
illustrating the methods used, the tests considered, as well as the robustness
of the results. We also discuss additional relations that can be deduced
when exploring the data and how these correlate with various properties
of the binary. Finally, we present a simple mechanical toy model that
explains the main spectral features of the post-merger signal and can
even reproduce analytically the complex waveforms emitted right after the
merger.
\end{abstract}


\maketitle


\section{Introduction}
\label{sec:intro}

As a series of advanced detectors such as LIGO~\citep{Harry2010},
Virgo~\citep{Accadia2011_etal}, and KAGRA~\citep{Aso:2013} is going to
become operational in the next five years, we are likely to witness soon
the first direct detection of gravitational waves (GWs). The sources that
are likely to be detected first are obviously the strongest ones and, in
particular, the inspiral and post-merger of neutron-star binaries or
neutron-star$-$black-hole binaries, and binary black holes. Although an
uncertainty as large as 3 orders of magnitude accompanies the
estimates coming from population-synthesis models, it is widely assumed
that binary neutron star (BNS) mergers will represent the most common
source, with an expected detection rate of $\sim 40\, \mathrm
{yr}^{-1}$~\cite{Abadie:2010_etal}.

In addition to being excellent sources of GWs, the merger of a BNS system
represents also the most attractive scenario to explain the phenomenology
associated with short gamma-ray bursts (SGRBs). While suggested already
in the 1980s~\cite{Narayan92,Eichler89}, the recent progress of numerical
simulations~\cite{Shibata99d, Baiotti08, Anderson2007, Liu:2008xy,
  Bernuzzi2011} and the wealth of additional astronomical simulations
(see \cite{Berger2013b} for a recent review) has put this scenario on 
firm ground. Over the years, these simulations have shown that the merger
of BNSs inevitably leads to the formation of a massive metastable object,
which can either collapse promptly or survive up to a fraction of a
second emitting large amounts of gravitational radiation. If the NSs
comprising the binary have large magnetic fields and extended
magnetospheres the inspiral can be accompanied by a precursor
electromagnetic signal~\cite{Palenzuela2013a}, while the merger can lead
to instabilities~\cite{Siegel2013,Kiuchi2014} and to the formation of
magnetically confined jet structures once a torus is formed around the
black hole (BH)~\cite{Rezzolla:2011,Paschalidis2014}. The coincident detection of an
electromagnetic and GW emission would represent the most convincing
evidence of the link between SGRBs and BNSs.

Any GW signal from a binary including a neutron star will contain
important signatures of the equation of state (EOS) of matter at nuclear
densities. A first signature is represented by the tidal corrections to
the orbital phase; these are reasonably well understood
analytically~\cite{Flanagan08, Baiotti:2010, Bernuzzi2012} and can be
tracked accurately with advanced high-order numerical
codes~\cite{Radice2013b,Radice2013c}. A second signature is instead
related to the post-merger phase, where the object formed by the merger
[most likely a hypermassive neutron star (HMNS)] can emit GWs in a narrow
frequency range before collapsing to a BH~\cite{Shibata05d,
  Oechslin07b, Baiotti08}.

The first detailed evidence that it is possible to extract information
about the EOS of nuclear matter by carefully investigating the spectral
properties of the post-merger signal was provided by Bauswein and
Janka~\cite{Bauswein2011,Bauswein2012}. After performing a large number
of simulations using a smoothed particle hydrodynamics code solving the
Einstein field equations assuming conformal flatness and employing a GW
backreaction scheme within a post-Newtonian approximation, they pointed
out the presence of a peak at high-frequency in the spectrum (dubbed
$f_{\rm peak}$) and showed it correlated with the properties of the EOS,
\eg with the radius of the maximum-mass nonrotating configuration (see
also Ref.~\cite{Hotokezaka2013c} for a subsequent general-relativistic
study). The correlation found was rather tight, but restricted to
binaries having all the same total mass of $2.7\,\Msun$.  It was later
recognized that $f_{\rm peak}$ corresponds to a fundamental fluid mode
with $m=2$ of the HMNS~\cite{Stergioulas2011b} and that the information
from this frequency could also be used to set constraints on the
maximum mass of the system and hence on the EOS~\cite{Bauswein2014}.

Following this bulk of work, we have recently presented in
Ref.~\cite{Takami:2014} a new approach in which the detection of GWs 
from merging BNSs can be used to determine the EOS of matter at
nuclear density and hence the structure of neutron stars. More
specifically, making use of a large number of accurate
numerical-relativity simulations of binaries with nuclear EOSs and
different masses, we have shown that the post-merger emission is
characterized by two distinct and robust spectral features. The
high-frequency of these two peaks is the same already discussed in
Refs.~\cite{Bauswein2011,Bauswein2012}, but we have also revealed that
the correlations with the stellar properties depend on the total mass of
the system and hence on the EOS. On the other hand, we have also shown
that the low-frequency peak is related to the merger process and related
to the total compactness of the stars in the binary. This relation is
essentially universal and provides a powerful tool to set tight
constraints on the EOS.

The purpose of this paper is to provide additional information on the
extensive analysis performed and presented in~\cite{Takami:2014},
illustrating the methods used, the tests considered, as well as the robustness
of the results. Besides describing in greater detail the spectral
properties discussed above, we also discuss additional correlations that 
either have been already discussed in the literature, \eg in
Refs.~\cite{Read2013,Bernuzzi2014}, or that are presented here for the
first time. An important aspect of this paper is also that of presenting
a simple mechanical toy model that can explain rather intuitively the
main spectral features of the post-merger signal and shed light on the
physical interpretation of the origin of the various peaks.  Despite its
crudeness, the toy model can even reproduce the complex waveforms emitted
right after the merger, hence possibly opening the way to an analytical
modeling of a part of the signal that has been so far considered
impossible to model but with numerical-relativity calculations.
The waveforms are freely available for download and further analysis~\cite{relastrowiki41}.

The paper is organized as follows. Section~\ref{sec:ns} is dedicated to
the numerical setup used to carry out the simulations and discusses our
mathematical formulation, the structure and extent of the numerical
domain, the EOSs used, and the set of BNSs
considered. Section~\ref{sec:aotgws} deals instead with the analysis of
the GW signal, concentrating in particular on the approach followed to
compute in an automated way the spectral properties of the post-merger
signal. The validation of the numerical results is instead discussed in
Sec.~\ref{sec:votnr}, where we consider how the grid resolution or the
thermal contribution to the EOS influence the
results. Section~\ref{sec:results} is the scientific core of the paper,
and after a global overview of the results (Sec.~\ref{sec:go}), we
present a simple explanation for the origin of the $f_1$ and $f_3$
frequency peaks (Sec.~\ref{sec:f1_f3_orig}), as well as a number of
correlations that the relevant frequencies in the system, \ie $f_{\rm
  max}$ (the frequency at amplitude maximum) $f_1, f_2,$ and $f_3$, have
with the stellar properties
(Sec.~\ref{sec:corr_fmax}--\ref{sec:corr_other}). We also discuss in
Sec.~\ref{sec:constr_EOS} how to exploit the knowledge of the spectral
properties to constrain the EOS of nuclear matter, either in idealized
conditions or in more realistic ones
(Sec.~\ref{sec:detectability}). While the conclusions are reserved for
Sec.~\ref{sec:conclusions}, we dedicate Appendix~\ref{appendix_a} to
illustrate the details of the toy model.

We use a spacelike signature $(-,+,+,+)$ and a system of units in which
$c=G=\Msun=1$ unless stated differently.

\section{Numerical Setup}
\label{sec:ns}

Much of the numerical setup used in these simulations has been presented
in greater detail in other
papers~\cite{Baiotti08,Baiotti:2009gk,Baiotti:2010ka}, and for
compactness we will review here only the basic aspects, referring the
interested reader to the papers above for additional information. All of
our simulations have been performed in full general relativity using a
fourth-order finite-differencing code
\texttt{McLachlan}~\cite{Brown:2008sb, Loffler:2011ay}, which solves a
conformal traceless formulation of the Einstein
equations~\cite{Nakamura87, Shibata95, Baumgarte99}, with a ``$1+\log$''
slicing condition and a ``Gamma-driver'' shift
condition~\citep{Alcubierre02a,Pollney:2007ss}. At the same time, the
general-relativistic hydrodynamics equations are solved using the
finite-volume code \texttt{Whisky}~\citep{Baiotti04}, which has been
extensively tested in simulations involving the inspiral and merger of
BNSs~\cite{Baiotti08, Baiotti:2009gk, Rezzolla:2010, Baiotti:2010}.

We note that differently from previous works of ours, \eg
\cite{Read2013}, but also from those of other authors, \eg \cite{Tsatsin2013}, we
have not employed as approximate Riemann solver the Marquina flux
formula~\cite{Donat96}, which has been revealed to be problematic when used in
conjunction with nuclear-physics EOSs approximated as piecewise
polytropes (see discussion in Sec.~\ref{sec:EOS}). Instead, we have
employed the Harten-Lax-van Leer-Einfeldt (HLLE)~\cite{Harten83}
approximate Riemann solver~\cite{Harten83}, which is less accurate but
more robust, in conjunction with a piecewise parabolic method (PPM) for
the reconstruction of the evolved variables~\cite{Colella84}. We also
note that although we have recently developed a new code, \ie
\texttt{WhiskyTHC}, able to reach high-order
convergence~\cite{Radice2013b,Radice2013c}, we have not employed it here
mostly because our focus is on the post-merger dynamics, where the large
scale shocks reduce the accuracy of both codes to first order, and for
which \texttt{Whisky} appears to be slightly more robust.

For the time integration of the coupled set of the hydrodynamic and
Einstein equations we have used the method of lines (MOL) in conjunction
with an explicit fourth-order Runge-Kutta
method~\cite{Rezzolla_book:2013}.  In all our simulations we prescribe a
Courant-Friedrichs-Lewy (CFL) factor of $0.35$ to compute the size of the
time step.

\subsection{Grid structure and extent}

To place the outer boundary at sufficiently large distances and guarantee
at the same high spatial resolution near the neutron stars, we employ an
adaptive-mesh refinement (AMR) approach which follows closely the one
adopted in~\cite{Baiotti08, Kastaun2013}. In practice, the grid hierarchy
is handled by the \texttt{Carpet} mesh-refinement
driver~\cite{Schnetter-etal-03b}. It implements vertex-centered mesh
refinement, also known as the box-in-box method, and allows for
regridding during the calculation as well as multiple grid centers. As
discussed in Ref.~\cite{Schnetter-etal-03b}, within this approach a
number of grids with different resolutions, called refinement levels,
overlay each other, and are nested in such a way that the coarsest grid
has the largest extent and the finest grid the smallest extent. The
time step on each grid is set by the Courant condition and so by the
spatial grid resolution for that level. Boundary data for finer grids are
calculated with spatial prolongation operators employing fifth-order
polynomials and with prolongation in time employing second-order
polynomials.

During the inspiral phase, the positions of the centers of the two
highest-resolution grids are determined by the location of the maximum of
the rest-mass density. One grid is placed at the grid point correspondent
to \ensuremath{\rho_\mathrm{max}} and the other at the $\pi$-symmetric
point, namely the point obtained by a rotation of $180$ deg around
the \ensuremath{z} axis for equal-mass binaries.  The grid hierarchy is
composed of six refinement levels and a $2:1$ refinement factor for
successive levels. The grid resolution varies from $\ensuremath{\Delta
  h_5 = 0.15\,\Msun}$ (\ie $\simeq 221\,{\rm m}$) for the finest level 
to $\ensuremath{\Delta h_0 = 4.8\,\Msun}$ (\ie $\simeq 7.1\,{\rm km}$)
for the coarsest level, whose outer boundary is at $514\,\Msun$ (\ie
$\simeq 759\,{\rm km}$). Initially, the number of grid points across the
linear dimension of a star is of the order of $100$, while these become
roughly twice when the merger has taken place and a HMNS has been formed.

The whole grid is set up to be symmetric with respect to the
\ensuremath{(x,y)} plane both for equal- and unequal-mass binaries, with
a reflection symmetry across the $z=0$ plane to reduce computational
costs. Additionally, a $\pi$-symmetry condition across the $x=0$ plane is
adopted in the case of equal-mass binaries to further reduce the
cost. The boundary conditions are chosen to be radiative for the metric
to prevent gravitational waves from scattering back into the grid, and
static for the hydrodynamical variables. No matter reaches the outer
boundary over the time scale of the simulations.

\subsection{Equations of state}
\label{sec:EOS}

We model the stars with nuclear-physics EOSs by considering five different
EOS models: \ie APR4~\cite{Akmal1998a}, ALF2~\cite{Alford2005},
SLy~\cite{Douchin01}, H4~\cite{GlendenningMoszkowski91}, and
GNH3~\cite{Glendenning1985}. We note that all of these EOSs satisfy the
current observational constraint on the observed maximum mass in neutron
stars, \ie $2.01\pm0.04M_\odot$ obtained for the pulsar PSR
J0348+0432~\cite{Antoniadis2013}. In order to cover a range in stellar
compactness that is larger than the one spanned by the above five nuclear
EOSs, we have also considered an additional EOS given by a single
polytrope, with pressure expressed as $p_\mathrm{c}=K \rho^\Gamma$, where
$\rho$, $K$, and $\Gamma$ are the rest-mass density, the polytropic
constant, and the polytropic exponent, respectively (see, \eg
Ref.~\cite{Rezzolla_book:2013} for details).

Instead of using tables for the various EOSs, we have found it simpler
and more convenient to reproduce each EOS in terms of a number $N$ of
piecewise polytropes~\cite{Read:2009a}. More specifically, the ``cold''
and nuclear-physics contribution to each EOS is obtained after expressing
the pressure and specific internal energy $\epsilon_\mathrm{c}$ in the
rest-mass density range $\rho_{i-1} \leq \rho < \rho_i$ as
\begin{align}
\label{EOS:cold_a}
p_\mathrm{c} &= K_i \rho^{\Gamma_i}\,,\\
\label{EOS:cold_b}
\epsilon_\mathrm{c} &= \epsilon_i + K_i
\frac{\rho^{\Gamma_i-1}}{\Gamma_i -1}\,,
\end{align}
In addition, we impose the
continuity of pressure and specific internal energy through the following
conditions 
\begin{align}
\label{eq:pwp_cont}
K_{i+1} &= K_i ~ \rho_i^{ \Gamma_i-\Gamma_{i+1} }\,,\\
\epsilon_{i+1} &= \epsilon_{i} +
K_i \frac{\rho_i ^{\Gamma_i-1}}{\Gamma_i-1} - 
K_{i+1} \frac{\rho_i^{\Gamma_{i+1}-1}}{\Gamma_{i+1}-1}\,, 
\end{align}
for $i=1,\ldots,N-1$. 

\begin{figure}
\begin{center}
\vskip -0.75cm
\includegraphics[width=\columnwidth]{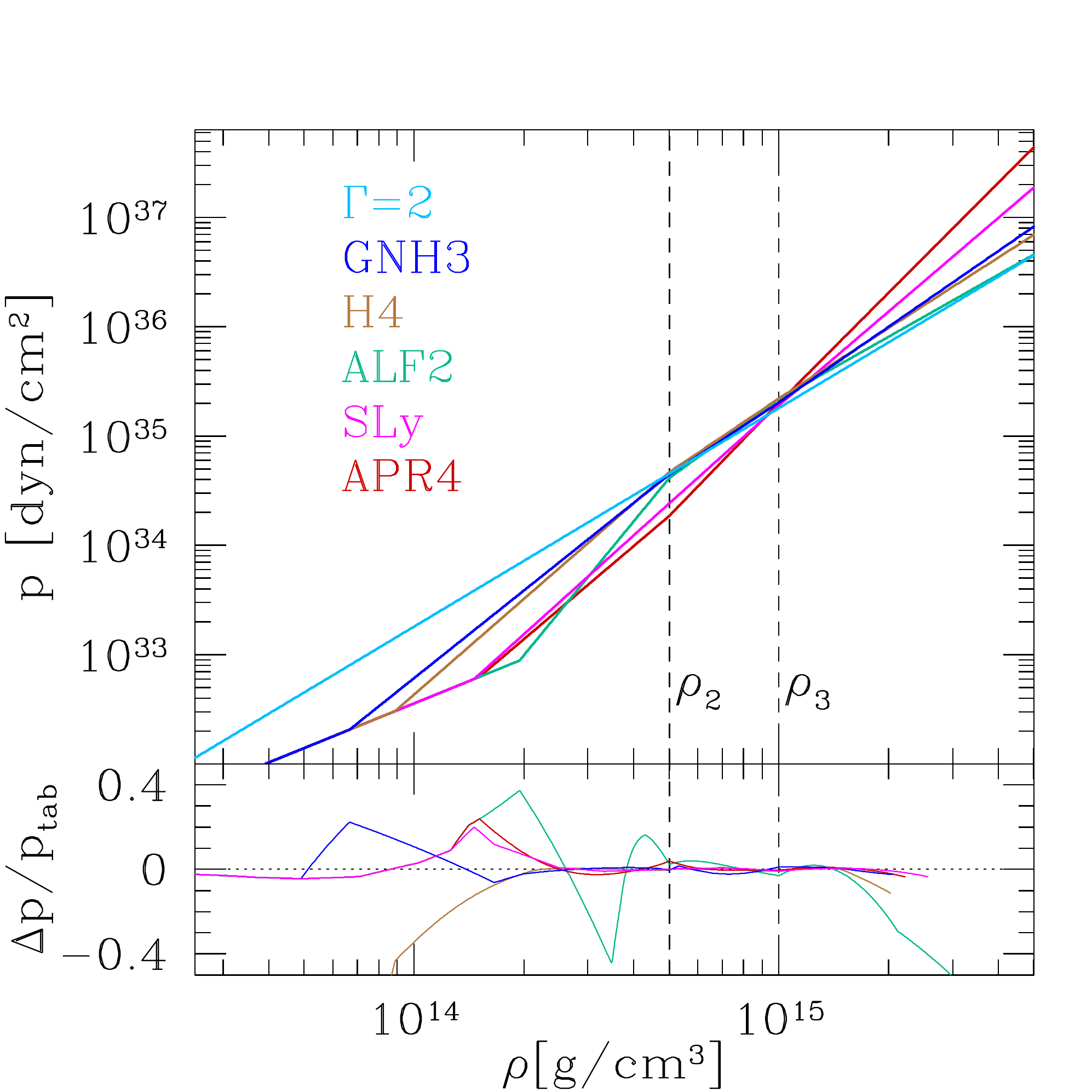}
\caption{\label{fig:EOSs} \textit{Top panels:} pressure as a function of
  the rest-mass density for the EOSs in Table \ref{tab:EOSs} when
  represented in terms of piecewise polytropes. \textit{Bottom panel:}
  the corresponding relative error $\Delta p/p_\mathrm{tab}$, where
  $\Delta p \equiv p_\mathrm{tab}-p$ and $p_\mathrm{tab}$ is the pressure
  in tabulated form. The tabulated EOSs for H4 and ALF2 are
    provided by~\cite{Read:2009a}, while the others are from
    the \texttt{LORENE} library~\cite{lorene41}.}
\end{center}
\end{figure}

As shown in Ref.~\cite{Read:2009a}, $N=4$ pieces, three of which
describe the high-density core and one the crustal region, are
sufficient to reproduce to good precision most of the EOSs.  In the top
panel of Fig.~\ref{fig:EOSs} we show the behavior of the various EOSs
used as well as when represented as piecewise polytropes, while the bottom panel
measures the relative error. Also reported in Table \ref{tab:EOSs} are
the various parameters used for the piecewise polytropes.

The cold nuclear-physics EOSs are adequate to represent the state of the
neutron-star matter prior to the merger. After contact, however,
large shocks will develop, considerably increasing the internal
energy. To account for this additional heating we could use hot
nuclear-physics EOSs, which provide information about the dependence of the
pressure and density also as a function of temperature. In practice,
however, the number of such EOSs is still pretty limited, thus preventing
us from doing a systematic investigation. As a result, we
follow~\cite{Bauswein2011,Bauswein2012} and account for thermal effects
by using a so-called ``hybrid EOS''~\cite{Rezzolla_book:2013}, that is,
by adding to the cold part an ideal-fluid component that accounts for the
shock heating~\cite{Janka93}. In practice the total pressure and specific
internal energy are expressed as
\begin{align}
\label{EOS:full_a}
 p &= p_\mathrm{c} + p_\mathrm{th}\,,\\
\label{EOS:full_b}
\epsilon &= \epsilon_\mathrm{c} + \epsilon_\mathrm{th}\,,
\end{align}
where $p_\mathrm{c}, \epsilon_\mathrm{c}$ are given by
Eqs.~\eqref{EOS:cold_a} and \eqref{EOS:cold_b}, while the ``thermal'' part
is given by
\begin{align}
\label{EOS:hot}
 p_\mathrm{th} &= \rho \epsilon_\mathrm{th}
\left(\Gamma_\mathrm{th} -1 \right)\,,\\
\epsilon_\mathrm{th} & = \epsilon - \epsilon_\mathrm{c} \,,
\end{align}
where $\Gamma_\mathrm{th}$ is arbitrary but constrained mathematically to
be $1 \leq \Gamma_\mathrm{th} \leq 2$. Note that this hybrid-EOS
approach is possible because the total specific internal energy
$\epsilon$ is one of the evolved variables; hence its value is computed
independently at each time level.

\begin{table}[tb]
\begin{tabular}{c|c|ccc|c|cc}
\hline
EOS & $N$  
 & $\Gamma_2$ & $\Gamma_3$ & $\Gamma_4$ 
 & $\rho_1$ 
 & $M_\mathrm{max}$ & $R_\mathrm{max}$ \\
 &   
 & & &  
 & [$\mathrm{g}/\mathrm{cm}^3$] 
 & $[M_\odot]$ & $[\mathrm{km}]$ \\
\hline
$\Gamma=2$ & $1$ 
 & $\cdots$ & $\cdots$ & $\cdots$ 
 & $\cdots$ 
 & $1.8206$ & $12.536$\\
GNH3 & $4$ 
 & $2.664$ & $2.194$ & $2.304$ 
 & $6.66038 \times 10^{13}$ 
 & $1.9768$ & $11.266$\\
H4 & $4$ 
 & $2.909$ & $2.246$ & $2.144$
 & $8.87824 \times 10^{13}$ 
 & $2.0282$ & $11.603$\\
ALF2 & $4$ 
 & $4.070$ & $2.411$ & $1.890$ 
 & $1.94771\times 10^{14}$ 
 & $1.9911$ & $11.308$\\
SLy & $4$ 
 & $3.005$ & $2.988$ & $2.851$
 & $1.46231 \times 10^{14}$ 
 & $2.0606$ & $9.9349$ \\
APR4 & $4$ 
 & $2.830$ & $3.445$ & $3.348$
 & $1.51201 \times 10^{14}$ 
 & $2.2000$ & $9.8733$ \\
\hline
\end{tabular}
\caption{Parameters used to represent the nuclear-physics EOSs. In
  addition to the values reported in the table, we use the following
  values for the polytropic constants and indices: $K=123.647$, $K_1 /
  c^2 = 3.99873692 \times 10^{-8} [\mathrm{(g/cm^3)^{1-\Gamma_1}}]$,
  $\Gamma_1=1.35692395$. Similarly, different polytropic pieces are
  joined at the rest-mass densities $\rho_2=5.01187234 \times
  10^{14}[\mathrm{g}/\mathrm{cm}^3]$, and $\rho_3 =
  10^{15}[\mathrm{g}/\mathrm{cm}^3]$. The last two columns in the table
  report the gravitational mass and the radius of the maximum-mass model in
  the corresponding Tolman–Oppenheimer–Volkoff (TOV) sequence.
  \label{tab:EOSs}}
\end{table}

Before concluding, an important remark should be made about the use of
piecewise polytropes in relativistic hydrodynamic simulations, especially
when adopting higher-order reconstruction methods or sophisticated
approximate Riemann solvers~\cite{Rezzolla_book:2013}. The
prescription~\eqref{EOS:cold_a} for each piece of the polytrope, together
with the junction conditions~\eqref{eq:pwp_cont} across different pieces,
guarantee that the pressure is everywhere a continuous function of the
rest-mass density. Of course these conditions cannot guarantee that the
derivatives of the pressure are themselves continuous. The accepted
wisdom is that these discontinuities are harmless as they involve only a
very limited number of cells during the evolution. However, whether this
is actually true in practice depends crucially on the accuracy of the
reconstruction algorithm and on the robustness of the Riemann solver
employed. More importantly, the negative impact that these
discontinuities may have on the evolution is often difficult to diagnose
and hence avoid.

\begin{figure}
\begin{center}
\includegraphics[width=\columnwidth]{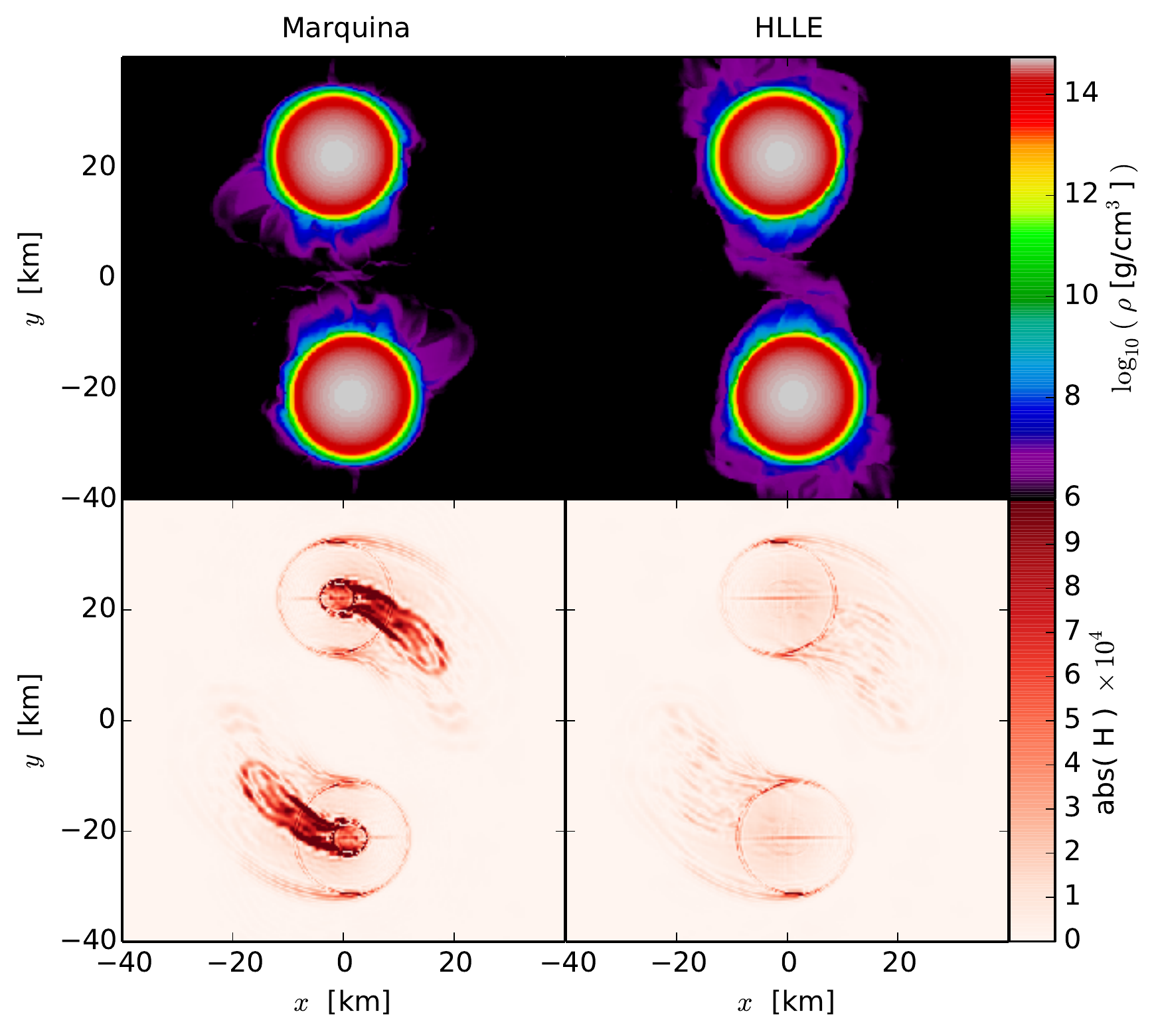}
\caption{\textit{Top panels:} Snapshots of the rest-mass density on the
  $(x,y)$ plane at $t\simeq-12.50\,\mathrm{ms}$ in the evolution of the
  binary \mn{H4-q10-M1350} with $\Gamma_\mathrm{th}=1.36$, computed using
  either the Marquina approximate Riemann solver (top left panel) or the
  HLLE Riemann solver (top right panel). Note that the two rest-mass
  distributions are rather similar. \textit{Bottom panels:} Snapshots of
  the corresponding violations of the Hamiltonian constraints on the
  $(x,y)$ plane. Note that the Marquina solver leads to violations that
  are 1 or more orders of magnitude
  larger. \label{fig:marquina_vs_hlle} }
\end{center}
\end{figure}

Figure \ref{fig:marquina_vs_hlle} offers a very clear example of the
risks that can accompany the use of a piecewise polytropic EOS by showing
in the top panels two snapshots of the rest-mass density on the $(x,y)$
plane at a representative time of the evolution
($t\sim-12.50\,\mathrm{ms}$) of the binary \mn{H4-q10-M1350} with
$\Gamma_\mathrm{th}=1.36$, computed using either the Marquina approximate
Riemann solver~\cite{Donat96} (top left panel) or the HLLE Riemann
solver~\cite{Harten83} (top right panel). In both cases the
reconstruction is made using the PPM prescription.

A quick comparison between the two distributions of the rest-mass density
does not reveal any problematic behavior and indeed the solutions seem
very much comparable, with differences appearing mostly in the outer
layers of the stars, close to the atmosphere and hence of little
relevance in terms of the dynamics of the binary. This comparison is,
however, rather deceptive and the conclusion is actually incorrect. The
evolution with the Marquina solver, in fact, suffers from a much larger
violation of the Hamiltonian constraint, with norm-2 values that can be
1 order of magnitude larger. This is illustrated in the bottom panels,
which show the distributions of the corresponding violations of the
Hamiltonian constraints on the $(x,y)$ plane. While both evolutions show
sources of violations near the stellar surface (\cf outer rings in both
solvers), the solution with the Marquina formula shows much larger
violations in the central regions of the stars and, in particular, around
the rest-mass density $\rho_2$, where the two most-central pieces of the
EOS are joined.

A one-dimensional inspection of the rest-mass density would indeed reveal
that the stars present rather macroscopic discontinuities around $\rho_2$
in the case of the solution with the Marquina solver, which are instead
absent with the less accurate but more robust HLLE solver. As expected,
the discontinuities appearing with the Marquina solver in the case of
piecewise EOSs, disappear when a single polytropic EOS is used, with
violations in the Hamiltonian constraint that are comparable to those of
the HLLE solver.

\begin{table*}
\begin{tabular}{l|c|c|c|c|c|c|c|c|c|c|c|c|c}
\hline
\hline
Model & EOS & $q$ & $\bar{M}$ & $\bar{R}$ 
& $M_\mathrm{ADM}$ & $\bar{M_\mathrm{b}}$
& $\bar{M}/\bar{R}$  
& $f_\mathrm{orb}$ & $J$
& $\bar{I}/\bar{M}^3$ & $\bar{k_2}$ & $\lambda/\bar{M}^5$ & $f_{\rm cont}$ \\
& & & $[\Msun]$ & $[\mathrm{km}]$ 
& $[\Msun]$ & $[\Msun]$
& & $[\mathrm{Hz}]$ & $[\Msun^2]$ 
&&&& $[\mathrm{Hz}]$ \\
\hline
\mn{GAM2-q10-M1350} & $\Gamma=2$ & $1$ & $1.350$ & $16.774$ & $2.6748$ & $1.4336$ & $0.11883$ & $282.47$ & $7.3039$ & $21.158$ & $0.097945$ & $2755.3$ & $980.58$ \\
\mn{GAM2-q10-M1375} & $\Gamma=2$ & $1$ & $1.375$ & $16.664$ & $2.7239$ & $1.4623$ & $0.12183$ & $284.37$ & $7.5274$ & $20.203$ & $0.095091$ & $2361.6$ & $999.43$ \\
\mn{GAM2-q10-M1400} & $\Gamma=2$ & $1$ & $1.400$ & $16.550$ & $2.7730$ & $1.4911$ & $0.12490$ & $286.25$ & $7.7540$ & $19.296$ & $0.092229$ & $2023.0$ & $1018.8$ \\
\mn{GAM2-q10-M1425} & $\Gamma=2$ & $1$ & $1.425$ & $16.434$ & $2.8221$ & $1.5200$ & $0.12803$ & $288.10$ & $7.9838$ & $18.434$ & $0.089356$ & $1731.5$ & $1038.9$ \\
\mn{GAM2-q10-M1450} & $\Gamma=2$ & $1$ & $1.450$ & $16.313$ & $2.8712$ & $1.5491$ & $0.13125$ & $289.93$ & $8.2167$ & $17.612$ & $0.086470$ & $1480.3$ & $1059.7$ \\
\hline
\mn{GNH3-q10-M1250} & GNH3 & $1$ & $1.250$ & $13.817$ & $2.4780$ & $1.3464$ & $0.13358$ & $273.29$ & $6.4067$ & $18.890$ & $0.11753$  & $1842.4$ & $1262.1$ \\
\mn{GNH3-q10-M1275} & GNH3 & $1$ & $1.275$ & $13.810$ & $2.5271$ & $1.3756$ & $0.13632$ & $275.38$ & $6.6187$ & $18.237$ & $0.11531$  & $1633.0$ & $1275.7$ \\
\mn{GNH3-q10-M1300} & GNH3 & $1$ & $1.300$ & $13.801$ & $2.5763$ & $1.4050$ & $0.13908$ & $277.44$ & $6.8340$ & $17.614$ & $0.11305$  & $1448.1$ & $1289.4$ \\
\mn{GNH3-q10-M1325} & GNH3 & $1$ & $1.325$ & $13.790$ & $2.6255$ & $1.4345$ & $0.14187$ & $279.53$ & $7.0538$ & $17.019$ & $0.11075$  & $1284.7$ & $1303.2$ \\
\mn{GNH3-q10-M1350} & GNH3 & $1$ & $1.350$ & $13.777$ & $2.6746$ & $1.4641$ & $0.14468$ & $281.58$ & $7.2766$ & $16.450$ & $0.10841$  & $1139.9$ & $1317.3$ \\
\mn{GNH3-q09-M1300} & GNH3 & $0.92593$ & $1.300$ & $13.797$ & $2.5763$ & $1.4052$ & $0.13912$ & $277.51$ & $6.8255$ & $17.607$ & $0.11297$ & $1445.2$ & $1289.9$ \\
\hline
\mn{H4-q10-M1250}   & H4   & $1$ & $1.250$ & $13.533$ & $2.4780$ & $1.3506$ & $0.13638$ & $273.25$ & $6.4058$ & $18.610$ & $0.12361$  & $1746.5$ & $1302.1$ \\
\mn{H4-q10-M1275}   & H4   & $1$ & $1.275$ & $13.539$ & $2.5271$ & $1.3799$ & $0.13904$ & $275.40$ & $6.6191$ & $18.004$ & $0.12147$  & $1558.3$ & $1314.1$ \\
\mn{H4-q10-M1300}   & H4   & $1$ & $1.300$ & $13.544$ & $2.5763$ & $1.4094$ & $0.14172$ & $277.52$ & $6.8356$ & $17.426$ & $0.11930$  & $1391.4$ & $1326.1$ \\
\mn{H4-q10-M1325}   & H4   & $1$ & $1.325$ & $13.548$ & $2.6255$ & $1.4390$ & $0.14440$ & $279.60$ & $7.0552$ & $16.873$ & $0.11708$  & $1243.1$ & $1338.3$ \\
\mn{H4-q10-M1350}   & H4   & $1$ & $1.350$ & $13.550$ & $2.6746$ & $1.4687$ & $0.14711$ & $281.61$ & $7.2770$ & $16.344$ & $0.11483$  & $1111.1$ & $1350.6$ \\
\hline
\mn{ALF2-q10-M1225} & ALF2 & $1$ & $1.225$ & $12.252$ & $2.4288$ & $1.3373$ & $0.14762$ & $271.00$ & $6.1923$ & $16.975$ & $0.13290$  & $1263.8$ & $1496.2$ \\  
\mn{ALF2-q10-M1250} & ALF2 & $1$ & $1.250$ & $12.276$ & $2.4779$ & $1.3672$ & $0.15034$ & $273.16$ & $6.4014$ & $16.455$ & $0.13049$  & $1132.6$ & $1507.0$ \\  
\mn{ALF2-q10-M1275} & ALF2 & $1$ & $1.275$ & $12.298$ & $2.5271$ & $1.3971$ & $0.15307$ & $275.12$ & $6.6103$ & $15.957$ & $0.12803$  & $1015.6$ & $1517.9$ \\  
\mn{ALF2-q10-M1300} & ALF2 & $1$ & $1.300$ & $12.319$ & $2.5763$ & $1.4272$ & $0.15582$ & $277.26$ & $6.8274$ & $15.480$ & $0.12552$  & $911.00$ & $1529.0$ \\  
\mn{ALF2-q10-M1325} & ALF2 & $1$ & $1.325$ & $12.337$ & $2.6254$ & $1.4574$ & $0.15858$ & $279.36$ & $7.0475$ & $15.021$ & $0.12297$  & $817.44$ & $1540.1$ \\  
\hline
\mn{SLy-q10-M1250}  & SLy  & $1$ & $1.250$ & $11.469$ & $2.4779$ & $1.3720$ & $0.16092$ & $273.04$ & $6.3977$ & $14.000$ & $0.10266$  & $634.27$ & $1668.8$ \\
\mn{SLy-q10-M1275}  & SLy  & $1$ & $1.275$ & $11.470$ & $2.5271$ & $1.4024$ & $0.16413$ & $275.20$ & $6.6110$ & $13.545$ & $0.10025$  & $561.11$ & $1685.3$ \\
\mn{SLy-q10-M1300}  & SLy  & $1$ & $1.300$ & $11.469$ & $2.5763$ & $1.4330$ & $0.16736$ & $277.34$ & $6.8275$ & $13.113$ & $0.097835$ & $496.81$ & $1701.9$ \\
\mn{SLy-q10-M1325}  & SLy  & $1$ & $1.325$ & $11.468$ & $2.6254$ & $1.4637$ & $0.17060$ & $279.36$ & $7.0455$ & $12.702$ & $0.095415$ & $440.20$ & $1718.5$ \\
\mn{SLy-q10-M1350}  & SLy  & $1$ & $1.350$ & $11.465$ & $2.6745$ & $1.4946$ & $0.17386$ & $281.34$ & $7.2663$ & $12.309$ & $0.092993$ & $390.29$ & $1735.2$ \\
\mn{SLy-q09-M1300}  & SLy  & $0.92593$ & $1.300$ & $11.467$ & $2.5763$ & $1.4333$ & $0.16739$ & $277.23$ & $6.8154$ & $13.115$ & $0.097828$ & $496.28$ & $1702.3$ \\
\hline
\mn{APR4-q10-M1275} & APR4 & $1$ & $1.275$ & $11.060$ & $2.5271$ & $1.4090$ & $0.17022$ & $275.22$ & $6.6107$ & $12.821$ & $0.097595$ & $455.30$ & $1779.9$ \\
\mn{APR4-q10-M1300} & APR4 & $1$ & $1.300$ & $11.067$ & $2.5763$ & $1.4399$ & $0.17344$ & $277.24$ & $6.8245$ & $12.436$ & $0.095396$ & $405.19$ & $1795.5$ \\
\mn{APR4-q10-M1325} & APR4 & $1$ & $1.325$ & $11.073$ & $2.6254$ & $1.4709$ & $0.17667$ & $279.31$ & $7.0437$ & $12.070$ & $0.093194$ & $360.93$ & $1811.1$ \\
\mn{APR4-q10-M1350} & APR4 & $1$ & $1.350$ & $11.079$ & $2.6746$ & $1.5020$ & $0.17992$ & $281.37$ & $7.2665$ & $11.720$ & $0.090990$ & $321.78$ & $1826.7$ \\
\mn{APR4-q10-M1375} & APR4 & $1$ & $1.375$ & $11.084$ & $2.7237$ & $1.5334$ & $0.18317$ & $283.39$ & $7.4924$ & $11.387$ & $0.088786$ & $287.10$ & $1842.3$ \\
\hline
\hline
\end{tabular}
\caption{ All models and their properties. The various columns denote the
  gravitational mass ratio $q\equiv M_1/M_2$ at infinite separation, the
  average gravitational mass $\bar{M}$ at infinite separation, the
  average radius $\bar{R}$ at infinite separation, the
  Arnowitt-Deser-Misner (ADM) mass
  $M_\mathrm{ADM}$ of the system at initial separation, the baryon mass
  $\bar{M}_\mathrm{b}$, the compactness ${\mathcal C}\equiv
  \bar{M}/\bar{R}$, the orbital frequency $f_\mathrm{orb}$ at the initial
  separation, the total angular momentum $J$ at the initial separation,
  the dimensionless moment of inertia $\bar{I}/\bar{M}^3$ at infinite
  separation, the $\ell=2$ dimensionless tidal Love number $\bar{k}_2$ at
  infinite separation, the dimensionless tidal deformability
  $\lambda/\bar{M}^5$ defined by $\lambda\equiv 2 \bar{k}_2 \bar{R}^5/3$,
  and the contact frequency $f_{\rm cont}=\mathcal{C}^{3/2}/(2\pi
  \bar{M})$~\cite{Damour:2012}. The quantities with a bar are defined as
  averages, \ie $\bar{A} \equiv (A_1+A_2)/2$.
\label{tab:models}}
\end{table*}

To the best of our knowledge this is the first time that such a
comparison is made and that these problems have been discussed in the
context of BNS evolutions. It is possible that the violations of the
constraints are made more severe by the fact that the violations have
zero propagation velocities and that once formed they cannot leave the
computational domain. In this case, the use of alternative formulations
of the equations, such as the recently proposed conformal and covariant
Z4 formulation (CCZ4) \cite{Alic:2011a}, which has been shown to lead
to reduced violations of the constraints~\cite{Alic2013}, are likely to
yield improved evolutions in the case of higher-order methods or Riemann
solvers. A more systematic investigation is needed to address this point.

\subsection{Initial data}

In order to generate the initial data for quasiequilibrium irrotational
BNSs with these EOSs, we use the multidomain spectral-method code
\texttt{LORENE}~\citep{Gourgoulhon-etal-2000:2ns-initial-data} under the
assumption of a conformally flat spacetime metric. We consider a rather
small initial coordinate separation of the stellar centers of
$45\,\mathrm{km}$ for all models, since we here focus mostly on the GW
emission from the merger and post-merger stages.

Clearly, if the HMNS is short-lived, it is hard to obtain reliable
statistical properties of the GWs from the HMNS, and thus construct
empirical correlations. As a result, we have selected initial-data models
with masses such that the HMNS has a lifetime $t_\mathrm{HMNS} \geq
5000\,\Msun \approx 24.63\ms$. For each EOS, we have considered five
equal-mass binaries with average (gravitational) mass at infinite
separation in the range $\bar{M} \equiv (M_1+M_2)/2 = (1.275-1.375)\Msun$
for the APR4 EOS, $(1.225-1.325)\Msun$ for the ALF2 EOS,
$(1.250-1.350)\Msun$ for the GNH3, H4, and SLy EOSs, and
$(1.350-1.450)\Msun$ for the $\Gamma=2$ polytrope. Higher masses would
lead to HMNSs that collapse to a BH-torus system well before a time
$5000\,\Msun \approx 25\,{\rm ms}$. In addition, we have also considered
two unequal-mass binaries with $\bar{M}=1.300\,\Msun$ and mass ratio $q
\simeq 0.93$ for the GNH3 and SLy EOSs. Detailed information on all the
models and the properties used in our evolutions is collected in Table
\ref{tab:models}, but it is worth remarking that all in all we have
considered 32 different binaries with different EOSs and mass
ratios. Together with the work of Ref.~\cite{Hotokezaka2013c}, this
represents the largest sample of simulations of BNSs with nuclear-physics
EOSs.

\section{Analysis of the gravitational-wave signal}
\label{sec:aotgws}

\subsection{Definitions and conventions}

We extract the GW signal at different surfaces of constant coordinate
radius using the Newman-Penrose formalism, so that the GW polarization
amplitudes $h_+$ and $h_\times$ are related to the Weyl curvature scalar
$\psi_4$ by (see Sec.~IV of Ref.~\cite{Baiotti08} for details)
\begin{equation}
\ddot{h}_+ - {\rm i} \ddot{h}_\times =
\psi_4 = \sum_{\ell=2}^{\infty}\sum_{m=-\ell}^{\ell} 
\psi_4^{\ell m}\;_{-2}Y_{\ell m}(\theta,\varphi),
\end{equation} 
where the overdot indicates a time derivative and we have introduced the
(multipolar) expansion of $\psi_4$ in spin-weighted spherical
harmonics~\cite{Goldberg:1967} of spin weight $s=-2$. In our analysis we
consider only the $\ell=m=2$ mode, which is the dominant one both in the
inspiral and after the merger; \ie we assume\footnote{Note that for
  simplicity in Ref. \cite{Takami2014} we have used $h_{+,\times} =
  h^{22}_{+,\times}$.}
\begin{equation}
h_{+,\times} 
= \sum_{\ell=2}^{\infty}\sum_{m=-\ell}^{\ell}
 h^{\ell m}_{+,\times} \,{}_{-2}Y_{\ell m}\left(\theta,\varphi\right)
\approx h^{22}_{+,\times} \,{}_{-2}Y_{22}\left(\theta,\varphi\right)\,,
\end{equation}
where ${}_{s}Y_{\ell m}\left(\theta,\varphi\right)$ are the spin-weighted
spherical harmonics. Unless explicitly stated, we consider the optimally
oriented case, for which
${}_{-2}Y_{22}\left(\theta=0,\varphi=0\right)\approx0.63$. The GWs are
sampled in time at a rate of $\Delta t=1.68\,M_{\odot} \simeq
8.27\times10^{-3}\ms$, equivalent to a sampling rate of $\simeq
121\,\mathrm{kHz}$.

Following previous work~\cite{Read2013}, and in order to evaluate the GWs
in the same phase for all models, we align the waveforms at the time of
merger, which we set to be $t=0$ and define to be correspondent to the
time when the GW amplitude
\begin{equation}
|h| \equiv ( {h^2_+} + {h^2_{\times}} )^{1/2}
\end{equation}
reaches its first maximum. As a result, for all binaries we consider GW
signals in the time interval $t \in [-1500,5000]\, \Msun \approx
[-7.39,24.63] \ms$, which includes GWs from approximately 3$-$4 orbits
before merger.

The instantaneous frequency of the GW is then computed as~\cite{Read2013}
\begin{equation}
f_{_\mathrm{GW}} \equiv \frac{1}{2\pi}\frac{d{\phi}}{d{t}}\,,
\end{equation}
where $\phi = \arctan( h_{\times} / h_{+} )$ is the phase of the complex
gravitational waveform. As a consequence, the GW frequency at the time of
merger, or equivalently the frequency at maximum amplitude\footnote{In
  Ref.~\cite{Read2013} this frequency was denoted as $f_\mathrm{peak}$.},
is given by
\begin{align}
f_\mathrm{max} \equiv f_{_\mathrm{GW}}(t=0)\,.
\end{align}

We next define the power spectral density (PSD) of the effective
amplitude as
\begin{eqnarray}
&& \tilde{h}(f) \equiv \sqrt{ \frac{ |\tilde{h}_{+}(f)|^2 +
      |\tilde{h}_{\times}(f)|^2 }{2} } \,,
\end{eqnarray}
with
\begin{eqnarray}
&& \tilde{h}_{+,\times}(f) \equiv 
\left\{
\begin{array}{ll}
\displaystyle
\int 
h_{+,\times}(t)\,e^{-i 2\pi f t} dt & ( f \ge 0 )\\
\displaystyle
0 & ( f < 0 )
\end{array}
\right.\,,
\end{eqnarray}
and where the $+$ and $\times$ indices refer to the two polarization
modes. Finally, the signal-to-noise ratio (SNR) is computed as
\begin{equation}
\mathrm{SNR} \equiv \left[~ \int ^\infty _0 
\frac{\bigl| 2~\tilde{h}(f) f^{1/2}\bigr|^2}{S_{h}(f)}
~\frac{df}{f}~\right]^{1/2}\,,
\end{equation}
where $S_{h}(f)$ is the noise PSD of the GW detector under consideration
(\ie Advanced LIGO~\cite{url:adLIGO_Sh_curve}, or the Einstein Telescope
  (ET)~\cite{Punturo:2010,Punturo2010b}).

\begin{figure}
\begin{center}
\includegraphics[width=\columnwidth]{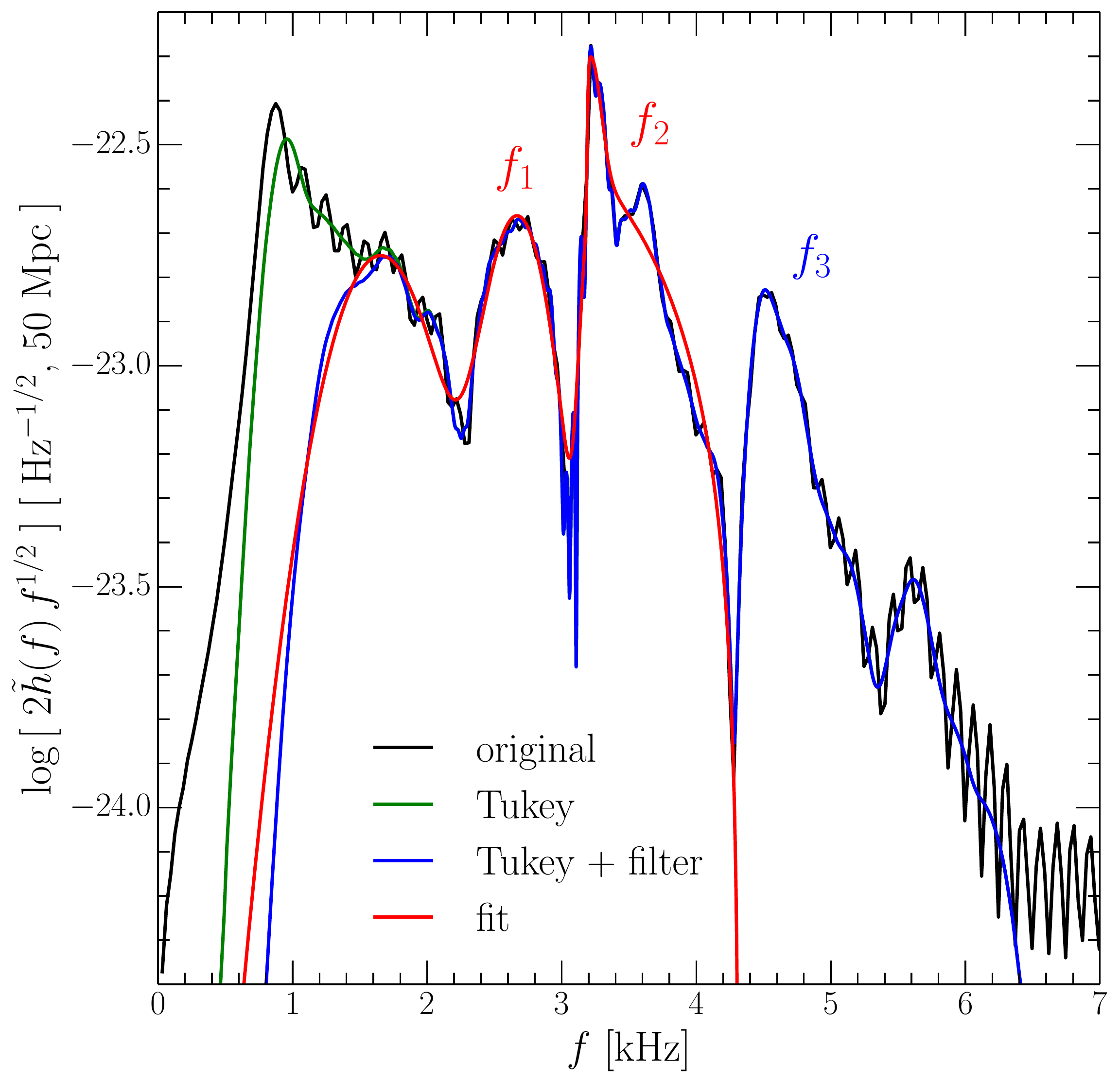}
\caption{\label{fig:PSD_fit} PSD $2\tilde{h}(f)f^{1/2}$ relative to the
  binary \mn{APR4-q10-M1375}. Indicated with a black line is the full
  PSD, while the green line shows the PSD after the application of a
  Tukey window, and the blue line the PSD filtered with a high-pass
  Butterworth filter. Finally, shown in red is the fit made to capture
  the peak frequencies $f_1$ and $f_2$.}
\end{center}
\end{figure}

\subsection{Spectral-properties detection}
\label{sec:spec-det}

An important aspect of our analysis consists in extracting from the PSD
of the post-merger signal those spectral features that can be correlated
with the physical properties of the binary system. Such an extraction
could be done by a simple visual inspection of the spectra as the peaks
are generally quite evident. Yet, doing so for computing the exact
position of the various frequencies would introduce an inevitable bias
that would be difficult to reproduce systematically. As a result,
following the spirit of Ref. \cite{Messenger2013}, we adopt a more
algorithmic approach in which the PSDs are first fitted using a
prescribed functional and then estimated from the fit. The identification
of the peak obtained in this way is a robust measure in the sense that it
matches very well the one that would be done ``by eye'' and guarantees
that other authors can reproduce it straightforwardly. More specifically,
our detection of the spectral properties can be summarized with the
following steps:

\begin{itemize}
\item[(1)] We first apply to the GW time series a symmetric time-domain
  Tukey window function with alpha parameter
  $0.25$~\cite{Harris78,McKechan2010}. This window affects the initial
  and final $\sim 4\ms$ of the waveform and is applied to compute a PSD
  without the artificial noise due to the cutting off of the waveform. A
  representative example of the application of this window is shown in
  Fig.~\ref{fig:PSD_fit}, which reports the PSD of the binary
  \texttt{APR4-q10-M1375}. Shown with a black line is the PSD relative to
  the full waveform, while the PSD computed after the Tukey window is
  shown with a green line. Clearly, no differences are visible for $f
  \gtrsim 1.5\,{\rm kHz}$, and a cutoff to the power is introduced for
  $f \gtrsim 6.5\,{\rm kHz}$.

\item[(2)] We apply a fifth-order high-pass Butterworth
  filter~\cite{Butterworth1930} with a cutoff frequency at
  $f_\mathrm{cut}\sim f_{_\mathrm{GW}} (t=-1500\,\Msun) +
  0.5\,\mathrm{kHz} \sim [(0.65$-$0.75) + 0.5]\,\mathrm{kHz}$. This is
  done to remove the contribution coming from the inspiral and thus
  provide a PSD which is easier to model analytically. The resulting
  filtered PSD is shown in Fig.~\ref{fig:PSD_fit} with a blue line.

\item[(3)] We fit the filtered PSD $\tilde{h}(f)$ obtained after step (2)
  with the analytic function
\begin{align}
\mathcal{S}(f) = &\phantom{+} 
A_{1}e^{-(f-F_{1})^2/W_{1}^{2}} + \mathcal{S}_2(f) \nonumber \\
&+A_{0}e^{-(f-F_{0})^2/W_{0}^{2}} \,,
\label{eq:fit}
\end{align}
where the first term captures the low-frequency peak at the merger, the
second term captures the high-frequency peak at HMNS, while the last term
is used to model the very-low frequency peak corresponding to the
inspiral. Although Ref.~\cite{Messenger2013} used a trapezoid function to
approximate the high-frequency peak in an ideal-fluid EOS, we use here a
combination of a trapezoid and a Gaussian function as it produces a
better match to the functional form of the PSD in the case of
nuclear-physics EOSs. The explicit form of this part is given by
\begin{equation}
\qquad \qquad \mathcal{S}_2(f) = A_{2G}e^{-(f-F_{2G})^2/W_{2G}^{2}} +A(f)~\gamma(f)\,,
\end{equation}
where
\begin{align}
A(f) \equiv 
&\frac{1}{2W_{2}}
 \bigl[~\left(A_{2b}-A_{2a}\right)\left(f-F_{2}\right) \nonumber \\
& \hskip 0.8cm + W_2\left(A_{2b}+A_{2a}\right)~\bigr]\,, \\
\gamma(f) \equiv 
& \left(~ 1+e^{-(f-F_{2}+W_{2})/s} ~\right)^{-1} \nonumber \\ 
& ~\times  \left(~ 1+e^{ (f-F_{2}-W_{2})/s} ~\right)^{-1}\,.
\end{align}

The 14 parameters, $A_1$, $F_1$, $W_1$, $A_{2a}$, $A_{2b}$, $F_2$, $W_2$,
$s$, $A_{2G}$, $F_{2G}$, $W_{2G}$, $A_0$, $F_0$, $W_0$, are computed via
a nonlinear least-squares fitting. An example of the resulting fit is
shown also in Fig.~\ref{fig:PSD_fit} with a red line. Clearly, the
fitting procedure has provided a very good representation of the low- and
high-frequency peaks. Similarly good results are obtained also with other
EOSs.
\item[(4)] Finally, we associate the two spectral features of the PSD
  which are our interest here, \ie $f_1$ and $f_2$, with the values of
  the coefficient $F_1$ and with the frequency average of the function
  $\mathcal{S}_2$\footnote{Because the notation in the literature has
    been evolving and a bit confusing, we remark that what we indicate as
    $f_2$ was marked as $f_\mathrm{peak}$ in Ref.~\cite{Bauswein2011} and
    as $f_2$ in Ref.~\cite{Stergioulas2011b}. This latter notation is
    becoming the standard one.}
\begin{align}
f_1 \equiv F_{1}\,, &&
f_2 \equiv \frac{\int ~\mathcal{S}_2(f) ~f~df}{\int ~\mathcal{S}_2(f) ~df}\,.
\end{align}

\end{itemize}

\section{Validation of the numerical results}
\label{sec:votnr}

As mentioned above, the purpose of this work is mainly to highlight the
relation between the spectral properties of the post-merger signal and
the physical properties of the merging binary. In this sense, here we are
not particularly interested to make sure that this relation is
quantitatively accurate; far more important to us is that our analysis
captures the correct qualitative behavior. Nevertheless, it is important
to assess how the results obtained also depend on choices that are under
our control, such as the grid resolution and the choice of the polytropic
index $\Gamma_{\rm th}$ for the thermal part of the EOS. While a
discussion of these aspects is rarely done in the literature, in what
follows we discuss both of these aspects in detail.

\begin{figure}
\begin{center}
\includegraphics[width=\columnwidth]{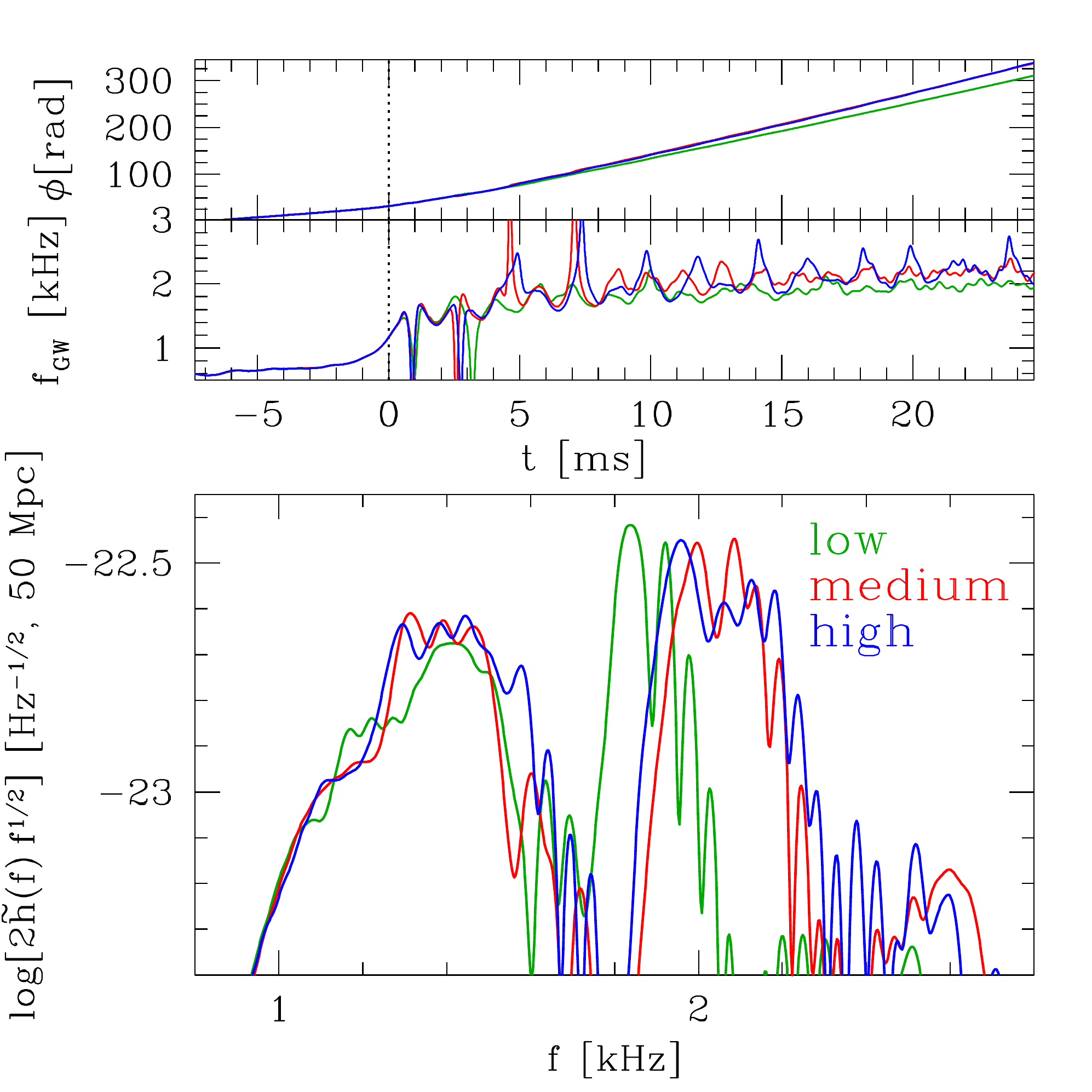}
\caption{\textit{Top panel:} evolution of the instantaneous GW phase
  $\phi$ for the binary \mn{GAM2-q10-M1400} when computed at low~(green),
  medium~(red), and high~(blue) resolution, respectively. \textit{Middle
    panel:} instantaneous frequency $f_{\mathrm{GW}}$ of the GW signal
  for the same binary and at the same resolutions. \textit{Bottom panel:}
  corresponding PSDs $2\tilde{h}(f)f^{1/2}$ for optimally oriented
  sources at a distance of $50\,{\rm Mpc}$. \label{fig:grid_res}}
\end{center}
\end{figure}

\subsection{Dependence on grid resolution}

Performing accurate measurements of the convergence order as the ones
presented in~\cite{Radice2013c} or in~\cite{Bernuzzi2011} is clearly not
possible when considering the very large number of binaries and EOSs
considered here. Hence, rather than determining the convergence order, we
have here explored whether our results are in a regime of consistency,
\ie the truncation error decreases with increasing
resolution~\cite{Rezzolla_book:2013}. While a much weaker requirement
than convergence, an overall consistency guarantees that the numerical
errors are bound and decreasing. We have therefore considered as a
representative example the evolution of the single-polytropic BNS
\mn{GAM2-q10-M1400} when performed at three different spatial resolutions
of $(\Delta h_5)_{\mathrm{low}}=0.175\,\Msun\simeq 258\,\mathrm{m}$ (low
resolution), $(\Delta h_5)_{\mathrm{med}}=0.150\,\Msun\simeq
221\,\mathrm{m}$ (medium resolution), and $(\Delta
h_5)_{\mathrm{high}}=0.125\,\Msun\simeq 185\,\mathrm{m}$ (high
resolution).

The results of this comparison are shown in Fig.~\ref{fig:grid_res},
where the top panel reports evolution of the GW phase, with the green,
red, and blue lines referring to the low, medium, and high resolutions,
respectively. Shown instead in the middle panel is the instantaneous
frequency $f_{\mathrm{GW}}$ of the GW signal. It is apparent that the
phase evolution is already well captured by the middle resolution, with
differences with respect to the high resolution that are $\Delta \phi
\simeq 0.9\,{\rm rad}$ at the end of simulation, that is, with a
fractional change $\Delta \phi/\phi \simeq 0.3\%$. Similarly, the
evolutions of the three instantaneous frequencies are essentially
indistinguishable up to the merger (\cf dotted vertical line), so that
the peak frequency $f_\mathrm{max}$ varies at most of $\simeq 12
\,\mathrm{Hz}$ (\ie $\simeq 1\%$) when going from the low to the high
resolution and with differences of $\lesssim 7\,\mathrm{Hz}$ (\ie $\simeq
0.6 \%$) between the medium and high resolutions. On the other hand, the
three instantaneous frequencies are clearly different after the
merger. This behavior, which cannot be removed by a simple time shift, is
a clear indication that the resolutions used are not sufficiently high to
guarantee that the instantaneous frequency is a robust quantity after the
merger.

The bottom panel of Fig.~\ref{fig:grid_res} shows a different type of
information as it focuses on the time-integrated spectral density
$2\tilde{h}(f)f^{1/2}$ for the three different resolutions and is relative
to a source at $50\,{\rm Mpc}$. A rapid inspection of the figure is
sufficient to conclude that while the low resolution may be sufficient to
capture the behavior of the low-frequency ($f_1$) peak around $1.4\,{\rm
  kHz}$, it does not provide a consistent description of the
high-frequency ($f_2$) peak around $2.1\,{\rm kHz}$. Fortunately,
however, the medium resolution is sufficient to obtain a reliable measure
of $f_2$; more precisely the differences in the estimates of the $f_1$
and $f_2$ frequencies between the medium and high resolution runs vary by
$\lesssim 20\,\mathrm{Hz}$ and $\lesssim 35\,\mathrm{Hz}$,
respectively. Because these uncertainties are below the ones at which our
results are valid, \ie $\sim 200\,{\rm Hz}$ (see discussion in
Sec.~\ref{sec:corr_f1}), we conclude that the medium resolution of
$(\Delta h_5)_{\mathrm{med}}=0.150\,\Msun$ is sufficient for our
scopes. As a side remark we note that this choice is also a forced
one. The results presented here, in fact, have been collected over more
than one year of calculations with an expenditure of $\sim3\times
10^{6}$ CPU hours. Computational costs larger than these are simply not
available to us at the moment.

\subsection{Dependence on the thermal component}

As discussed in Sec.~\ref{sec:EOS}, because we are not using here hot
nuclear-physics EOSs, the thermal effects are accounted for by hybrid
EOS in which an ideal-fluid contribution is added to the total
pressure. The choice of the corresponding adiabatic index
$\Gamma_\mathrm{th}$ is somewhat arbitrary, with only a
\emph{mathematical} constraint that it lays in the range $1 \leq
\Gamma_\mathrm{th} \leq 2$ (but see discussion
in~\cite{Rezzolla_book:2013}), while finite-temperature EOSs such as the
Shen EOS~\cite{shen98} suggest that values
$\Gamma_\mathrm{th}{\sim}1.6$-$2.0$ are
preferable~\cite{Bauswein:2010dn}. Because $\Gamma_\mathrm{th}$
regulates the amount of thermal pressure produced after the merger, and
hence the survival time of the HMNS, it is natural to ask how the
spectral properties of the post-merger signal depend on the value chosen
for $\Gamma_\mathrm{th}$.

\begin{figure}
\begin{center}
\includegraphics[width=\columnwidth]{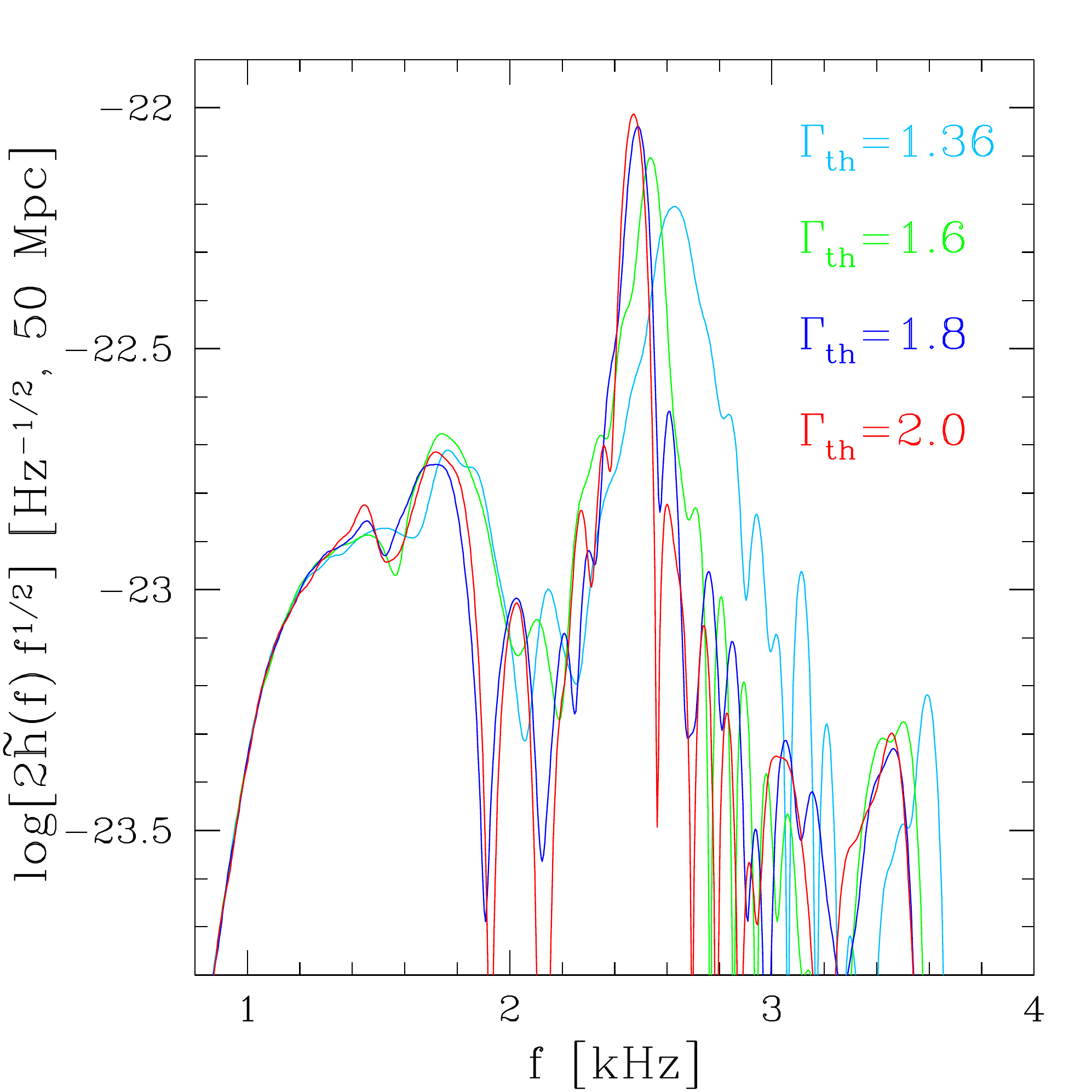}
\caption{Behavior of the PSDs $2\tilde{h}(f)f^{1/2}$ when the adiabatic
  exponent of the thermal contribution to the EOS is set to be
  $\Gamma_\mathrm{th}=1.36,1.6,1.8$, and $2.0$ for the binary
  \mn{H4-q10-M1300}. The PSDs have been computed for waveforms over the
  time interval $t \in [-1500,2800]\,\Msun \approx[-7.39,13.79] \,
  \mathrm{ms}$. \label{fig:gamma_th}}
\end{center}
\end{figure}

Figure~\ref{fig:gamma_th} collects the results of this analysis and in
particular it shows the PSDs relative to the binary \mn{H4-q10-M1300}
when four different values of the thermal component, \ie
$\Gamma_\mathrm{th}=1.36,1.6,1.8$, and $2.0$ are used. Note that because
smaller values of $\Gamma_\mathrm{th}$ favor the collapse to a black
hole, the PSDs are computed over the common interval in which a HMNS is
present, \ie $t \in [-1500,2800]\, \Msun \approx [-7.39,13.79]\,
\mathrm{ms}$.

\begin{figure*}
\begin{center}
\includegraphics[width=1.99\columnwidth,angle=0]{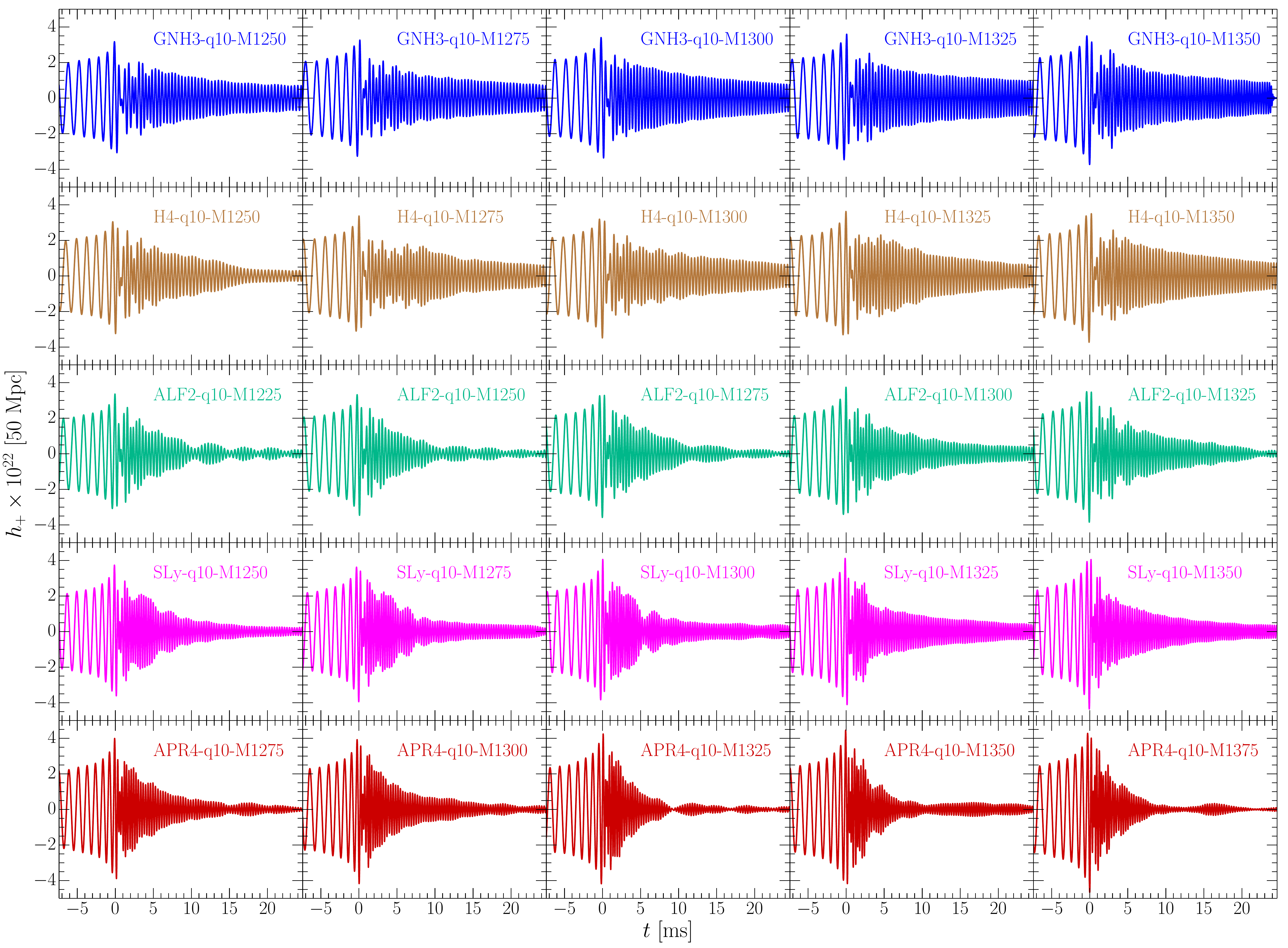}
\caption{Gravitational waveforms for all the binaries with equal masses
  and nuclear-physics EOSs as evolved at the reference medium
  resolution. Each row refers to a given EOS, while each column
  concentrates on a given initial mass. The different EOSs are
  distinguished by different colors, and we will adopt this color coding
  also for all the subsequent plots; more details on the various binaries
  are shown in Table \ref{tab:models}. 
\label{waveform_5x5}}
\end{center}
\end{figure*}

As one would expect given that the $f_1$ peak is built right after the
merger, its value is not much affected by the choice for
$\Gamma_\mathrm{th}$, with the differences that are $\lesssim
90\,\mathrm{Hz}$. The value of the $f_2$ peak, on the other hand, is
somewhat more sensitive for small values of $\Gamma_\mathrm{th}$, but is
remarkably robust for $\Gamma_\mathrm{th} \gtrsim 1.8$, that is, in the
most realistic range. To be more quantitative, the frequency of the $f_2$
peak varies by $\lesssim 50\,\mathrm{Hz}$ when $1.6 \le
\Gamma_\mathrm{th} \le 2.0$, with a maximum variation of $\lesssim
130\,\mathrm{Hz}$ when the more extreme case of $\Gamma_\mathrm{th} =
1.36$ is included. In the light of these results, but also to maintain a
consistency with the results presented in Ref.~\cite{Bauswein2011} and
with our single-polytrope sequence \texttt{GAM2-q10}, we have assumed
$\Gamma_\mathrm{th}=2$ as the standard thermal contribution in all of our
simulations.

\begin{figure*}
\begin{center}
\includegraphics[width=1.99\columnwidth,angle=0]{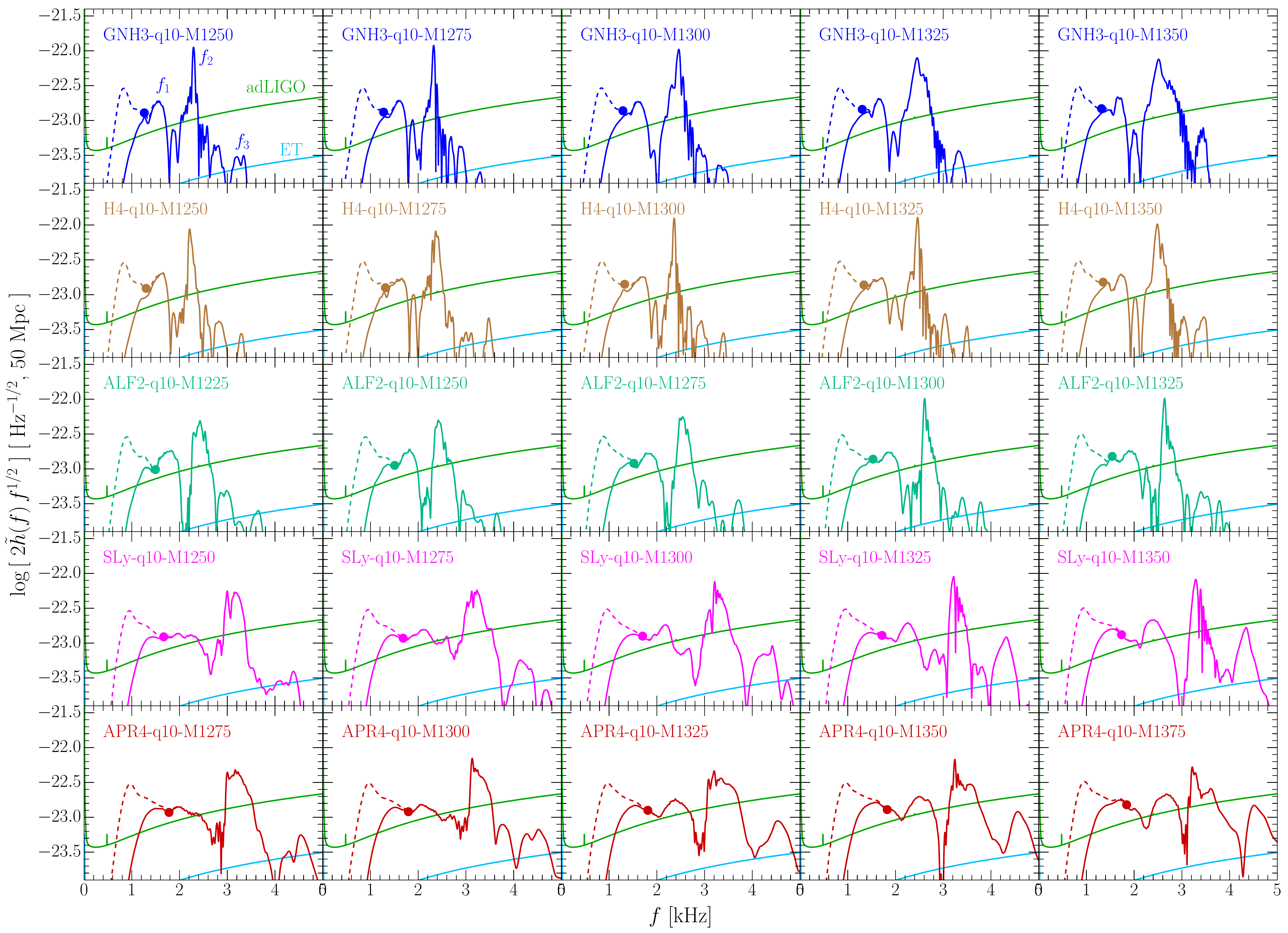}
\caption{PSDs $2 \tilde{h}(f) f^{1/2}$ for the equal-mass binaries with
  nuclear-physics EOS shown in Fig.~\ref{waveform_5x5}. Solid lines of
  different colors refer to the high-passed waveforms, while the dashed
  lines refer to the full waveforms. Indicated with colored circles are
  the various contact frequencies $f_{\rm cont}$, while the curves of
  Advanced LIGO and ET are shown as green and light-blue lines,
  respectively. \label{PSD_5x5_all}}
\end{center}
\end{figure*}

\section{Results}
\label{sec:results}

\subsection{Global overview}
\label{sec:go}

As discussed above and as listed in Table \ref{tab:models}, we have
computed a total of 32 binaries having six different EOSs and five
different masses. Before making a detailed discussion on the correlations
between the spectral features, \ie $f_1$ and $f_2$, and the physical
properties of the merging stars, it is useful to have a global overview
of the results so that it is easier to capture the general behaviors of
the HMNS and the most obvious features of the gravitational-wave signal.

Figure~\ref{waveform_5x5} collects all of the waveforms for the
equal-mass models with nuclear-physics EOSs as evolved at the reference
medium resolution. The waveforms are organized in a tabular form in which
each row refers to a given EOS, while each column concentrates on a given
initial mass. The different EOSs are distinguished by different colors
and we will adopt this color coding also for all the subsequent plots.

When scrolling through the different panels, some common features become
quite evident. First, the pre-merger signal and the post-merger one are
in frequency ranges that are considerably different. This is not
surprising given the significant difference in compactness in the system
before and after the merger, but points to the fact that the
phenomenology of the signal will necessarily have to be split into two
parts. Second, a transient period of $1$-$3\ms$ separates these two signals
and represents the time needed by systems to readjust itself from the
quasiperiodic inspiral over to the quasiperiodic rotation of the
HMNS. As we will comment later on, despite being very short, this
transient contains very important information and is where the power in
the $f_1$ peak is built. Third, the transient is followed by a period of
$\sim 10\ms$ during which oscillations at different frequencies are
present and which anticipates the stage in which the HMNS has reached a
``stationary'' state and is rotating at a frequency that increases only
slowly with time as a result of GW losses. Fourth, the amplitude of the
signal in the post-merger phase increases with mass. This is rather easy
to understand as larger masses will also lead to larger changes in the
quadrupole moment; the only exception to this behavior is represented by
the APR4 EOS, which is particularly soft at high densities and hence
yields an HMNS which rapidly becomes almost axisymmetric. Finally, the
amplitude in the post-merger decays and the rate at which this happens
depends inversely on the mass and on the stiffness, so that soft EOSs
show a large damping in amplitude (\eg binary \mn{APR4-q10-M1325}), while
stiff ones show a much smaller damping (\eg binary
\mn{GNH3-q10-M1300}). This too is easy to understand and reflects the
fact that HMNSs with larger masses (and not ultrasoft) will tend to
maintain the bar deformation for longer times.

Next, we can explore an equivalent global view of the GW signal, but when
measured in the frequency domain. More specifically, we show in
Fig.~\ref{PSD_5x5_all} the PSDs $2 \tilde{h}(f) f^{1/2}$ for the same
binaries considered in Fig.~\ref{waveform_5x5} and organized in the same
manner when the binaries are at a distance of $50\,{\rm Mpc}$.  Note that
the colored dashed lines refer to the full time series (\ie including the
inspiral), while the solid lines refer to the high-passed filtered data
as discussed in Sec.~\ref{sec:spec-det}. In each panel we indicate
with a filled circle the ``contact frequency'' $f_{\rm cont} \equiv
\mathcal{C}^{3/2}/(2\pi \bar{M})$~\cite{Damour:2012}, where $\mathcal{C}
\equiv \bar{M}/\bar{R}$ is the average compactness, $\bar{R} \equiv
(R_1+R_2)/2$, and $R_{1,2}$ are the radii of the nonrotating stars
associated with each binary~\cite{Takami:2014}. Finally, reported with
solid lines are the sensitivity curves of Advanced LIGO and ET (green and
light-blue lines, respectively).

Also in the frequency domain, a rapid scan of the panels allows one to
discern the most important features. First, and as discussed by several
authors~\cite{Shibata05d, Oechslin07b, Baiotti08, Bauswein2011,
  Bauswein2012, Hotokezaka2013c, Takami2014, Messenger2013}, all PSDs
show a clear and strong peak, \ie the $f_2$ peak, which, at these
distances, can be 1 order of magnitude or more above the sensitivity
curve of the Advanced LIGO detectors. This peak is clearly related to the
rotation of the bar-deformed HMNS and corresponds, in a corotating frame,
to a (quadrupolar) $\ell=m=2$ mode moving at a positive pattern speed in
the prograde direction~\cite{Stergioulas2011b}\footnote{As customary, the
  prograde direction is the direction of rotation of the HMNS as seen in
  an inertial frame.}. As we will comment later in
Sec.~\ref{sec:corr_f2}, this mode can be seen to correlate with a
number of properties of the stars comprising the binary, although this
dependence is different for different EOSs and is ``universal'' only at a
fixed mass.

All of the panels also show the presence of a low-frequency peak, \ie the
$f_1$ peak, which has already been discussed in detail in
Ref.~\cite{Stergioulas2011b}, where it was indicated as $f_-$. This peak
always has a power smaller than that of $f_2$ and it can happen that if
the EOS is particularly soft (\eg as for the binary \mn{APR4-q10-M1275})
or if the mass is particularly small (\eg as for the binary
\mn{SLy-q10-M1250}), it is hard to distinguish it from the
background. However, because the peak is also sitting in a region where
the sensitivity of detectors is higher, it will be detectable at these
distances with a SNR smaller but comparable to that of the $f_2$ peak
(\cf Table~\ref{tab:SNRs}). As remarked in~\cite{Takami:2014}, this peak
is is produced by the nonlinear oscillations of the two stellar cores
that collide and bounce repeatedly right after the merger. More
important, as we will comment later in Sec.~\ref{sec:corr_f1},
this mode correlates tightly with the stellar compactness $\mathcal{C}$
in a way that is essentially universal, that is independent of the
EOS.

In addition to the $f_1$ and $f_2$ peaks, the PSDs also show the presence
of an additional peak at frequencies higher than $f_2$ (see top left
panel of Fig.~\ref{PSD_5x5_all}). We have dubbed this peak as $f_3$ (in
Ref.~\cite{Stergioulas2011b} it was instead indicated as $f_+$) and its
value is approximated as $f_3 \sim 2f_2 - f_1$ with a precision of about
$10\%$. While equally interesting and potentially containing additional
information on the merging system, this peak is the one with the least
power of the three and is usually located at very high frequencies,
always below the sensitivity curve of Advanced LIGO. Hence, more
sensitive detectors, such as ET, will be needed to observe this spectral
feature even at moderate distances.

\subsection{On the origin of the $f_1$ and $f_3$ peaks}
\label{sec:f1_f3_orig}

It has so far been unclear what is the actual physical origin of the two
frequency peaks $f_1$ and $f_3$. It is possible to attribute $f_1$ to a
nonlinear interaction between the quadrupole and quasiradial
modes~\cite{Stergioulas2011b}; similarly, it is possible that $f_3$ is an
overtone or the result of the nonlinear interaction of the $f_2$ mode
with other nonquasiradial modes~\cite{Stergioulas2011b}. These
perturbative suggestions are given substance by the fact that the $f_2$
peak is, to first approximation, the average of the $f_1$ and $f_3$
frequencies, and it is well known that if a perturbed system has
eigenfrequencies $f_i$, the nonlinearity of the equations will also
produce modes at frequencies $f_i \pm f_j$ (see Sec.~28 of
Ref.~\cite{Landau-Lifshitz1}). On the other hand, the amplitudes of these
nonlinear couplings are usually found to be considerably smaller than the
originating eigenfrequencies (see the discussion in~\cite{Zanotti05}), and
our PSDs show instead that the amplitudes in the $f_1-f_3$ peaks vary by
a factor of few and not of orders of magnitude.

\begin{figure*}
\begin{center}
\includegraphics[width=2.0\columnwidth]{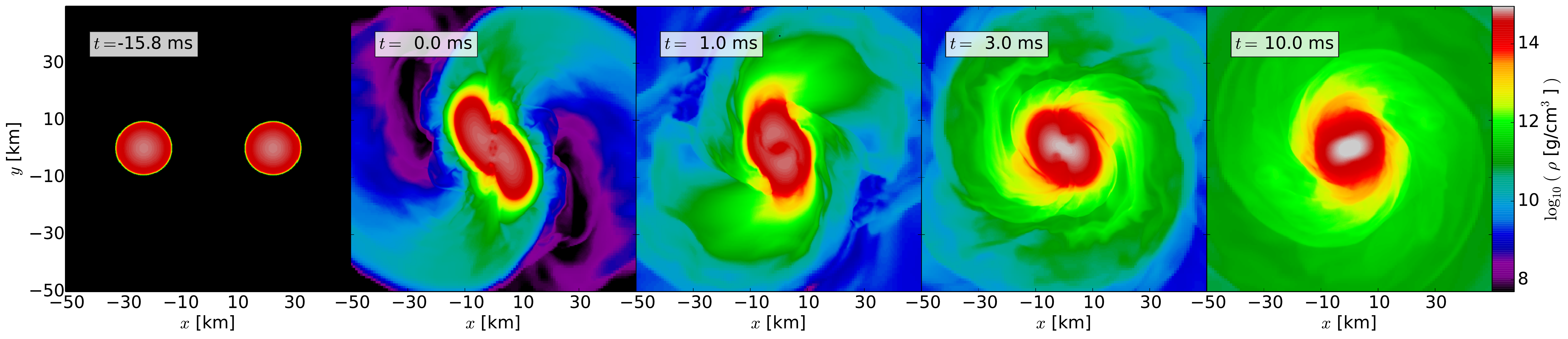}
\caption{Snapshots of the rest-mass density on the $(x,y)$ plane for the
  binary \mn{ALF2-q10-M1325}. From left to right, the panels refer to
  five characteristic times: the initial time, the time of the merger,
  the time right after the merger (\ie at $t=1.0\ms$), when the $\ell = m
  = 2$ deformation in the HMNS starts to develop (\ie at $t=3.0\ms$), and
  a later time (\ie at $t=10.0\ms$). Note that only in the last panel is
  the bar-deformed HMNS well defined and quasistationary.
 \label{fig:snaps_xy}}
\end{center}
\end{figure*}

On the other hand, a different interpretation is possible on the origin
of these modes. In this interpretation, which we suggest here, they are
simply produced by the GW emission due to the nonlinear oscillations of
the two stellar cores that collide and bounce repeatedly. Animations of
the few milliseconds following the instant when the stars get in contact,
in fact, show that the HMNS attains a quasistationary configuration with
a marked bar-mode deformation only $\sim 5\ms$ after the merger. On the
other hand, the object produced after the contact is far more irregular
and the two stellar cores collide and bounce repeatedly as a result of
the strong rotation and very high densities. This is shown in
Fig.~\ref{fig:snaps_xy} for the representative binary
\mn{ALF2-q10-M1325}. The figure contains four different panels
reproducing the rest-mass density on the $(x,y)$ plane at five
characteristic times: the initial time, the time of the merger, the time
right after the merger (\ie at $t=1.0\ms$), when the stellar core stops
oscillating and an $\ell = m = 2$ deformation in the HMNS starts to
develop (\ie at $t=3.0\ms$), and then when the bar-deformed HMNS (\cf
region in white) is well defined and with a quasistationary core (\ie at
$t=10.0\ms$).

Following this phenomenology, it is possible to build a mechanical toy
model, whose mathematical details are presented in
Appendix~\ref{appendix_a}, in which the object produced right after the
stellar contact is composed of an axisymmetric disk rotating rapidly at a
given angular frequency, say $\Omega(t)$, to which two spheres are
connected (\eg via a shaft) but are also free to oscillate via a spring
that connects them (see Fig.~\ref{fig:toy} in
Appendix~\ref{appendix_a}). In such a system, the two spheres will either
approach each other, decreasing the moment of inertia of the system, or
move away from each other, increasing the moment of inertia. Because the
total angular momentum is essentially conserved, the system's angular
frequency will vary between a minimum value $\Omega_1$ (corresponding to
the time when the two spheres are at the largest separation) and a
maximum value $\Omega_3$ (corresponding to the time when the two spheres
are at the smallest separation). The values of $\Omega_1$ and $\Omega_3$
depend nonlinearly on the properties of the system (\ie the mass and
radius of the disk, and the mass of the spheres) but are such that
$\Omega_2 = \tfrac{1}{2}(\Omega_1+\Omega_3)$, just as $f_2 \approx
\tfrac{1}{2}(f_1+f_3)$ in the PSDs we have computed. Stated differently,
the mechanical toy model considered here will rotate with an angular
frequency that is a function of time and bounded by $\Omega_1$ and
$\Omega_3$. Because the time spent at a given frequency is $\tau_{\Omega}
\equiv \Omega/(d\Omega/dt)$, more time is obviously spent at the
frequencies $\Omega(t)=\Omega_1$ and $\Omega(t)=\Omega_3$, where
$d\Omega/dt \simeq 0$. As a result, more power is expected to appear in
the GW signal at these frequencies, hence producing a low-frequency peak
around $\Omega_1$ and a high-frequency peak around $\Omega_3$. If
dissipative processes are present, \eg if the spring is not ideal and the
oscillations are damped, then the angular frequency will tend secularly
to $\Omega_2$, \ie $\Omega(t)=\Omega_2$ for ${t\to \infty}$ (\cf
Fig.~\ref{fig:toy_omega} below). As a result, most of the power in the PSD will
appear around $\Omega_2$, with two main sidebands at $\Omega_1$ and
$\Omega_3$. Conversely, if dissipative processes are not present, then
the GW signal will have contributions at frequencies $\Omega_2$ and at
its overtones $\Omega_n \simeq (n/2)\Omega_2$, such that $\Omega_2 \simeq
\tfrac{1}{2} (\Omega_1 + \Omega_3)$. (Note that in the presence of
dissipative processes a $\simeq$ sign is needed in the estimate of
$\Omega_2$ because the asymptotic frequency is only approximately the
average of $\Omega_1$ and $\Omega_3$; this is shown in the middle panel
of Fig.~\ref{fig:toy_omega} and reflects the fact that the system is not
perfectly balanced.) Overall, and as we will discuss in more detail in
Appendix~\ref{appendix_a}, this toy model can therefore account for both
the presence of the main peak $f_2$ and for the two equally distant 
sidebands at $f_1$ and $f_3$.

\begin{figure*}
\begin{center}
\includegraphics[width=1.99\columnwidth,angle=0]{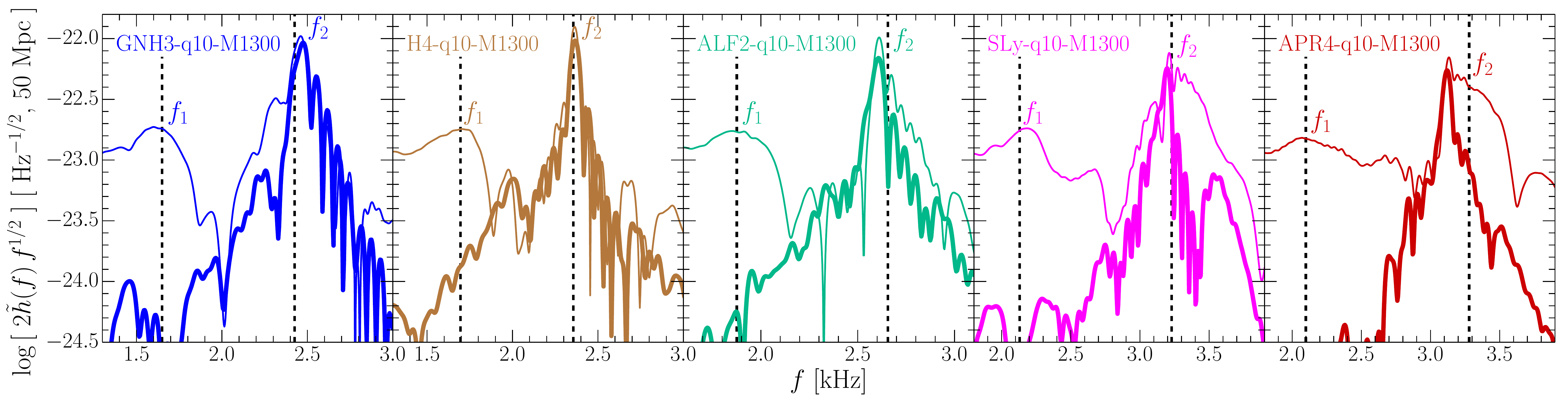}
\caption{Selected PSDs $2 \tilde{h}(f) f^{1/2}$ relative to binaries with
  average mass $\bar{M}=1.30\,\Msun$. The thin solid lines refer to the
  PSDs already shown in Fig.~\ref{PSD_5x5_all}, while the thick solid
  lines refer to a time interval excluding the first $3\,\rm{ms}$ after
  the merger, \ie $t \in [609,5000]\, \Msun\approx[3.00,24.63]\ms$. The
  vertical dashed lines indicate the positions of the $f_1$ and $f_2$
  frequencies and highlight that the power in the $f_1$ peak is built
  right after the merger and essentially disappears if the first
  $3\,\rm{ms}$ are excluded. \label{PSD_5x5_HMNS}}
\end{center}
\end{figure*}

There is a simple way of testing whether these modes are coming just from
the immediate post-merger phase or are produced on longer time scales in
terms of nonlinear couplings. This is shown in Fig.~\ref{PSD_5x5_HMNS},
which reports again the PSDs for the five EOSs and for a representative
value of the mass, \ie $\bar{M}=1.30\,\Msun$. Thin solid lines of
different colors show the same PSDs as in Fig.~\ref{PSD_5x5_all}, with
the two vertical dashed lines marking the positions of the peak
frequencies $f_1$ and $f_2$. Shown instead with thick solid lines of the
same colors are the PSDs when the waveforms are restricted to the
interval $t \in [609,5000]\,\Msun\approx[3.00,24.63]\ms$, that is, when
the first $3\ms$ after the merger are cut from the
time series. Remarkably, in this case the $f_1$ and $f_3$ peaks
essentially disappear, while the $f_2$ peaks remain very strong and
without considerable changes in frequency apart for the very soft
EOSs. We find this result a convincing validation of the correctness of
the toy model and a strong evidence that most of the power in the $f_1$
and $f_3$ peaks is built essentially over $2$-$3\ms$ after the merger.

We should also note that when the dominant contribution from the initial
$f_1$ and $f_3$ peaks is removed, and hence one is able to measure the
power produced by the long evolution of the HMNS, smaller peaks do appear
on either side of $f_2$, and they are close to the $f_1$ or $f_3$
frequencies. It is then possible that \emph{these} smaller-amplitude
peaks represent the manifestation of the nonlinear couplings mentioned
above and are therefore carriers of interesting information on the
properties of the HMNS. Clearly, more work is needed to validate these
results and explore the long-term spectrum of the HMNS.

\begin{figure}
\begin{center}
\includegraphics[width=0.75\columnwidth]{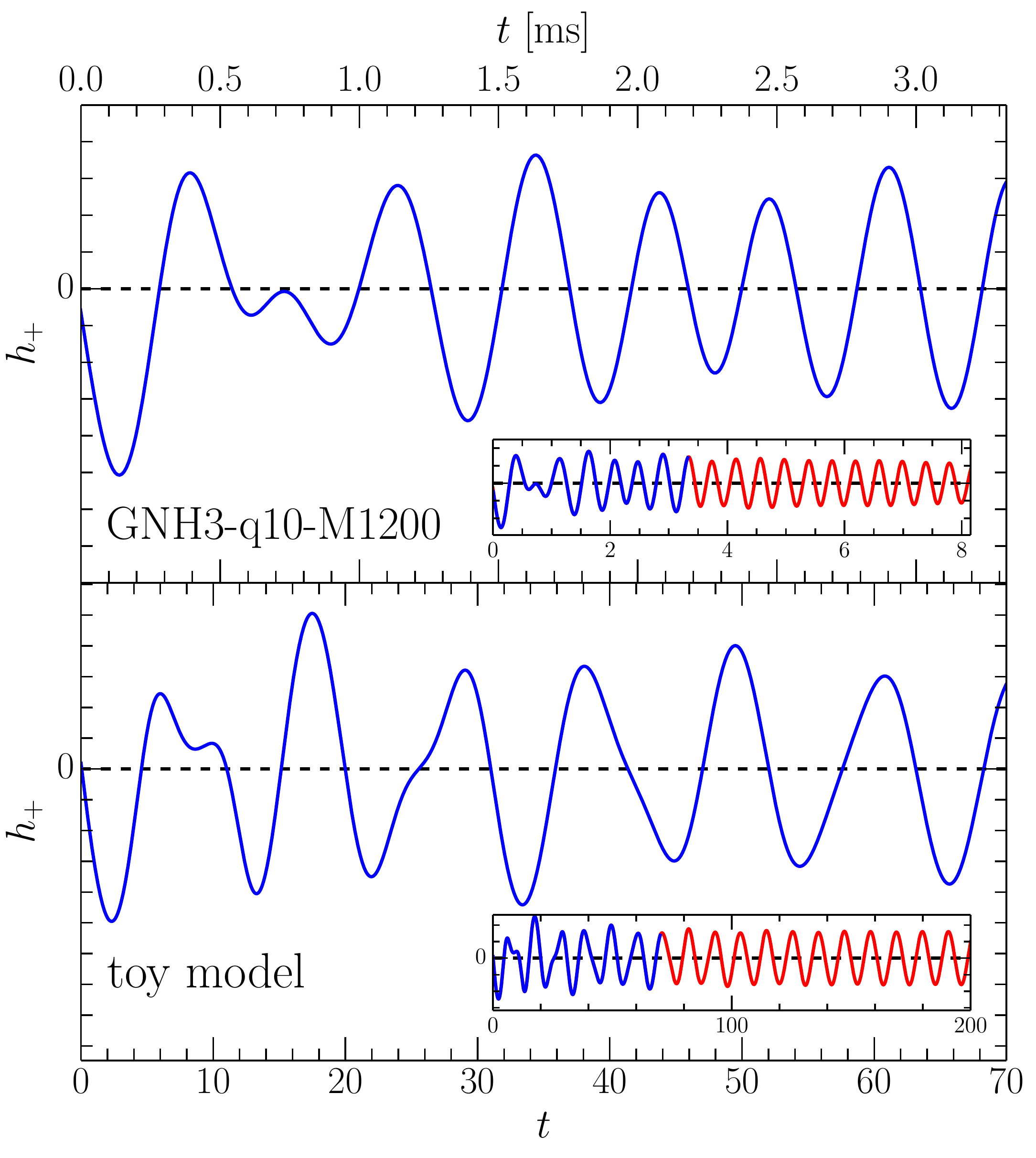}
\caption{\textit{Top panel:} Full numerical-relativity strain in the $+$
  polarization computed for the binary \mn{GNH3-q10-M1300} for the first
  $3.3\,{\rm ms}$ after the merger. \textit{Bottom panel:} Strain in the $+$
  polarization as computed from the mechanical toy model discussed in
  Appendix~\ref{appendix_a}. The similarities are quite remarkable,
  especially when considering that the signal shown is representative of
  the most complex part of the full post-merger signal (\cf
  insets). \label{fig:h_plus_cf}}
\end{center}
\end{figure}

Another concrete indication that the toy model provides a good
description of the dynamics right after the merger is offered by
Fig.~\ref{fig:h_plus_cf}, whose top panel shows the full
numerical-relativity strain in the $+$ polarization as computed for the
binary \mn{H4-q10-M1300} for the first $3.3\,{\rm ms}$ after the
merger. This is to be compared with the GW strain in the $+$ polarization
as computed from the mechanical model (see Appendix~\ref{appendix_a} for
details). The similarities are quite remarkable, especially when
considering the crudeness of the toy model and the fact that the signal
shown is the most complex one portion of the full post-merger signal,
which is far more regular once the HMNS has reached a stationary state
(\cf insets). The ability of the toy model to describe the GW emission in
the merger phase depends on how well the two stellar cores can be
described as two isolated spheres oscillating quasiharmonically. This is
obviously a rather good approximation for stiff EOSs, but the toy model
can provide similarly good qualitative agreements essentially for all the
waveforms shown in Fig.~\ref{waveform_5x5}. Hence, at least in principle,
the toy model provides a convenient tool to improve the description of
the post-merger signal proposed in Ref.~\cite{Clark2014}, possibly
opening the way to the fully analytic construction of the post-merger
signal for template building.

\subsection{Correlations of the $f_\mathrm{max}$ frequency}
\label{sec:corr_fmax}

In this section and in the two following it, we will discuss in more
detail how to use some of the frequencies in the GW signal discussed
above to extract physical information on the stars comprising the binary
and hence on their EOS. 

\begin{figure}
\begin{center}
\includegraphics[width=\columnwidth]{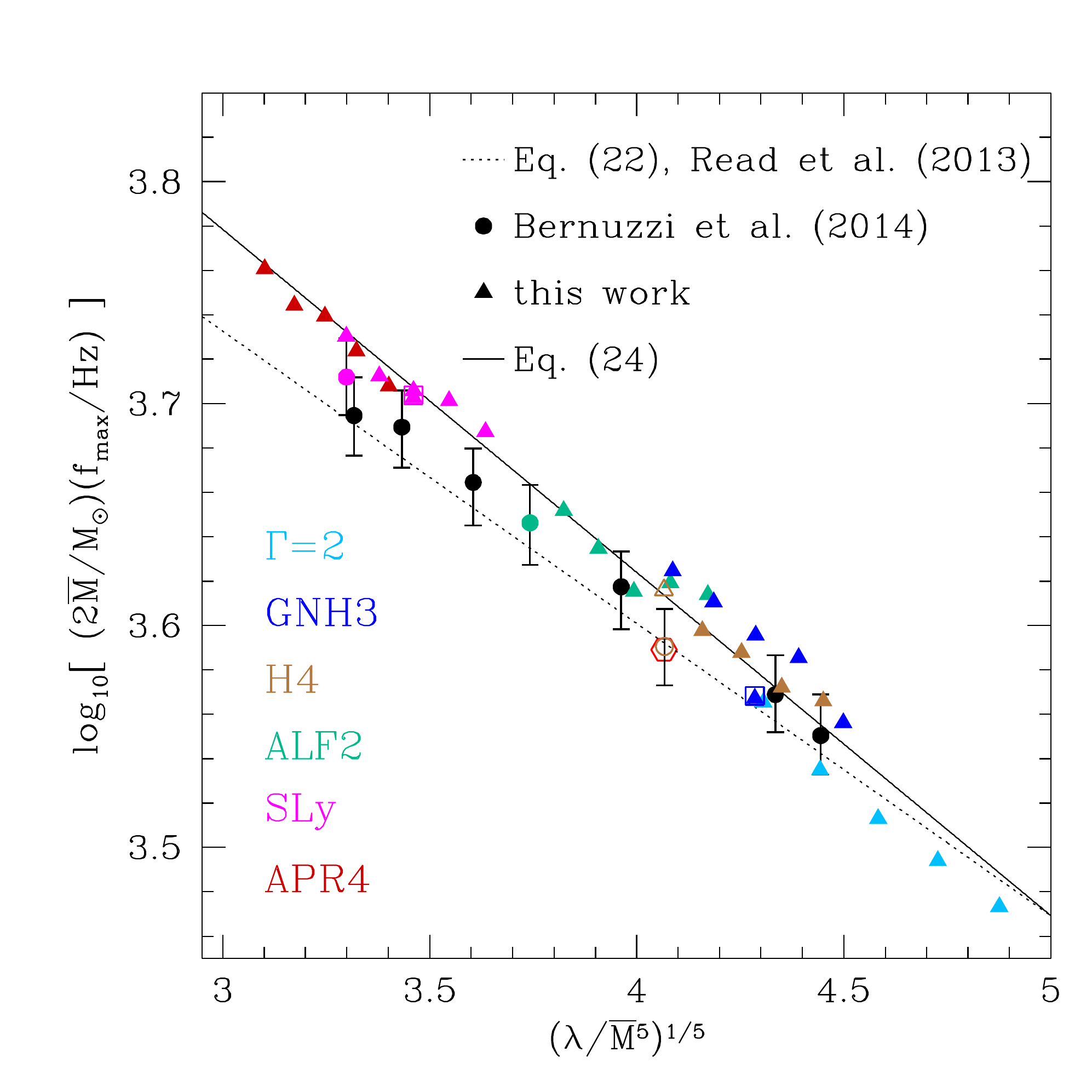}
\caption{Mass-weighted frequencies at amplitude maximum $f_\mathrm{max}$
  shown as a function of the dimensionless tidal deformability
  $\lambda/\bar{M}^5$. Filled circles refer to the data
  from~\cite{Bernuzzi2014}, while colored triangles indicate the data
  from our binaries (triangles in boxes refer to the unequal-mass
  binaries~\mn{GNH3-q09-M1300} and \mn{SLy-q09-M1300}). The dotted line
  shows the relation suggested in~\cite{Read2013} [\ie
    Eq.~\eqref{eq:read2013}], while the solid line represents the best
  fit to our data [\ie Eq.~\eqref{eq:newfit}]. Note that the systematic
  differences between circles and triangles are due to the small differences
  in the definition of the time of merger; indeed the red hexagon
  represents the new position to which the value for $\bar{M}f_{\rm max}$
  (empty brown triangle) moves to when the difference in the time of
  merger is taken into account and is very close to the one in
  Ref.~\cite{Bernuzzi2014} (empty brown circle).
\label{fig:fp_f1}}
\end{center}
\end{figure}

We will start by considering the frequency at peak amplitude $f_{\rm
  max}$ that, in contrast with the $f_1-f_3$ frequencies, is an
instantaneous frequency and not a feature of the PSDs. We then recall
that when restricting the analysis to BNSs having stars with equal masses
and set to be $\bar{M}=1.35\,\Msun$, Read et al.~\cite{Read2013} have
recently found that $f_\mathrm{max}$ can be effectively characterized
only in terms of the tidal deformability $\lambda$ through a relation
that is essentially independent of the EOS and was found to be expressed
by the relation
\begin{equation}
\label{eq:read2013}
\log_{10} \left(\frac{f_\mathrm{max}}{\mathrm{Hz}}\right)= 
3.69652 - 0.131743\, \Lambda^{1/5}\,,
\end{equation}
where the dimensionless tidal deformability $\Lambda$ is defined as
\begin{equation}
\Lambda \equiv \frac{\lambda}{\bar{M}^5} =
\frac{2}{3} \bar{k}_2 
\left(\frac{\bar{R}}{\bar{M}}\right)^5\,,
\end{equation}
and $\bar{k}_2$ is the $\ell=2$ dimensionless tidal Love number. [The
  empirical relation~\eqref{eq:read2013} is shown as a dotted line in
  Fig.~\ref{fig:fp_f1}.] Already in Ref.~\cite{Read2013} it was noted
that this behavior is reminiscent of the so-called universal
``I-Love-Q'' relations~\cite{Yagi2013a, Yagi2013b} that have been pointed
out for isolated neutron stars and have since been the subject of intense
research~\cite{Urbanec2013, Doneva2014a, Haskell2014, Pappas2014,
  Chakrabarti2014, Stein2014, Maselli2013, Chatziioannou2014, Sham2014a,
  Pani2014, Doneva2014b, Yagi2014, Sham2014b}. A similar behavior has
been found also within an effective-one-body (EOB) description of the
tidal effects in BNSs~\cite{Bernuzzi2014}. More specifically, the EOB
analytic approach has revealed quasiuniversal relations of the
mass-rescaled GW frequency and of the binding energy at the time of
merger when expressed as functions of $\ell=2$ dimensionless tidal
coupling constant $\kappa^T_2$, which is related to tidal deformability
by $\lambda=16 \kappa^T_2 \bar{M}^5/3$ in an equal-mass case. Together
with semianalytic calculations, Ref.~\cite{Bernuzzi2014} also reported
the results of numerical-relativity simulations of equal-mass BNSs for
nine different EOSs finding both a good match with the results
of~\cite{Read2013} and with those of the EOB approximation (the data
relative to the numerical simulations in Ref.~\cite{Bernuzzi2014} are
shown with filled circles in Fig.~\ref{fig:fp_f1}).

Because the numerical-relativity results both of Ref.~\cite{Read2013} and
of Ref.~\cite{Bernuzzi2014} were restricted to the analysis of equal-mass
systems with average mass $\bar{M}=1.35\,\Msun$, it is interesting to
reconsider these quasiuniversal relations when also the mass of the
system is allowed to vary. Of course this was already explored
in~\cite{Bernuzzi2014} within the EOB approximation, but we can now
extend it also to a fully nonlinear regime.

A summary of the correlation of the $f_{\rm max}$ frequency with the
tidal deformability is shown in Fig.~\ref{fig:fp_f1}, where we indicate
with a dotted line the empirical relation suggested in~\cite{Read2013}
[\ie Eq.~\eqref{eq:read2013}], with filled circles the data
from~\cite{Bernuzzi2014}, and with colored triangles the data from our
binaries (triangles in boxes refer to the unequal-mass
binaries~\mn{GNH3-q09-M1300} and \mn{SLy-q09-M1300}). Clearly, also the
results presented here support the universality of $f_\mathrm{max}$ as a
function of the dimensionless tidal deformability $\Lambda$. At the same
time, since expression~\eqref{eq:read2013} does not have any information
about the mass of the system $\bar{M}$, this can be introduced rather
simply through the expression
\begin{equation}
\label{eq:newfit}
\log_{10} \left(\frac{f_\mathrm{max}}{\mathrm{Hz}}\right) \approx
 4.2423  - 0.1546\, \Lambda^{1/5} - \log_{10}
 \left(\frac{2\bar{M}}{\Msun}\right)
\,.
\end{equation}
Stated differently, the universality is really in the quantity $\bar{M}
f_{\rm max}$, as already pointed out in
Ref.~\cite{Bernuzzi2014}. Expression~\eqref{eq:newfit} is shown as a
solid line in Fig.~\ref{fig:fp_f1} and provides a very good fit of the
data, with a maximum relative difference between the simulations and
fitted values that is only $\sim 0.7\%$.

Before concluding this section we should remark about the amount of
scattering in the numerical data. There could be a number of reasons
behind this scatter (\eg different codes, different initial data, etc.)
but we believe that the largest source of difference is systematic and
comes from a slightly different definition of the merger time. For
example, while we define it to be the time of the first maximum in the
$\ell=m=2$ mode amplitude $|h|=( {h_+}^2 + {h_{\times}}^2 )^{1/2}$,
Ref.~\cite{Read2013} (and possibly Ref.~\cite{Bernuzzi2014}) defined it
as the time with the maximum amplitude of $|{h_+}|$. Indeed, if we adopt
the same definition as in~\cite{Read2013,Bernuzzi2014}, then the
systematic difference disappears. This is illustrated with the binary
\mn{H4-q10-M1350}, for which we use the same initial data as in
Ref.~\cite{Bernuzzi2014}; the red hexagon represents the new position to
which the value for $\bar{M}f_{\rm max}$ (empty brown triangle) moves 
when the difference in the time of merger is taken into account and
clearly the new value corresponds closely to the one in
Ref.~\cite{Bernuzzi2014} (empty brown circle).

\subsection{Correlations of the $f_1$ frequency}
\label{sec:corr_f1}

\begin{figure}
\begin{center}
\includegraphics[width=\columnwidth]{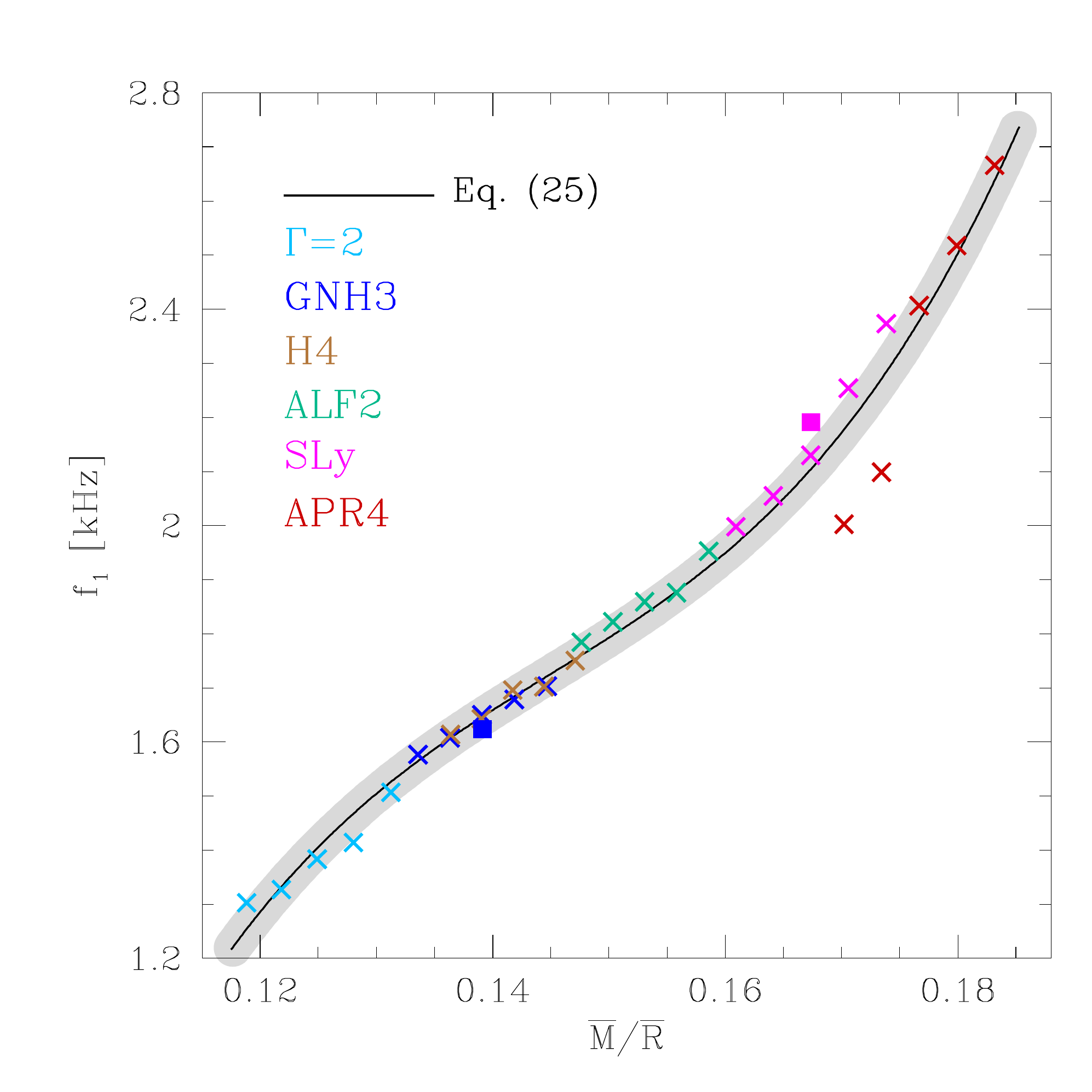}
\caption{Values of the low-frequency peaks $f_1$ shown as a function of
  the stellar compactness for the six different EOSs considered; note the
  universal behavior exhibited also by unequal-mass binaries (filled
  squares). Shown as a solid black line is the cubic fit, while the grey
  band is the estimate of the total errors (\cf left panel of Fig.~2
  of~\cite{Takami:2014}).
 \label{fig:corr_f1}}
\end{center}
\end{figure}

As already discussed in Ref.~\cite{Takami:2014}, possibly one of the most
interesting aspects of our spectral analysis is that there is a very
clear correlation between the low-frequency peak $f_1$ and the stellar
compactness ${\mathcal C}\equiv \bar{M}/\bar{R}$. This is shown in
Fig.~\ref{fig:corr_f1}, which reports the values of the $f_1$ frequencies
plotted as a function of $\mathcal{C}$ for the various EOSs. Each cross
refers to a given mass and the frequencies grow with mass; \ie for each
EOS the smallest $f_1$ frequency corresponds to the smallest mass and the
largest frequency to the largest mass. Also shown as a shaded grey band
is the estimate of the total error, which is effectively dominated by the
fitting procedure of the PSD, since the average numerical error from the
simulations is estimated to be $0.06$ kHz, while the average uncertainty
in the fitting procedure of the PSD is of $0.2$ kHz (see also
Ref.~\cite{Messenger2013}).

The behavior of the low-frequency peak is remarkably consistent with a
simple cubic polynomial of the type
\begin{equation}
\frac{f_1}{\mathrm{kHz}} \approx a_0 + a_1~{\mathcal C} +
a_2~{\mathcal C}^2 + a_3~{\mathcal C}^3 \,.
\label{eq:empirical_f1}
\end{equation}
where 
\begin{align}
&a_0 = -22.0717 \pm 6.477 \,, 
&a_1 = 466.616 \pm 131.2 \,, \nonumber\\
&a_2 = -3131.63 \pm 878.7 \,, 
&a_3 = 7210.01 \pm 1947 \,.
\label{eq:empirical_f1_coeffs}
\end{align}
The fit yields then a $\chi$-squared of $0.09$, with a fitting
uncertainty $\lesssim 0.06$~kHz and a maximum difference between data
from the simulations and the polynomial function that is $\sim
0.17\,\mathrm{kHz}$. The very tight fit in Fig.~\ref{fig:corr_f1} is
somewhat spoiled by the behavior of two binaries with the APR4 EOS,
namely, \mn{APR4-q10-M1275} and \mn{APR4-q10-M1300}. The reason behind
the large scatter in these two models is rather simple and is due to the
fact that for these binaries the power in the $f_1$ peak is very small
because the EOS is very soft (\cf Fig.~\ref{PSD_5x5_all}). Stated
differently, the softness of the EOS and the smallness of the mass in
these binary prevent the large bounce of the stellar cores after the
first collision. As a result, the GW emission at low frequencies is
considerably reduced. Excluding these two points from our data set of the
$f_1$ frequencies would reduce the maximum difference from the fitting
expression~\eqref{eq:empirical_f1} to be only $\sim 0.09\,\mathrm{kHz}$.

We should note that the $f_\mathrm{max}$ and $f_1$ frequencies are
clearly distinct, both numerically (the actual values in Hz are
different) and physically (while the first is an instantaneous frequency,
the second is a feature of the PSD). Yet, these frequencies tend to blend
in the limit of very soft EOSs and small masses. For example, the values
of $f_\mathrm{max}$ and $f_1$ have a difference of $\lesssim 0.3\%$ and
$\lesssim 0.07\%$ for the binaries \mn{APR4-q10-M1275} and
\mn{APR4-q10-M1300}, respectively.

A concluding remark: we have shown that also the $f_1$ frequency exhibits
a quasiuniversal behavior with the stellar compactness, and hence that,
together with the $f_{\rm max}$ frequency, it can be added to the
increasingly large set of stellar properties that have an EOS-independent
behavior. Yet, this does not add much to our understanding of why such a
behavior should take place in nature. Two papers have recently
proposed interesting explanations for this behavior, suggesting that
universality arises as an emergent approximate symmetry~\cite{Yagi2014},
or because of an almost incompressible behavior of compact stars at high
densities~\cite{Sham2014b}. Interestingly, whatever leads to the $f_1$
universal behavior cannot be attributed to an emergent symmetry (which
is definitely lost at the merger) or to an incompressible behavior
(large shocks are produced at the merger). Rather, it suggests that the
universal behavior is preserved also in the absence of symmetries or
when large compressions take place.

\begin{figure}
\begin{center}
\includegraphics[width=\columnwidth]{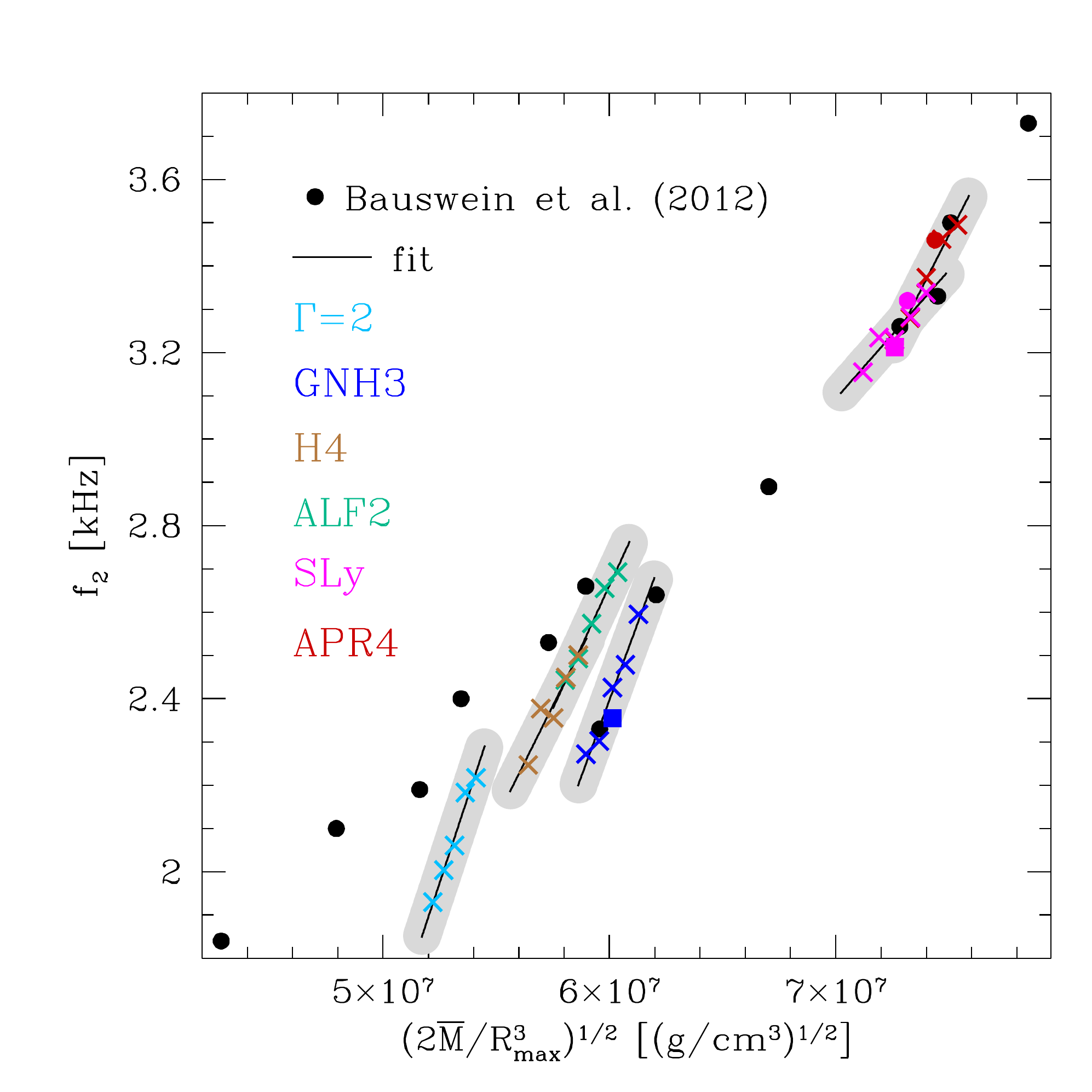}
\caption{Values of the $f_2$ peak frequencies shown as a function of the
  pseudoaverage rest-mass density $\left(2\bar{M}/R^3_{\rm
    max}\right)^{1/2}$, where colored crosses and squares refer to
  equal- and unequal-mass binaries, respectively. Shown as solid black
  lines are the linear fits within each EOS, while the grey band is the
  estimate of the total errors. We also report as filled circles the
  results presented in Ref.~\cite{Bauswein2011} for $\bar{M}=1.35\,\Msun$
  binaries, where we use magenta and red circles for the binaries that
  correspond to our models \mn{SLy-q10-M1350} and \mn{APR4-q10-M1350},
  respectively.
\label{fig:corr_f2} }
\end{center}
\end{figure}
%
\begin{figure}
\begin{center}
\includegraphics[width=\columnwidth]{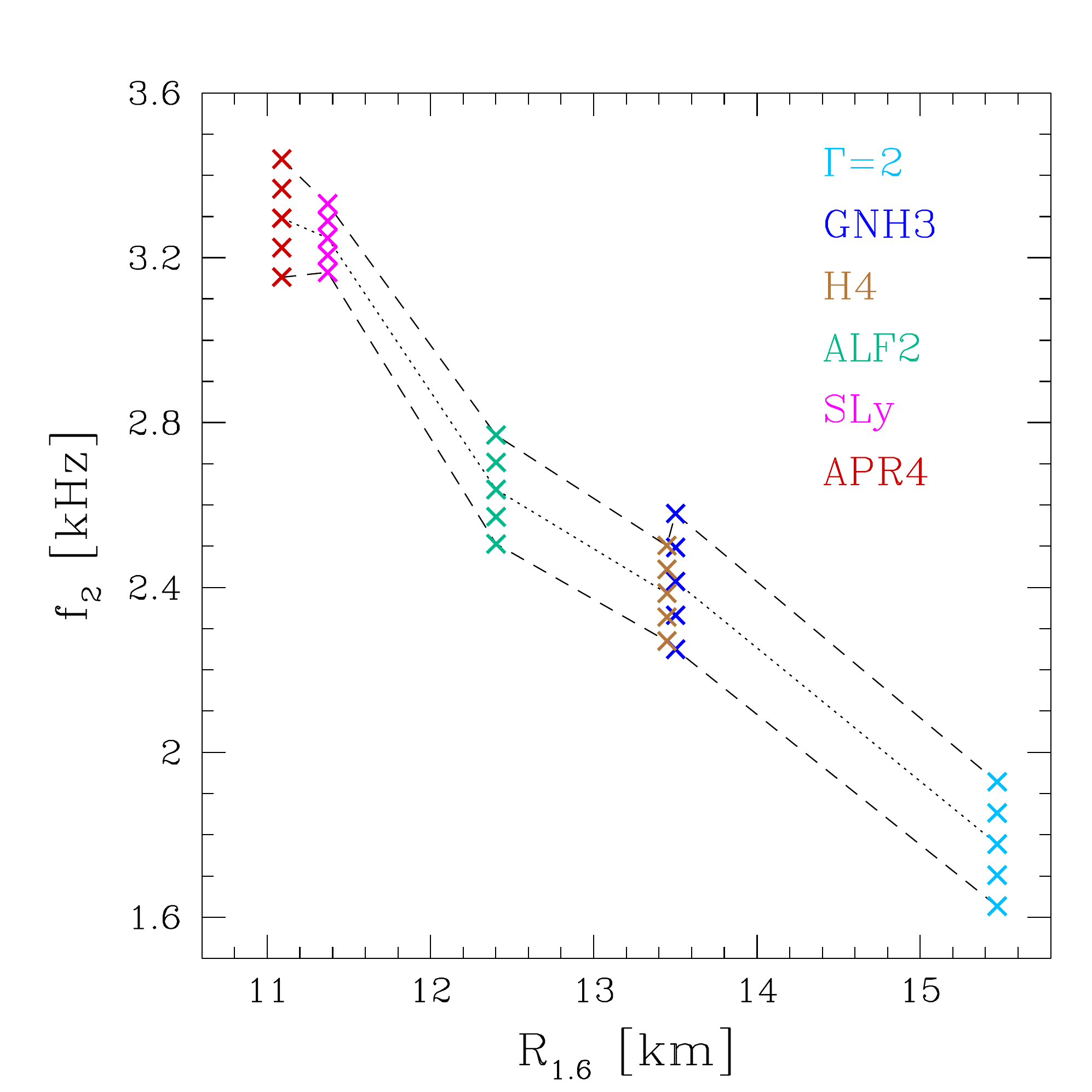}
\caption{Values of the $f_2$ peak frequencies shown in terms of
  $R_{1.6}$, the radius of a nonrotating NS with
  $\bar{M}=1.6\,M_\odot$. The various crosses refer to the $f_2$
  frequencies at masses $\bar{M}/\Msun=1.250, 1.275, 1.300, 1.325$, and
  $1.350$; these values are either available directly, or interpolated
  from the mass values used in the simulations. The dashed and dotted
  lines connect, respectively, the largest, the smallest, and the
  intermediate masses, showing that it is difficult to find a bilinear
  fitting across different EOSs. \label{fig:corr_f2_R16}}
\end{center}
\end{figure}
%
\begin{figure*}
\begin{center}
\includegraphics[width=\columnwidth]{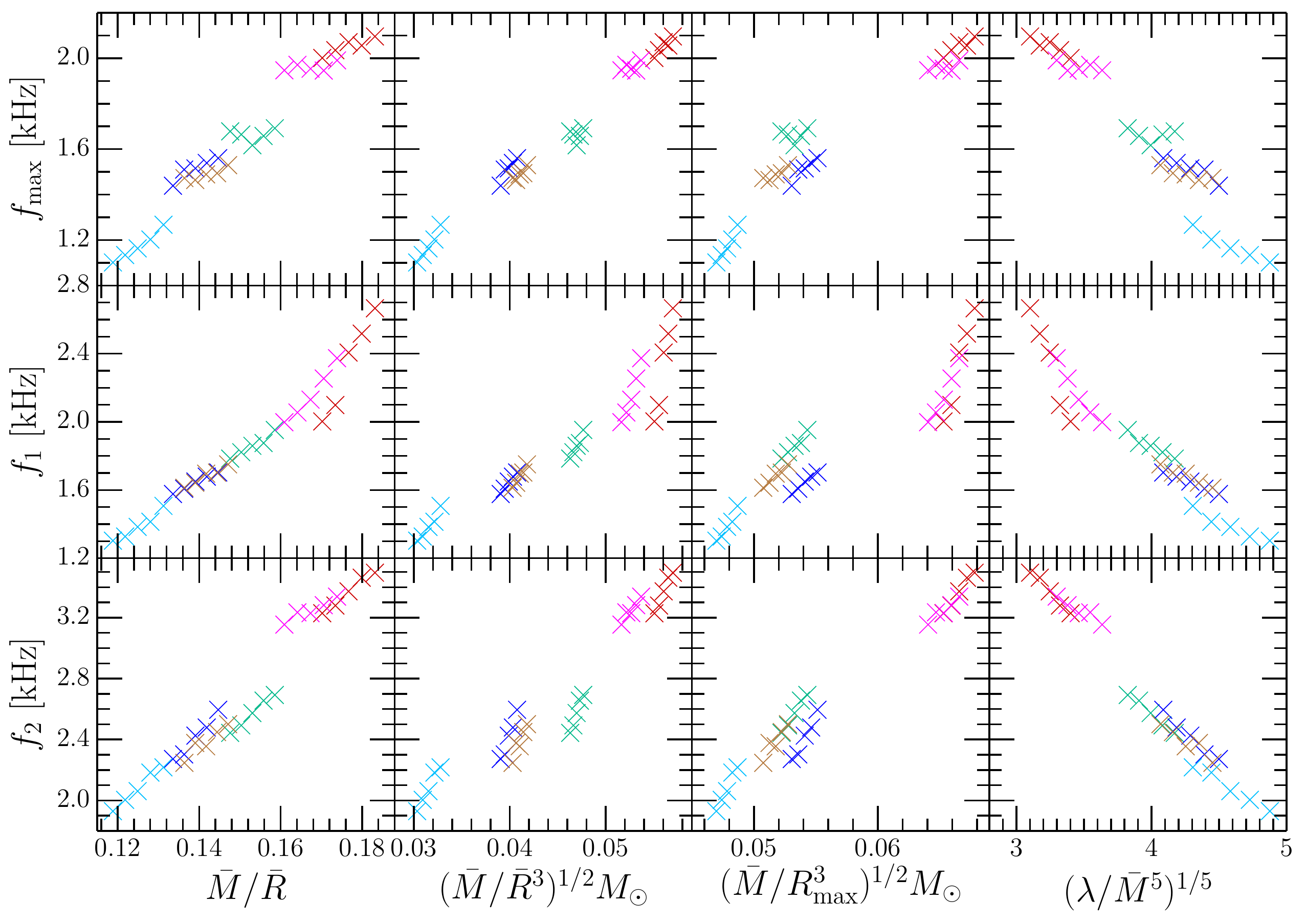}
\hskip 0.25cm
\includegraphics[width=1.025\columnwidth]{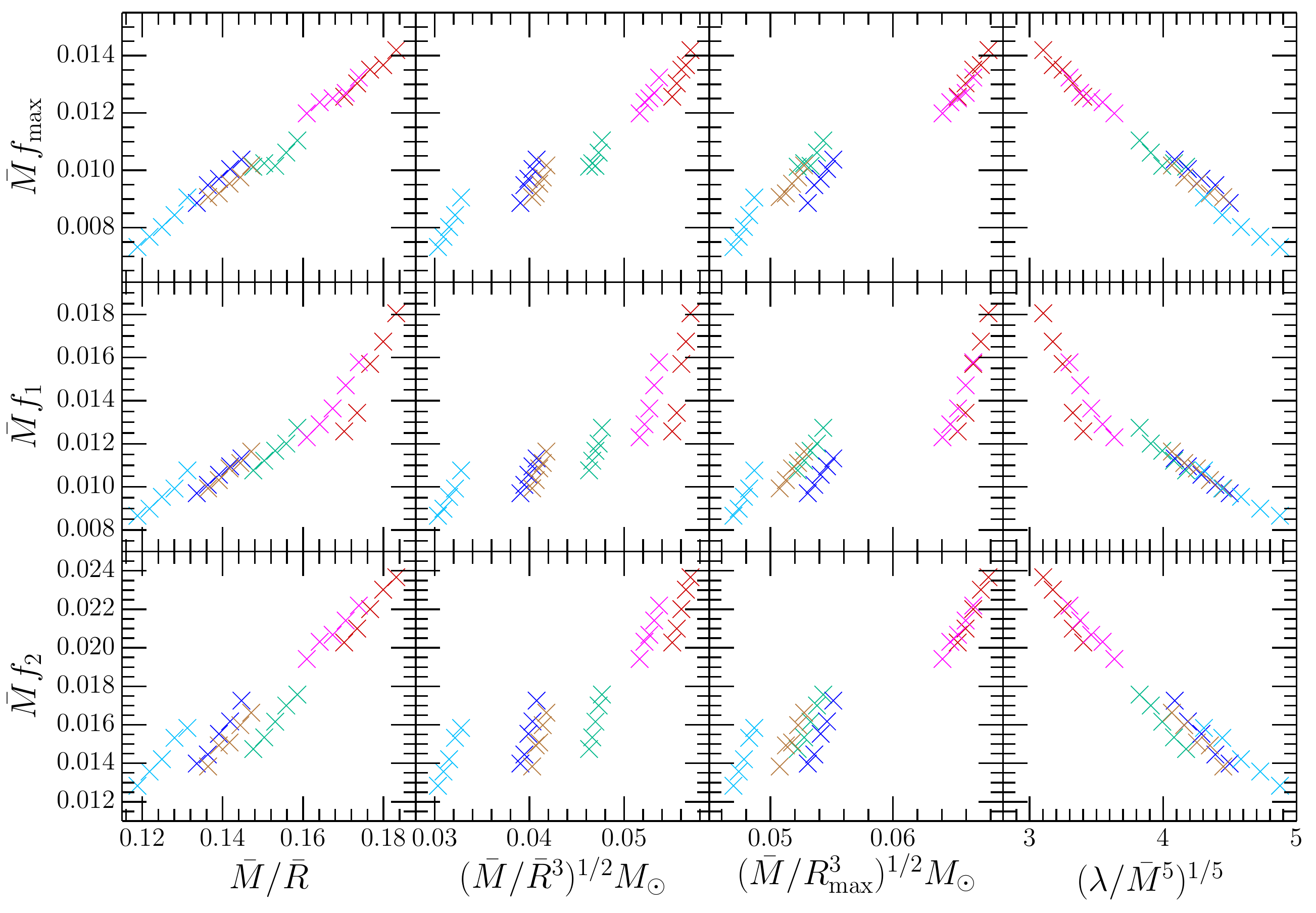}
\caption{Other empirical correlations between the $f_{\rm max}, f_1$, and
  $f_2$ frequencies and the physical quantities of the binary system, \eg
  the compactness $\bar{M}/\bar{R}$, the average density
  $(\bar{M}/\bar{R}^3)^{1/2}$, the pseudoaverage rest-mass density
  $(\bar{M}/R^3_{\rm max})^{1/2}$, or the dimensionless tidal
  deformability $(\lambda/\bar{M}^5)^{1/5}$. The left panel refers to
  frequencies expressed in kHz, while the right one refers to dimensionless
  frequencies expressed in terms of the average mass $\bar{M}$.
  \label{fig:corrFreq_3x4}}
\end{center}
\end{figure*}

\subsection{Correlations of the $f_2$ frequency}
\label{sec:corr_f2}

Historically the first spectral feature to have been detected and
discussed~\cite{Shibata05d, Oechslin07b, Baiotti08, Bauswein2011}, the
high-frequency $f_2$ peak has been shown to correlate with a number of
physical quantities of the progenitor stars. The most representative
example is shown in Fig.~\ref{fig:corr_f2}, which reports the value of
the $f_2$ frequency as a function of the pseudoaverage rest-mass density
$\left(2\bar{M}/R^3_{\rm max}\right)^{1/2}$ for the different EOSs, where
$R_{\mathrm{max}}$ is the radius of the maximum-mass nonrotating
configuration for the EOS under consideration. Indicated with crosses and
filled squares are the equal- and unequal-mass binaries,
respectively. Also reported as solid circles are the results presented in
Fig.~4 of Ref.~\cite{Bauswein2011} and relative to $\bar{M}=1.35\,\Msun$
binaries, where we have indicated with magenta and red circles those
models that have the same EOS considered here (\ie \mn{SLy-q10-M1350} and
\mn{APR4-q10-M1350}, respectively). Overall, the mass dependence in $f_2$
(\ie what distinguishes different points of the same color) does not
suggest a tight universal correlation in our data, although one may
conclude differently if only binaries of one mass were considered as in
Ref.~\cite{Bauswein2011}. The situation does not improve if instead of
using the fitted value for $f_2$ following the algorithm in
Sec.~\ref{sec:spec-det}, we simply use the local maximum value of the
high-frequency peak. However, within each EOS, the data are rather well
reproduced with a linear fit (the $\chi$-squared value is $\lesssim
0.004$). Although EOS is dependent, these fits still provide a set of
relations that can be used to constrain the EOS, as we anticipated
in~\cite{Takami:2014} and will further discuss in
Sec.~\ref{sec:constr_EOS}.

Finally, shown in Fig.~\ref{fig:corr_f2_R16} are the values of the $f_2$
peak frequencies in terms of $R_{1.6}$, that is, the radius of a
nonrotating NS with $\bar{M}=1.6\,M_\odot$ (\cf Fig.~23 of
Ref.~\cite{Bauswein2012}). Because it is relevant here to correlate
frequencies with the same mass, the various crosses refer to the $f_2$
frequencies at masses $\bar{M}/\Msun=1.250, 1.275, 1.300, 1.325$, and
$1.350$; these values are either available directly, or interpolated from
the mass values used in the simulations (see Table~\ref{tab:models}). As
it is apparent after connecting the different crosses with the same mass,
\eg with the dashed and dotted lines that connect, respectively, the
largest, the smallest, and the intermediate masses, it is difficult to
find a bilinear fitting function across the different EOSs as the one
suggested by Eq.~(3) of Ref.~\cite{Bauswein2012}.

\subsection{Other potential correlations}
\label{sec:corr_other}

In addition to the correlations of $f_{\rm max}, f_1$, and $f_2$ discussed,
respectively, in Secs.~\ref{sec:corr_fmax}, \ref{sec:corr_f1}, and
\ref{sec:corr_f2}, we have explored whether other empirical correlations
can be found between these frequencies and physical quantities of the
binary system, \eg the compactness $\bar{M}/\bar{R}$, the average density
$(\bar{M}/\bar{R}^3)^{1/2}$, the pseudoaverage rest-mass density
$(\bar{M}/R^3_{\rm max})^{1/2}$, or the dimensionless tidal deformability
$(\lambda/\bar{M}^5)^{1/5}$.

The results of this exploration are collected in
Fig.~\ref{fig:corrFreq_3x4}, where the left panel refers to the absolute
values of the frequencies in kHz, \eg $f_{\rm max}$, while the right one
refers to the corresponding normalized expressions, \eg $\bar{M} f_{\rm
  max}$. As in the previous figures, the different EOSs are distinguished
with different colors and the various symbols within each EOS refer to
binaries with different masses; quite generally, the highest value of the
frequency is the one corresponding to the largest mass in the sequence of
the same EOS.

Note that when restricted to a single binary mass, the correlation of the
$f_2$ frequency with the average rest-mass density appears to be much
tighter than it first appeared~\cite{Bauswein2011}, indicating that it is
essential to consider, in addition to a large set of EOSs, also a wide
range of masses (and mass ratios). It is also interesting to note that
the correlations can become ``tighter'' depending on what is the
normalization used. In other words, the ``universality'' of the
max-amplitude frequency with the average compactness $\mathcal{C}$ is
different whether one is considering $f_{\rm max}$ (left panel of
Fig.~\ref{fig:corrFreq_3x4}) or $\bar{M}f_{\rm max}$ (right panel of
Fig.~\ref{fig:corrFreq_3x4}). Conversely, a larger spread is seen in
the relation of the low-frequency peak with compactness if $\bar{M} f_1$
is used instead of $f_1$. We do not have a simple explanation for this
behavior, which, however, has been observed repeatedly within the
framework of quasiuniversal relations (see discussions in
Refs.~\cite{Doneva2014a} and \cite{Chakrabarti2014}).

Finally, the general overview of Fig.~\ref{fig:corrFreq_3x4} allows us to
appreciate the presence of at least two new ``quasiuniversal''
correlations in addition to the ones already discussed [\ie
  $\bar{M}f_{\rm max}$ vs. $(\lambda/\bar{M}^5)^{1/5}$ and $f_{1}$
  vs $\bar{M}/\bar{R}$]. Particularly interesting are the correlation
of $f_{\rm max}$ with the average density $(\bar{M}/\bar{R}^3)^{1/2}$
(right column of the left panel), but also the correlations of $f_{1}$
and $f_2$ with the dimensionless tidal deformability
$(\lambda/\bar{M}^5)^{1/5}$ (left column of the left panel). These
correlations appear also when expressing the results in terms of the
corresponding normalized frequencies $\bar{M} f_{\rm max}, \bar{M} f_1$,
and $\bar{M} f_2$ (right panel), although the scattering can also
increase in this case. In the following Sec.~\ref{sec:constr_EOS} we
will discuss how to translate two of these correlations into potential
measurements of the EOS if a post-merger signal with sufficiently large
SNR is detected. At the same time, Fig.~\ref{fig:corrFreq_3x4} highlights
that much more information can be extracted from these frequencies, and
hence more work is needed to fully exploit the spectral information on
the post-merger GW signal from BNSs.

\subsection{Constraining the EOS}
\label{sec:constr_EOS}

After having discussed in detail the spectral features of the post-merger
signal, some considerations should now be made on a new approach that
uses the correlations between the PSDs and the stellar properties to set
tight constraints on the EOS of nuclear matter. This approach has already
been presented in some detail in Ref.~\cite{Takami:2014}, but we recall
it here for completeness.

Essential for the success of this approach is a ``clear identification''
of the peaks in the PSD of the post-merger signal, which, in turn, will
require a high SNR and will depend on the EOS. More concretely, a measure
by Advanced LIGO of the $f_1$ peak with a $\mathrm{SNR}=5$ will require a
binary at a distance of about $25\,{\rm Mpc}$ for the APR4 EOS and of
about $40\,{\rm Mpc}$ for the (GNH3) EOSs. On the other hand, larger
distances of, respectively, about $50$ or $115\,{\rm Mpc}$ can be afforded
to achieve a measurement of the $f_2$ peak with a $\mathrm{SNR}=5$. These
distances are clearly smaller than the typical horizon scales of BNSs
expected for the advanced detectors, limiting the probability of
detection of such events. Yet, as we discuss below, the measurement would
be so significant that a single detection would be sufficient to
constrain the EOS, at least in principle.

\begin{figure*}
\begin{center}
\includegraphics[width=1.0\linewidth,angle=0]{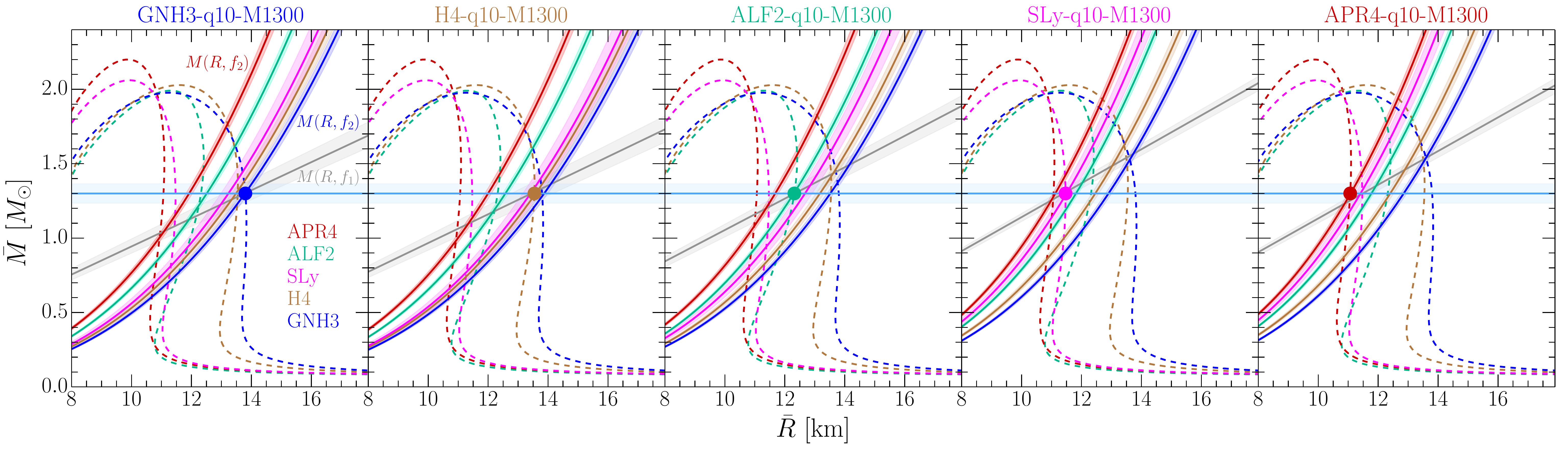}
\caption{\label{fig:validation} Examples of use of the spectral features
  to constrain the EOS. Once a detection is made, the relations
  $\bar{M}=\bar{M}(\bar{R},f_1)$ and
  $\bar{M}=\bar{M}(\bar{R},f_2;\mathrm{EOS})$ (colored solid lines) will
  cross at one point the curves of equilibrium configurations (colored
  dashed lines). Knowledge of the mass of the system (horizontal line)
  will provide a fourth constraint, removing possible degeneracies. The
  leftmost panel refers to the GNH3 EOS, while the rightmost one refers
  to the APR4 EOS. All panels are specific to a binary with $1.3\,\Msun$,
  but essentially identical plots can be shown also for other masses and
  we limit ourselves to a single mass only for compactness.}
\end{center}
\end{figure*}

Let us therefore assume that the GW signal from a BNS has been detected
and that the source is sufficiently close that all of the spectral
features are clearly identifiable. Using the measured values of $f_1$ and
$f_2$ we can draw a series of curves in the $(\bar{M},\bar{R})$
plane. The first of these curves will be the one expressing the relation
$\bar{M}=\bar{M}(\bar{R},f_1)$ (solid grey line), and, as discussed in
Sec.~\ref{sec:corr_f1}, this line will be the same for all EOSs. At
the same time, the measurement of $f_2$ will lead to as many lines as the
EOSs that are considered, where each curve will express the relation
$\bar{M}=\bar{M}(\bar{R},f_2;\mathrm{EOS})$ (solid colored lines). The
resulting set is shown in Fig.~\ref{fig:validation}, where the
leftmost panel refers to the GNH3 EOS, while the rightmost one refers to
the APR4 EOS. All panels are specific to a binary with $\bar{M} =
1.30\,\Msun$, but essentially identical plots can be shown also for other
masses, and we limit to one mass only for compactness. The uncertainties,
both theoretical and experimental, in the measurement of $f_{1}$ and
$f_2$ make the correlation curves $\bar{M}=\bar{M}(\bar{R},f_1)$ and
$\bar{M}=\bar{M}(\bar{R},f_2;\mathrm{EOS})$ be effectively ``bands'',
rather than thin lines, each with its own width representing the
corresponding probability distribution.

Concentrating, for example, on the left panel, which refers to the binary
\mn{GNH3-q10-M1300}, we can see that the $\bar{M}=\bar{M}(\bar{R},f_1)$
relation intersects each of the various equilibrium curves (colored
dashed lines) at one point (\eg at $\bar{M} \simeq
1.3\,M_{\odot},\,\bar{R} \simeq 13.8$ km for the GNH3 EOS), but also that
other ``crossings'' take place for the other EOSs. However, when using
also the relations $\bar{M}=\bar{M}(\bar{R},f_2; \mathrm{EOS})$, some
EOSs can be readily excluded (\eg APR4, ALF2, and GNH3, in this example)
and only the H4 and GNH3 EOSs have crossings (or ``near crossings'')
between the equilibrium-models curves and the frequency-correlations
curves. A more careful Bayesian analysis can associate with each of these
near crossings a certain probability depending on the width of the
different bands, thus providing additional support (at least from a
statistical point of view) to one single EOS over the others; the results
of this analysis will be presented in a forthcoming
paper~\cite{Messenger2015}. 

The ability to single out the correct EOS over the others is
significantly increased if the mass of the binary is known from the
inspiral signal\footnote{We note that in practice it will be difficult to
  distinguish the mass ratio from the chirp signal and hence the total
  mass of the system will represent a first reasonable
  approximation.}. In this way, an additional band in the
$(\bar{M},\bar{R})$ plane, shown as a horizontal light-blue stripe in the
panel, can be drawn. This stripe, whose width we have taken to correspond
to an error of 10\% in the determination of the mass, will intersect the
three bands relative to the GNH3 EOS in a small patch (ideally at one
point only if there were no errors in the measures of the frequencies;
\cf blue solid circle), while this will not be the case for the other
crossings. Hence, the simultaneous fulfilling of four distinct
constraints [\ie of four crossings at a point in the $(\bar{M},\bar{R})$
  plane] represents a new and effective approach to constrain the EOS of
nuclear matter.

At least in principle, this tool is so effective that a single detection
with sufficiently high SNR would be able to constrain the EOS. In
practice, even a high-SNR detection would be able to mark only one point
in the $(\bar{M},\bar{R})$ plane, which may well not be covered by any of
the presently known nuclear-physics EOSs. Furthermore, the detection
would only set a confidence level, possibly very high, on the probability
that the correct EOS should pass over that point in the
$(\bar{M},\bar{R})$ plane. Hence, it is clear that the success of this
approach relies on the ability of making a sufficient number of
detections with small theoretical and experimental uncertainties in the
measure of the peak frequencies. In this respect, the approach presented
here will set constraints on the EOS only in a statistical
sense. Unfortunately, as we will discuss in
Sec.~\ref{sec:detectability}, only a small fraction of the expected
detection rate of $\sim 40\, \mathrm {yr}^{-1}$~\cite{Abadie:2010_etal}
will be at sufficiently close distances.

Before closing this section we will make some additional
considerations. First, even if the measurement of the mass is not
available from the inspiral (this is possible, though not likely as the
chirp mass is one of the most robust measurements), the degeneracies in
the EOSs could be removed with a few positive detections, which would
tend to favor one EOS over the others. Second, even if none of the
presently known EOSs is the correct one to describe the properties of
nuclear matter, a number of detections at different masses would be able
to actually ``trace'' the curve of equilibrium models in the
$(\bar{M},\bar{R})$ plane; this track would then need to be reproduced by
any microphysical description of the EOS. Third, if only the $f_2$
frequency is measurable, the approach discussed above can still be used
as long as the mass is known; in this case three and not four curves will
have to cross at one point. Fourth, most of our simulations refer to
equal-mass binaries, but we expect that the $f_{1}$ and $f_{2}$
frequencies will not be very sensitive to the initial mass ratio; this
was already shown by Refs.~\cite{Bauswein2011, Hotokezaka2013c} and is
confirmed by the two unequal-mass binaries that are part of our
sample. Fifth, realistic values of the stellar spins are not likely to
influence the frequencies significantly given that the largest
contribution to the angular momentum of the HMNS comes from the orbital
angular momentum and not from the initial spins of the
stars~\cite{Kastaun2013, Bernuzzi2013}. Finally, because the $f_1$ peak
is produced soon after the merger, it should not be affected
significantly by magnetic fields and radiative effects, whose
modifications emerge on much larger
time scales~\cite{Giacomazzo:2010,Siegel2013}.

\subsection{Detectability of the peak frequencies}
\label{sec:detectability}

\begin{table}
\begin{tabular}{l|c|c|c|c|c|c|c}
\hline 
\hline
Binary
 & $f_{\rm max}$ & $f_1$ & $f_2$ & $\textrm{\scriptsize{SNR}}$ &
$\textrm{\scriptsize{SNR}}_{\rm cont}$ & 
$\textrm{\scriptsize{SNR}}_{f_1}$ & 
$\textrm{\scriptsize{SNR}}_{f_2}$\\
 & $[\mathrm{Hz}]$ & $[\mathrm{Hz}]$ & $[\mathrm{Hz}]$
 &  & & & \\
\hline
\mn{\small{GAM2-q10-M1350}}  & 1101 & 1303 & 1930 & 5.2 & 2.7 & 5 & 6 \\
\mn{\small{GAM2-q10-M1375}}  & 1134 & 1327 & 2004 & 5.2 & 2.4 & 6 & 5\\
\mn{\small{GAM2-q10-M1400}}  & 1163 & 1384 & 2061 & 5.2 & 2.3 & 5 & 5\\
\mn{\small{GAM2-q10-M1425}}  & 1202 & 1414 & 2183 & 5.4 & 2.4 & 6 & 5\\
\mn{\small{GAM2-q10-M1450}}  & 1267 & 1507 & 2127 & 5.5 & 2.3 & 5 & 5\\
\hline
\mn{\small{GNH3-q10-M1250}}  & 1439 & 1577 & 2272 & 4.5 & 2.1 & 4 & 14\\
\mn{\small{GNH3-q10-M1275}}  & 1510 & 1608 & 2302 & 4.6 & 2.1 & 4 & 15\\
\mn{\small{GNH3-q10-M1300}}  & 1515 & 1650 & 2425 & 4.8 & 2.2 & 3 & 14 \\
\mn{\small{GNH3-q10-M1325}}  & 1539 & 1679 & 2479 & 5.0 & 2.3 & 4 & 14\\
\mn{\small{GNH3-q10-M1350}}  & 1560 & 1703 & 2595 & 5.0 & 2.2 & 4 & 10\\
\mn{\small{GNH3-q09-M1300}}  & 1420 & 1624 & 2355 & 5.0 & 2.3 & 5 & 16\\
\hline
\mn{\small{ALF2-q10-M1225}}  & 1678 & 1785 & 2443 & 4.2 & 1.6 & 3 & 7\\  
\mn{\small{ALF2-q10-M1250}}  & 1664 & 1822 & 2493 & 4.3 & 1.5 & 3 & 5\\     
\mn{\small{ALF2-q10-M1275}}  & 1617 & 1859 & 2574 & 4.4 & 1.6 & 3 & 7\\  
\mn{\small{ALF2-q10-M1300}}  & 1658 & 1876 & 2655 & 4.6 & 1.7 & 3 & 5\\  
\mn{\small{ALF2-q10-M1325}}  & 1692 & 1953 & 2693 & 4.7 & 1.7 & 3 & 3\\  
\hline
\mn{\small{H4-q10-M1250}}    & 1473 & 1613 & 2247 & 4.5 & 2.0 & 4 & 9 \\
\mn{\small{H4-q10-M1275}}    & 1464 & 1643 & 2377 & 4.6 & 2.0 & 3 & 11 \\
\mn{\small{H4-q10-M1300}}    & 1489 & 1696 & 2356 & 4.8 & 2.1 & 4 & 12 \\
\mn{\small{H4-q10-M1325}}    & 1494 & 1702 & 2449 & 4.9 & 2.2 & 4 & 13 \\
\mn{\small{H4-q10-M1350}}    & 1529 & 1751 & 2501 & 5.0 & 2.3 & 3 & 14 \\
\hline
\mn{\small{SLy-q10-M1250}}   & 1947 & 1998 & 3154 & 4.3 & 1.4 & 2 & 6\\
\mn{\small{SLy-q10-M1275}}   & 1971 & 2055 & 3235 & 4.4 & 1.5 & 3 & 6\\
\mn{\small{SLy-q10-M1300}}   & 1954 & 2130 & 3229 & 4.6 & 1.4 & 3 & 9\\
\mn{\small{SLy-q10-M1325}}   & 1946 & 2254 & 3282 & 4.7 & 1.5 & 3 & 7\\
\mn{\small{SLy-q10-M1350}}   & 1991 & 2373 & 3338 & 4.8 & 1.5 & 3 & 9\\
\mn{\small{SLy-q09-M1300}}   & 1940 & 2191 & 3212 & 4.6 & 1.6 & 3 & 8 \\
\hline
\mn{\small{APR4-q10-M1275}}  & 2001 & 2003 & 3229 & 4.3 & 1.3 & 2 & 5\\
\mn{\small{APR4-q10-M1300}}  & 2035 & 2099 & 3279 & 4.6 & 1.4 & 4 & 4\\
\mn{\small{APR4-q10-M1325}}  & 2071 & 2407 & 3373 & 4.6 & 1.3 & 3 & 4\\
\mn{\small{APR4-q10-M1350}}  & 2056 & 2518 & 3462 & 4.7 & 1.3 & 3 & 3\\
\mn{\small{APR4-q10-M1375}}  & 2096 & 1953 & 2693 & 4.9 & 1.3 & 2 & 3\\
\hline
\hline
\end{tabular}
\caption{Values of the $f_{\rm max}, f_{1}$, and $f_2$ peak frequencies,
  where the last two are computed with the fitting procedure. For all
  waveforms, the SNR is computed from $7.4\,\ms$ before the merger and up
  to $24.6\,\ms$ after the merger, and ${\rm SNR_{cont}}$ is the value
  computed for the waveform as above but with $f\ge f_{\rm cont}$. The
  last two columns report rounded to the integer figure the ``effective
  SNR'' at the frequencies $f_1$ and $f_2$, that is, the ratio between
  the PSD of the signal and PSD of the detector noise at those
  frequencies. Note that for the effective SNR the values differ at most
  of a factor of 3 and are even comparable for some binaries. For all
  cases, we assumed GW detections for sources with optimal orientation at
  $50\,{\rm Mpc}$ and the noise spectrum of Advanced LIGO.
\label{tab:SNRs}}
\end{table}

\begin{table*}
\begin{tabular}{l|c|c|c|c|c}
\hline 
\hline
Binary &
\mn{GNH3-q10-M1300} &
\mn{ALF2-q10-M1300} &  
\mn{H4-q10-M1300}   &
\mn{SLy-q10-M1300}  &
\mn{APR4-q10-M1300}  \\
\hline
\mn{GNH3-q10-M1300} & 0   & {31} & {7} & {78} & {74} \\
\hline
\mn{ALF2-q10-M1300} & 18  & 0   & {25} & {32} & {32}  \\  
\hline
\mn{H4-q10-M1300}   & 10  & 42  & 0 & {64} & {85} \\
\hline
\mn{SLy-q10-M1300}  & 150 & 130 & 130  & 0 & {7} \\
\hline
\mn{APR4-q10-M1300} & 140 & 90  & 190  & 11  & 0 \\
\hline
\hline
\end{tabular}
\caption{Errors $\delta f_{1}$ and $\delta f_{2}$ in the parameter
  estimation of $f_{1}$ and $f_{2}$ computed through the Fisher
  information matrix (see~\cite{Read:2009a}). The SNRs used in this
  estimate are computed from the waveform starting 7.4 ms before the
  merger and ending 24.6 ms after the merger, assuming sources with
  optimal orientation at $50\,{\rm Mpc}$ and the noise spectrum of Advanced
  LIGO. All values are expressed in Hz, with numbers above the diagonal
  referring to $\delta f_{1}$, and those below the diagonal referring
  to $\delta f_{2}$.
\label{tab:Fisher_f1_f2}}
\end{table*}

Obviously, our method can work only if the frequencies of the peaks can
be measured in GW detectors, there is a sufficient number of detections,
and the uncertainties in the measure of the frequencies are sufficiently
small. As shown qualitatively in Fig.~\ref{PSD_5x5_all}, this is possible
for sufficiently close events. For more quantitative estimates, we have
computed and reported in Tables~\ref{tab:SNRs} and~\ref{tab:Fisher_f1_f2}
the SNRs and the error estimates for the models we simulated. In
particular, Table~\ref{tab:SNRs} lists, in addition to the peak
frequencies $f_{1}$, and $f_{2}$, the SNR computed from each waveform in
the time interval starting at $7.4\,\ms$ before the merger and ending at
$24.6\,\ms$ after the merger. The choice of a fixed time interval was
made to facilitate the comparison of the detectability across different
models. For most models, this interval includes most of the post-merger
signal, but only a few orbits of the inspiral.

The sixth column of Table~\ref{tab:SNRs} shows ${\rm SNR}_{\rm cont}$,
that is, the SNR computed using the waveforms in the same interval as
above, but by considering only frequencies larger than the contact
frequency $f_{\rm cont}$, in order to single out the contribution to the
detected GW signal of the post-merger phase. In this respect, ${\rm SNR}
- {\rm SNR}_{\rm cont}$ represents the SNR of the (short) inspiral only
and can be seen as a lower limit on the detectability of these
signals. The last two columns list the ``effective SNR'' at the peak
frequencies $f_{1}$, and $f_2$, that is, the ratio between the PSD of the
signal and the PSD of the detector at those frequencies, \ie
$\textrm{{SNR}}_{f_{1,2}} \equiv 2 \tilde{h}(f_{1,2})
\sqrt{f_{1,2}}/S_h(f_{1,2})$. Note that a smaller power in the $f_1$ peak
is compensated by a smaller detector noise at lower frequencies, so that
the SNRs at the frequencies $f_1$ and $f_2$ differ at most by a factor of
3 and are even comparable for some binaries. For all cases, we have
assumed GW detections for sources with optimal orientation at $50\,{\rm
  Mpc}$ and the noise PSD of Advanced LIGO.

Assuming $\mathrm{SNR}=5$ and considering only the post-merger signal,
these results yield a detection horizon of about $13$-$27\,{\rm Mpc}$,
which reduces to $\simeq 9$-$19\,{\rm Mpc}$ for randomly oriented
sources. As summarized, for example, in Ref.~\cite{Abadie:2010_etal}, the
detection rate is $\dot{N} = R N_G$, where $R$ is the coalescence rate of
binaries per galaxy and $N_G$ is the number of galaxies accessible with a
GW search. According to~\cite{Abadie:2010_etal}, $R\sim
100\ \rm{MWEG}^{-1} \rm{Myr}^{-1}$ (MWEG = Milky Way Equivalent Galaxy)
and $N_G$ can be estimated from Eq. (4) and Fig. 1
of~\cite{Abadie:2010_etal} to be approximately $200$ MWEG for horizons of
$13$-$27\,{\rm Mpc}$\footnote{In particular, we used the bottom curve of
  the two shown in Fig. 1 of~\cite{Abadie:2010_etal}, since it refers to
  the accessible blue-light luminosity for a given horizon distance,
  taking location and orientation averaging into account.}. In this way
we derive a strict event rate of $\sim 0.01\,\mathrm{yr}^{-1}$, which can,
however, be increased to $\sim 0.1\,\mathrm{yr}^{-1}$ if advanced
detection techniques, such as the coherent wave-burst algorithm, is
employed~\cite{Clark2014}.

We have also used the Fisher information matrix to estimate the
uncertainties in the determination of the peak frequencies when a GW
detection is made~\cite{Read:2009b,Bauswein2012}. Fisher matrix
techniques are often used to forecast the precision of future
experiments. Under certain standard assumptions, the Fisher matrix is the
inverse of the covariance matrix, so it can be used to estimate the
uncertainties on the parameters of the model. One of these assumptions is
that, given a {\it fiducial} model, the Fisher matrix does not change too
much if a different model is considered. Namely, the Fisher matrix error
estimate is valid for models near the fiducial model~\cite{Read:2009b}.

In practice, we have followed ~\cite{Read:2009b} and estimated the errors
in the determination of the peak frequency $f_1$ as (a similar expression
can be obtained for the frequency $f_2$ by replacing $1$ with $2$ below)
\begin{align}
\label{eq:delta_f1}
&\delta f_1 \equiv \sqrt{\overline{(\delta f_1)^2}} =
{|f_{1{\rm A}}-f_{1{\rm B}}|} \times \nonumber\\
&\left(\!4{\rm Re}\!\!\int_0^\infty
\frac{[\tilde{h}_{\rm A}(f')-
\tilde{h}_{\rm B}(f')][\tilde{h}_{\rm A}(f')-\tilde{h}_{\rm B}(f')]^*}{S_h(f')}
d f'\!\!\right)^{\!\!-1/2}\,,
\end{align}
where $f_{1{\rm A}}$ and $f_{1{\rm B}}$ are the $f_1$ frequencies of two
models ${\rm A}$ and ${\rm B}$, $\tilde{h}_{\rm A}$ and $\tilde{h}_{\rm
  B}$ are the Fourier transforms of the waveforms of models ${\rm A}$ and
${\rm B}$, respectively, and $S_h$ is the noise spectrum of Advanced
LIGO. The denominator of expression~\eqref{eq:delta_f1} is the SNR of the
difference of the signals of the two models.

Neglecting correlations between the peak frequencies and other
parameters, we list in Table~\ref{tab:Fisher_f1_f2} the errors estimated
in this way for selected models, namely equal-mass models with the
average gravitational mass at infinite separation $\bar{M}/\Msun=1.30$
and all the EOSs we simulated. As mentioned, the Fisher matrix method is
valid only for models near the {\it correct} model, so the values in
Table~\ref{tab:Fisher_f1_f2} for models whose frequencies are very
different are not particularly meaningful and are reported only for
completeness. In any case, the errors computed using only the post-merger
phase are of the order of $1$-$10\%$ and thus compatible with the estimates
of \cite{Clark2014}.

\section{Conclusions}
\label{sec:conclusions}

Making use of a large number of accurate and fully general-relativistic
simulations of the inspiral and merger of BNSs with nuclear EOSs, we have
illustrated in detail the spectral properties of the post-merger GW
signal, extending and complementing the information already provided
in~\cite{Takami:2014} and in other papers that have looked at these
properties~\cite{Bauswein2011, Bauswein2012, Stergioulas2011b, Read2013,
  Bernuzzi2014}. More specifically, we have shown that correlations exist
between the two largest peak frequencies in the PSD of the post-merger
signal (\ie $f_1, f_2$) and the physical properties of the NSs composing
the binary system (\ie mass, radius and their combinations). One of these
correlations, namely, that of the low-frequency peak $f_1$ with
compactness $\mathcal{C}$, is rather tight and has all the properties of
being quasiuniversal; that is, it has a functional behavior that does
not depend on the EOS. On the other hand, the high-frequency peak $f_2$
shows correlations, either with the average density or with the radius of
a nonrotating star with a mass of $1.6\,\Msun$, that are far less tight
and depend on the EOS.

We have shown that exploiting the information coming from both of these
frequencies, together with the knowledge of the total mass of the binary,
a potential detection of the post-merger GW signal at the large SNR ratio can
set tight constraints on the EOS of matter at nuclear densities. While in
principle a single observation would be sufficient, in practice the
uncertainties in the measurements of the peaks and the errors in the
fitting of the various correlations will require a number of positive and
high-SNR detections for the robust identification of the EOS.

An interesting side product of this work is also the construction of a
simple mechanical toy model that can explain rather intuitively the main
spectral features of the post-merger signal and shed light on the
physical interpretation of the origin of the various peaks. Despite its
crudeness, the toy model can reproduce reasonably well the complex
waveforms emitted right after the merger.

The exciting prospects opened in this work call for a number of future
improvements. First, it remains to be determined how robust the results
found here are when realistic spins in the NSs are accounted for, or when
the dynamics is extended to an ideal-magnetohydrodynamics (ideal-MHD)
description~\cite{Giacomazzo:2010}. Second, the phenomenology discussed
here will need to be studied also when fully three-dimensional EOSs are
considered, or in more extreme regimes of unequal-mass ratios. Third, the
potential of the toy model presented here in predicting the post-merger
signal needs to be further developed and validated. Last but not least, a
proper assessment of the errors associated with the measurements of the
stellar radii will require a complete Bayesian analysis of the
probability that the constraints coming from the measurement of the two
peak frequencies and of the total mass are satisfied simultaneously. Work
in this direction is already in progress~\cite{Messenger2015}.

\begin{acknowledgments}
We thank F. Galeazzi for help on the construction of the toy model and
N. Stergioulas for useful discussions. Partial support comes from the DFG
Grant SFB/Transregio~7 and by ``NewCompStar'', COST Action
MP1304. K.\,T. is supported by the LOEWE-Program in HIC for
FAIR. L.\,B. is supported in part by the JSPS Grant-in-Aid for Scientific
Research C, No. 26400274. The simulations were performed on SuperMUC
at LRZ-Munich, on Datura at AEI-Potsdam, and on LOEWE at CSC-Frankfurt.
\end{acknowledgments}

\appendix 

\section{A MECHANICAL TOY MODEL FOR THE POST-MERGER SIGNAL}
\label{appendix_a}

\begin{figure}
\begin{center}
\includegraphics[width=0.65\columnwidth]{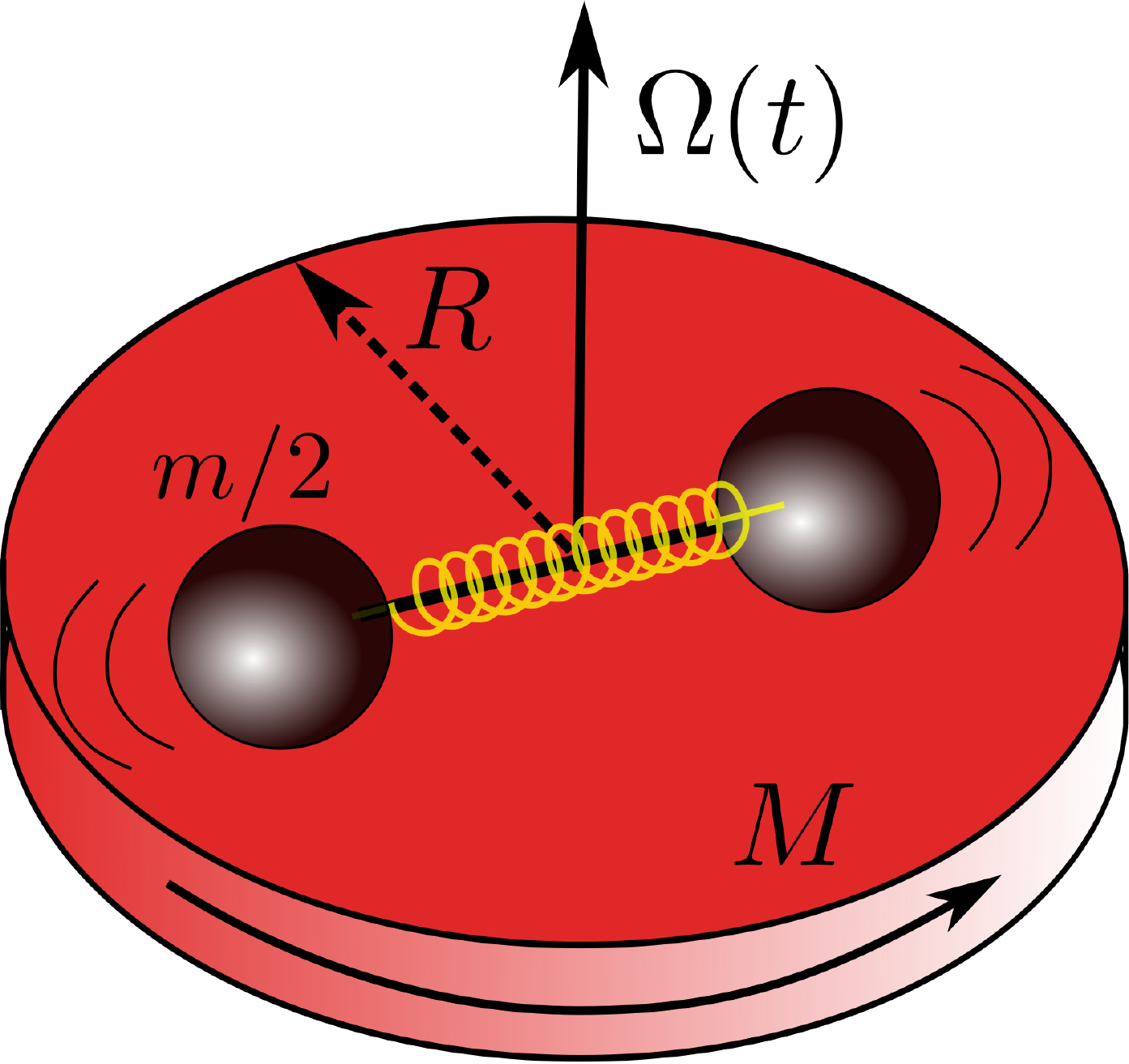}
\caption{Cartoon of the mechanical toy model composed of a disk of mass
  $M$ and radius $R$ rotating at frequency $\Omega(t)$. Two spheres, each
  of mass $m/2$ are connected to the disk, but are also free to oscillate
  via a spring that connects them. \label{fig:toy} }
\end{center}
\end{figure}

We illustrate here the properties of a mechanical toy model that can be
studied to reproduce many of the features of the spectral properties of
the post-merger signal. Consider therefore a mechanical system composed
of a disk of mass $M$ and radius $R$ rotating at a given angular
frequency $\Omega(t)$, to which two spheres, each of mass $m/2$, are
connected (\eg via a shaft), but are also free to oscillate via a spring
that connects them (see Fig.~\ref{fig:toy}). The spring has an elastic
constant $k$ such that the oscillation frequency $\sqrt{k/m}$ is
comparable with the rotation frequency $\Omega$.

In such a system, the two spheres will either approach each other,
decreasing the moment of inertia of the system, or move away from each
other, increasing the moment of inertia. Conservation of the total
angular momentum will then lead to a variation of the rotation speed of
the system and the generation of GWs with a precise spectrum.

Assuming that the spring will remain straight at all times and that the
canonical coordinates and momenta are, respectively, $q^i$ and $\dot{q}^i$,
the dynamics of the system with Lagrangian $L$ will be described by the
Euler-Lagrange equations
\begin{equation}
\frac{d}{dt}\left(\frac{\partial L}{\partial
  \dot{q}^i}\right)-\frac{\partial L}{\partial q^i} = \frac{\partial
  D}{\partial \dot{q}^i} \,,
\label{euler-lagr-damp}
\end{equation}
where, $L=T-V$, $T$, and $V$ are the kinetic and potential energies,
respectively, while $D$ accounts for dissipative effects. In a
cylindrical coordinate system $(\varpi, \varphi, z)$ these quantities are
expressed as 
\begin{align}
T & = \frac{1}{2} m
\left[\dot{\varpi}^2+\left(\varpi\dot{\varphi}\right)^2\right]+
\frac{1}{2}\left(\frac{MR^2}{2}\right)\dot{\varphi}^2\,,\\
V & =2 k (\varpi-\varpi_0)^2\,, \\
D & =-b \dot{\varpi}^2\,, 
\end{align}
where $\varpi_0$ is the natural displacement of the mass when $\Omega=0$,
and $b$ is the damping coefficient. As a result, the Lagrangian of the
system can be written as
\begin{align}
L= 
\frac{1}{2} m
\left[ \dot{\varpi}^2+\left(\varpi \dot{\varphi}\right)^2\right]
+\frac{1}{2}\left(\frac{MR^2}{2}\right)\dot{\varphi}^2 
-2k (\varpi-\varpi_0)^2\,.
\end{align}

We can then obtain the equation of motion for the radial displacement of
the sphere $\varpi=\varpi(t)$ after computing that 
\begin{align}
\frac{\partial L}{\partial \dot{\varpi}}&=m\dot{\varpi}\,,& 
\frac{d}{dt}\left(\frac{\partial L}{\partial \dot{\varpi}}\right)&
=m\ddot{\varpi}\,,\\
\frac{\partial L}{\partial \varpi}&=m\varpi\dot{\varphi}^2-4k(\varpi-\varpi_0)\,,&
\frac{\partial D}{\partial \dot{\varpi}}&=-2b\dot{\varpi}\,.
\end{align}
So as to obtain the following second-order differential equation:
\begin{equation}
\ddot{\varpi} +\frac{4k(\varpi-\varpi_0)}{m} -
\varpi\dot{\varphi}^2 +\frac{2b\dot{\varpi}}{m}=0\,,
\label{eq1}
\end{equation}

Similarly, after using 
\begin{align}
\frac{\partial L}{\partial \dot{\varphi}}
&=\left(\frac{MR^2}{2}+m\varpi^2\right)\dot{\varphi}\,,\\ 
\frac{d}{dt}\left(\frac{\partial L}{\partial \dot{\varphi}}\right)&=
\left(\frac{MR^2}{2}+m\varpi^2\right)\ddot{\varphi}+
2m\varpi\dot{\varpi}\dot{\varphi}\,,\\ 
\frac{\partial L}{\partial \varphi}&=0 = 
\frac{\partial D}{\partial \dot{\varphi}}\,,
\end{align}
we can express the equation of motion for the angular displacement of the
sphere via the differential equation
\begin{equation}
\ddot{\varphi} = -\frac{4m\varpi\dot{\varpi} \dot{\varphi}}{MR^2+2m\varpi^2}\,.
\label{eq2}
\end{equation}

Equation~(\ref{eq2}) can be rewritten as
\begin{equation}
\frac{d (\ln \dot{\varphi} )}{dt} = 
-\frac{d}{dt} \left[ \ln \left( \frac{MR^2}{2m}+\varpi^2 \right) \right]\,,
\end{equation}
and can be integrated analytically to obtain
\begin{equation}
\Omega(t) \equiv \dot{\varphi} = \frac{c_1}{\varpi^2+{MR^2}/{(2m)}}\,,
\label{theta_p}
\end{equation}
where $c_1$ is an integration constant related to the total angular
momentum $J_\mathrm{tot}$, \ie $c_1=J_\mathrm{tot}/m$. By replacing
$\dot{\varphi}$ in Eq.~(\ref{eq1}) with the expression given in
Eq.~(\ref{theta_p}), we can rewrite the differential equation~\eqref{eq1}
as
\begin{equation}
\label{eq:toy_model}
\ddot{\varpi} +\frac{4k(\varpi-\varpi_0)}{m}
-\left[\frac{c_1}{\varpi^2+{MR^2}/{(2m)}}\right]^2\varpi
+\frac{2b\dot{\varpi}}{m} =0\,.
\end{equation}
Equation~\eqref{eq:toy_model} can easily be integrated numerically to
obtain $\varpi=\varpi(t)$ and hence $\Omega=\Omega(t)$ from
Eq.~\eqref{theta_p}. 

\begin{figure}
\begin{center}
\includegraphics[width=\columnwidth]{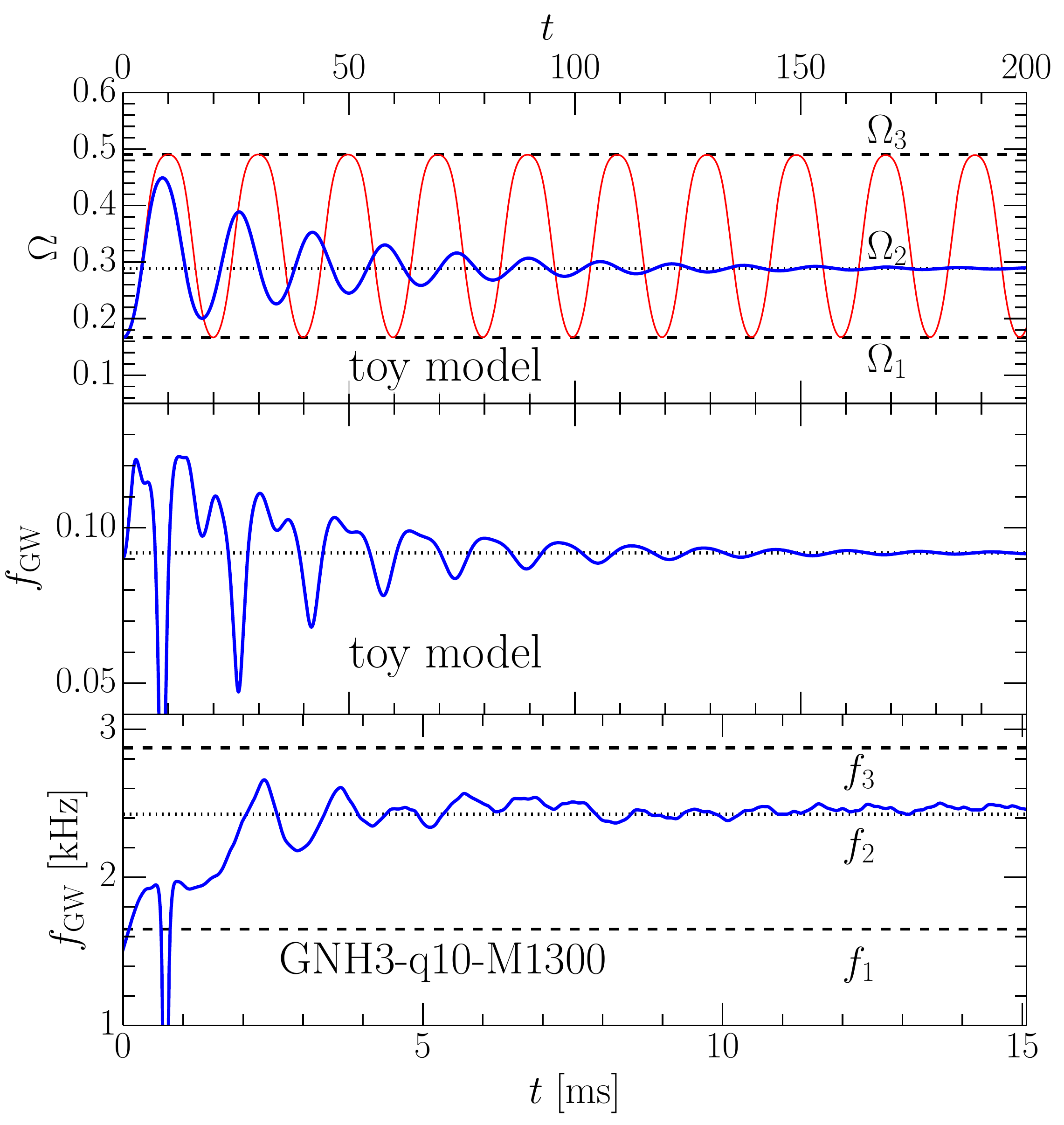}
\caption{\textit{Top panel:} Time evolution in arbitrary units of the
  angular frequency $\Omega(t)$ in the mechanical toy model for a typical
  solution obtained with $M=10, m=10$, and $k=0.25$; the red solid line
  refers to an undamped system (\ie with $b=0$), while the blue solid one
  to a damped system (\ie with $b=0.25$). Note that in the first case
  $\Omega$ oscillates between $\Omega_1$ and $\Omega_3$ (black dashed
  lines), while in the second case it asymptotes to $\Omega_2 \sim
  \tfrac{1}{2}(\Omega_1 + \Omega_3)$ (black dotted line). \textit{Middle
    panel:} Evolution of the instantaneous GW frequency as computed from
  the toy model. \textit{Bottom panel:} Evolution of the instantaneous GW
  frequency $f_{_{\rm GW}}$ for the binary \mn{GNH3-q10-M1300}; \cf
  Fig.~\ref{fig:h_plus_cf} for the actual GW strain.
\label{fig:toy_omega} }
\end{center}
\end{figure}

The top panel of Fig.~\ref{fig:toy_omega} shows a typical solution for
$\Omega(t)$ obtained with $M=10, m=10$, $k=0.25$, and $c_1=25$, where the
red solid line refers to an undamped system (\ie with $b=0$), while the
blue solid one to a damped system (\ie with $b=0.25$). 

As anticipated in Sec.~\ref{sec:f1_f3_orig}, conservation of the total
angular momentum implies that the evolution of the angular frequency of
the system is bounded by two frequencies $\Omega_1$ and $\Omega_3$ (\cf
dashed horizontal lines in the top panel of
Fig.~\ref{fig:toy_omega}). These frequencies correspond respectively to
the spinning frequency of the system when the spheres are at the largest
separation (\ie $\Omega_1$, when the moment of inertia is the largest)
and to the spinning frequency when the spheres are at the smallest
separation (\ie $\Omega_3$, when the moment of inertia is the
smallest). In the absence of dissipative forces, \ie for $b=0$, the
angular frequency simply oscillates between $\Omega_1$ and $\Omega_3$
(red solid line). Conversely, the oscillations are damped in the presence
of a dissipative term, \ie for $b\neq 0$, and the angular velocity
asymptotes to a constant value, \ie $\Omega(t) \to \Omega_2$ for ${t\to
  \infty}$, where $\Omega_2 \sim \tfrac{1}{2}(\Omega_1 + \Omega_3)$ for
the parameters chosen here (blue solid line).

Because the time spent at a given frequency is $\tau_{\Omega} =
\Omega/(d\Omega/dt)$, more time is obviously spent at those frequencies
where $d\Omega/dt \simeq 0$, \ie $\Omega_1$ and $\Omega_3$. As a result,
more power is expected to appear at these frequencies, hence producing a
low-frequency peak around $\Omega_1$ and a high-frequency peak around
$\Omega_3$. If dissipative processes are not present, the power is
expected at the frequency $\Omega_2$ and at its overtones $\Omega_n
\simeq (n/2)\Omega_2$ for our reference choice of parameters, from which
it follows that $\Omega_2 \simeq \tfrac{1}{2}(\Omega_1+\Omega_3)$. This
changes considerably if dissipative processes are present, as in this
case most of the time is spent by the system around $\Omega_2$, so that
most of the power in the PSD will appear around $\Omega_2$, with two main
sidebands at $\Omega_1$ and $\Omega_3$.

We can validate this picture by computing the GW signal of the toy model
in terms of the Newtonian quadrupole formula. To this end we need to
calculate the relevant components of the quadrupole-moment tensor, which
can be split into a time-independent part relative to the rotating disk
$\boldsymbol{I}^{\mathrm{d}}$ and a time-dependent part relative to the
oscillating spheres $\boldsymbol{I}^{\mathrm{s}}$, \ie
\begin{equation}
I_{ij}(t) = I^\mathrm{d}_{ij} + I^\mathrm{s}_{ij}(t)\,.
\end{equation}
The only nonzero components of $I^\mathrm{s}_{ij}(t)$ are
\begin{align}
I^\mathrm{s}_{xx}(t) & = \left(\frac{m}{2}\right)\varpi^2(t) \left[ 1 + \cos(2\varphi(t))\right]\,,\\
I^\mathrm{s}_{yy}(t) & = \left(\frac{m}{2}\right)\varpi^2(t) \left[ 1 - \cos(2\varphi(t))\right]\,,\\
I^\mathrm{s}_{xy}(t) & = \left(\frac{m}{2}\right)\varpi^2(t)\sin(2\varphi(t))\,.
\end{align}

Assuming an optimally oriented observer at distance $d$, the GW strain in
the two $+$ and $\times$ polarizations is then given by
\begin{align}
\label{eq:toy_hp}
\left(\frac{d}{2}\right) h_{+} &= 
\frac{\ddot{\Ibar}_{xx}-\ddot{\Ibar}_{yy}}{2} \nonumber\\
&= m \bigl[~ \left\{~ \dot{\varpi}^2 + \varpi ( \ddot{\varpi}-2\varpi\Omega^2 )
 ~\right\} \cos(2\varphi) \nonumber\\
&~~~~~~~~~~~~~~~ -\varpi ( 4 \dot{\varpi}\Omega+\varpi \dot{\Omega})\sin(2\varphi)
~\bigr]\,, \\
\label{eq:toy_hc}
\left(\frac{d}{2}\right) h_{\times} &= 
\ddot{\Ibar}_{xy} \nonumber\\
&= m \bigl[~ \left\{~ \dot{\varpi}^2 + \varpi ( \ddot{\varpi}-2\varpi\Omega^2 )
 ~\right\}\sin(2\varphi) \nonumber\\
&~~~~~~~~~~~~~~~ +\varpi ( 4 \dot{\varpi}\Omega+\varpi \dot{\Omega})\cos(2\varphi) 
~\bigr]\,,
\end{align}
where $\Ibar_{ij} \equiv I_{ij}-\frac{1}{3}\delta_{ij} \delta^{\ell
  m}I_{\ell m}$ is the reduced (trace-free) quadrupole moment tensor and
$\delta_{ij}$ is the Kronecker delta.

\begin{figure}
\begin{center}
\includegraphics[width=\columnwidth]{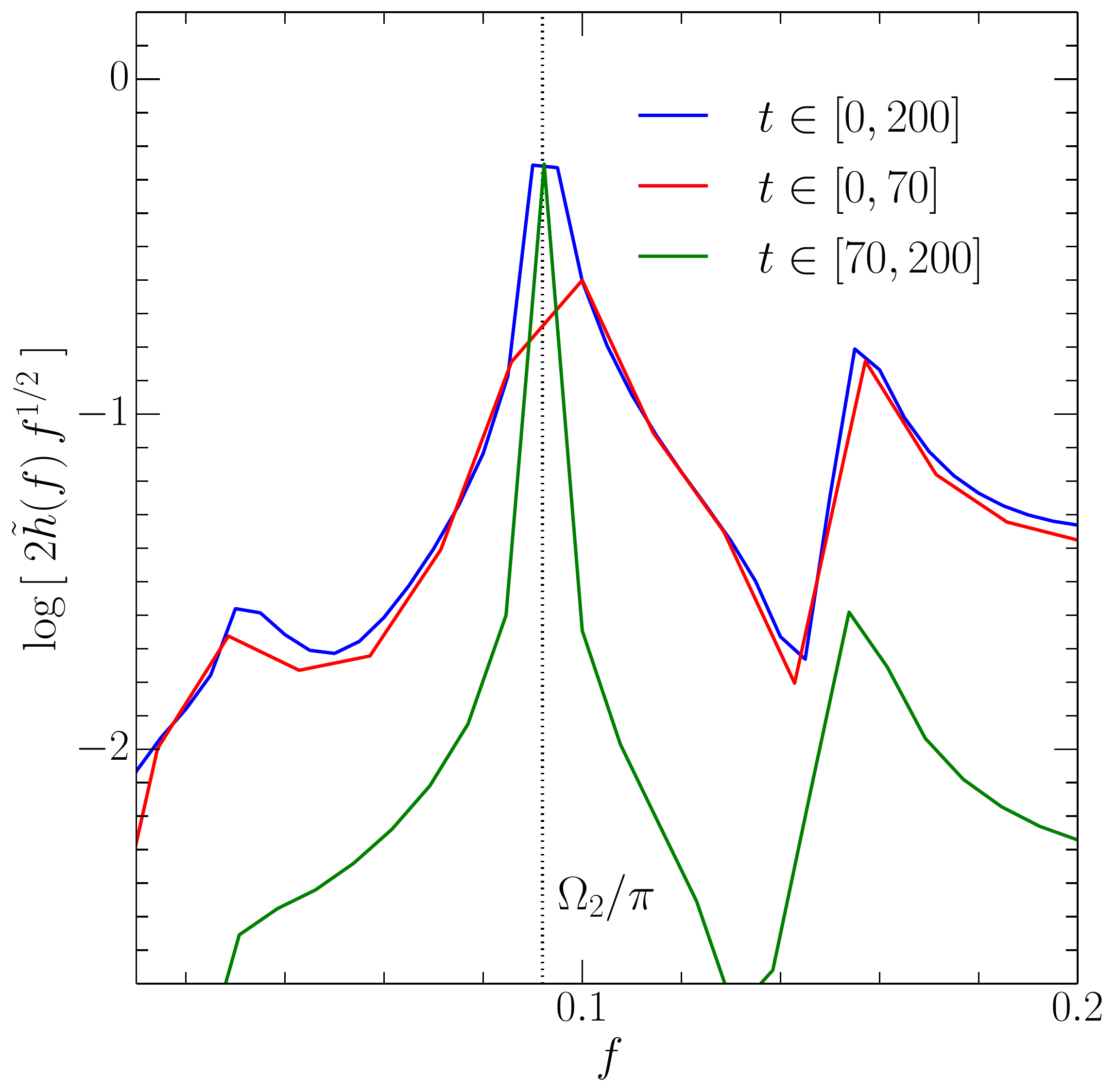}
\caption{PSDs $2\tilde{h}(f)f^{1/2}$ of the GW signal produced by the toy
  model. The blue solid line refers to the full time series, while the
  other lines refer to signals where either the final part has been
  removed (red solid line) or the final one (green solid line). Shown
  with a vertical dashed line is the frequency $F_2 = \Omega_2/\pi$. All
  units are arbitrary.
\label{fig:toy_PSD} }
\end{center}
\end{figure}

The results of the GW signal computed with the toy model were already
presented in Sec.~\ref{sec:f1_f3_orig}, where it was remarked that
despite the crudeness of the toy model, a good match appeared between the
strain computed from the fully numerical simulations and the one coming
from the toy model (\cf Fig.~\ref{fig:h_plus_cf}).
Figure~\ref{fig:toy_omega} offers another comparison between the GW
computed from the toy model and those calculated from the
numerical-relativity simulations. More specifically, the middle panel of
Fig.~\ref{fig:toy_omega} reports the instantaneous GW frequency $f_{_{\rm
    GW}}$ of the $+$ polarization~\eqref{theta_p} computed for the toy
model, while the bottom panel reports the corresponding instantaneous GW
frequency as computed for the binary \mn{GNH3-q10-M1300}. The
similarities in the two instantaneous frequencies are a measure of the
ability of the toy model to capture, especially for stiff EOSs, the
essence of the GW emission right after the merger.

Finally, Fig.~\ref{fig:toy_PSD} reports the PSDs $2\tilde{h}(f)f^{1/2}$
of the GW signal produced by the toy model via Eq.~\eqref{eq:toy_hp} and
for a damped oscillation with $b=0.25$. Adopting arbitrary units for the
time, the blue solid line refers to the full time series, \ie $t\in
[0,200]$, the red one refers to a signal in which the final part has been
removed, \ie $t\in [0,70]$, while the green one refers to a signal in which the
initial part has been removed, \ie $t\in [70,200]$. As expected, in the
case of the full time series, the PSD exhibits three main peaks at
frequencies $F_1, F_2$, and $F_3$, with $F_2 = \Omega_2/\pi \sim
\tfrac{1}{2}(F_1 + F_3)$. Furthermore, if the latter part of the signal
is removed, then the power in the $F_2$ frequency is significantly
reduced (red solid line). Similarly, for a signal that concentrates only
on the decaying phase of the oscillations, the $F_1, F_3$ peaks are
significantly suppressed (green solid line).

\bibliographystyle{apsrev4-1-noeprint}
\input{aeireferences.bbl}


\end{document}

%% file: aeireferences.bbl
%

%% file: manuscript.bbl
\begin{thebibliography}{97}%
\makeatletter
\providecommand \@ifxundefined [1]{%
 \@ifx{#1\undefined}
}%
\providecommand \@ifnum [1]{%
 \ifnum #1\expandafter \@firstoftwo
 \else \expandafter \@secondoftwo
 \fi
}%
\providecommand \@ifx [1]{%
 \ifx #1\expandafter \@firstoftwo
 \else \expandafter \@secondoftwo
 \fi
}%
\providecommand \natexlab [1]{#1}%
\providecommand \enquote  [1]{``#1''}%
\providecommand \bibnamefont  [1]{#1}%
\providecommand \bibfnamefont [1]{#1}%
\providecommand \citenamefont [1]{#1}%
\providecommand \href@noop [0]{\@secondoftwo}%
\providecommand \href [0]{\begingroup \@sanitize@url \@href}%
\providecommand \@href[1]{\@@startlink{#1}\@@href}%
\providecommand \@@href[1]{\endgroup#1\@@endlink}%
\providecommand \@sanitize@url [0]{\catcode `\\12\catcode `\$12\catcode
  `\&12\catcode `\#12\catcode `\^12\catcode `\_12\catcode `\%12\relax}%
\providecommand \@@startlink[1]{}%
\providecommand \@@endlink[0]{}%
\providecommand \url  [0]{\begingroup\@sanitize@url \@url }%
\providecommand \@url [1]{\endgroup\@href {#1}{\urlprefix }}%
\providecommand \urlprefix  [0]{URL }%
\providecommand \Eprint [0]{\href }%
\providecommand \doibase [0]{http://dx.doi.org/}%
\providecommand \selectlanguage [0]{\@gobble}%
\providecommand \bibinfo  [0]{\@secondoftwo}%
\providecommand \bibfield  [0]{\@secondoftwo}%
\providecommand \translation [1]{[#1]}%
\providecommand \BibitemOpen [0]{}%
\providecommand \bibitemStop [0]{}%
\providecommand \bibitemNoStop [0]{.\EOS\space}%
\providecommand \EOS [0]{\spacefactor3000\relax}%
\providecommand \BibitemShut  [1]{\csname bibitem#1\endcsname}%
\let\auto@bib@innerbib\@empty
\bibitem [{\citenamefont {{Harry}}\ \emph {et~al.}(2010)\citenamefont {{Harry}}
  \emph {et~al.}}]{Harry2010}%
  \BibitemOpen
  \bibfield  {author} {\bibinfo {author} {\bibfnamefont {G.~M.}\ \bibnamefont
  {{Harry}}} \emph {et~al.},\ }\href {\doibase 10.1088/0264-9381/27/8/084006}
  {\bibfield  {journal} {\bibinfo  {journal} {Class. Quantum Grav.}\ }\textbf
  {\bibinfo {volume} {27}},\ \bibinfo {pages} {084006} (\bibinfo {year}
  {2010})}\BibitemShut {NoStop}%
\bibitem [{\citenamefont {{Accadia}}\ \emph {et~al.}(2011)\citenamefont
  {{Accadia}} \emph {et~al.}}]{Accadia2011_etal}%
  \BibitemOpen
  \bibfield  {author} {\bibinfo {author} {\bibfnamefont {T.}~\bibnamefont
  {{Accadia}}} \emph {et~al.},\ }\href {\doibase
  10.1088/0264-9381/28/11/114002} {\bibfield  {journal} {\bibinfo  {journal}
  {Class. Quantum Grav.}\ }\textbf {\bibinfo {volume} {28}},\ \bibinfo {eid}
  {114002} (\bibinfo {year} {2011})}\BibitemShut {NoStop}%
\bibitem [{\citenamefont {{Aso}}\ \emph {et~al.}(2013)\citenamefont {{Aso}},
  \citenamefont {{Michimura}}, \citenamefont {{Somiya}}, \citenamefont
  {{Ando}}, \citenamefont {{Miyakawa}}, \citenamefont {{Sekiguchi}},
  \citenamefont {{Tatsumi}},\ and\ \citenamefont {{Yamamoto}}}]{Aso:2013}%
  \BibitemOpen
  \bibfield  {author} {\bibinfo {author} {\bibfnamefont {Y.}~\bibnamefont
  {{Aso}}}, \bibinfo {author} {\bibfnamefont {Y.}~\bibnamefont {{Michimura}}},
  \bibinfo {author} {\bibfnamefont {K.}~\bibnamefont {{Somiya}}}, \bibinfo
  {author} {\bibfnamefont {M.}~\bibnamefont {{Ando}}}, \bibinfo {author}
  {\bibfnamefont {O.}~\bibnamefont {{Miyakawa}}}, \bibinfo {author}
  {\bibfnamefont {T.}~\bibnamefont {{Sekiguchi}}}, \bibinfo {author}
  {\bibfnamefont {D.}~\bibnamefont {{Tatsumi}}}, \ and\ \bibinfo {author}
  {\bibfnamefont {H.}~\bibnamefont {{Yamamoto}}},\ }\href {\doibase
  10.1103/PhysRevD.88.043007} {\bibfield  {journal} {\bibinfo  {journal} {Phys.
  Rev. D}\ }\textbf {\bibinfo {volume} {88}},\ \bibinfo {eid} {043007}
  (\bibinfo {year} {2013})}\BibitemShut {NoStop}%
\bibitem [{\citenamefont {{Abadie}}\ \emph {et~al.}(2010)\citenamefont
  {{Abadie}} \emph {et~al.}}]{Abadie:2010_etal}%
  \BibitemOpen
  \bibfield  {author} {\bibinfo {author} {\bibfnamefont {J.}~\bibnamefont
  {{Abadie}}} \emph {et~al.},\ }\href {\doibase 10.1088/0264-9381/27/17/173001}
  {\bibfield  {journal} {\bibinfo  {journal} {Class. Quantum Grav.}\ }\textbf
  {\bibinfo {volume} {27}},\ \bibinfo {pages} {173001} (\bibinfo {year}
  {2010})}\BibitemShut {NoStop}%
\bibitem [{\citenamefont {{Narayan}}\ \emph {et~al.}(1992)\citenamefont
  {{Narayan}}, \citenamefont {{Paczynski}},\ and\ \citenamefont
  {{Piran}}}]{Narayan92}%
  \BibitemOpen
  \bibfield  {author} {\bibinfo {author} {\bibfnamefont {R.}~\bibnamefont
  {{Narayan}}}, \bibinfo {author} {\bibfnamefont {B.}~\bibnamefont
  {{Paczynski}}}, \ and\ \bibinfo {author} {\bibfnamefont {T.}~\bibnamefont
  {{Piran}}},\ }\href {\doibase 10.1086/186493} {\bibfield  {journal} {\bibinfo
   {journal} {Astrophysical Journal, Letters}\ }\textbf {\bibinfo {volume}
  {395}},\ \bibinfo {pages} {L83} (\bibinfo {year} {1992})}\BibitemShut
  {NoStop}%
\bibitem [{\citenamefont {{Eichler}}\ \emph {et~al.}(1989)\citenamefont
  {{Eichler}}, \citenamefont {{Livio}}, \citenamefont {{Piran}},\ and\
  \citenamefont {{Schramm}}}]{Eichler89}%
  \BibitemOpen
  \bibfield  {author} {\bibinfo {author} {\bibfnamefont {D.}~\bibnamefont
  {{Eichler}}}, \bibinfo {author} {\bibfnamefont {M.}~\bibnamefont {{Livio}}},
  \bibinfo {author} {\bibfnamefont {T.}~\bibnamefont {{Piran}}}, \ and\
  \bibinfo {author} {\bibfnamefont {D.~N.}\ \bibnamefont {{Schramm}}},\ }\href
  {\doibase 10.1038/340126a0} {\bibfield  {journal} {\bibinfo  {journal}
  {Nature}\ }\textbf {\bibinfo {volume} {340}},\ \bibinfo {pages} {126}
  (\bibinfo {year} {1989})}\BibitemShut {NoStop}%
\bibitem [{\citenamefont {{Shibata}}\ and\ \citenamefont {{Ury{\=
  u}}}(2000)}]{Shibata99d}%
  \BibitemOpen
  \bibfield  {author} {\bibinfo {author} {\bibfnamefont {M.}~\bibnamefont
  {{Shibata}}}\ and\ \bibinfo {author} {\bibfnamefont {K.}~\bibnamefont
  {{Ury{\= u}}}},\ }\href {\doibase 10.1103/PhysRevD.61.064001} {\bibfield
  {journal} {\bibinfo  {journal} {Phys. Rev. D}\ }\textbf {\bibinfo {volume}
  {61}},\ \bibinfo {eid} {064001} (\bibinfo {year} {2000})}\BibitemShut
  {NoStop}%
\bibitem [{\citenamefont {{Baiotti}}\ \emph {et~al.}(2008)\citenamefont
  {{Baiotti}}, \citenamefont {{Giacomazzo}},\ and\ \citenamefont
  {{Rezzolla}}}]{Baiotti08}%
  \BibitemOpen
  \bibfield  {author} {\bibinfo {author} {\bibfnamefont {L.}~\bibnamefont
  {{Baiotti}}}, \bibinfo {author} {\bibfnamefont {B.}~\bibnamefont
  {{Giacomazzo}}}, \ and\ \bibinfo {author} {\bibfnamefont {L.}~\bibnamefont
  {{Rezzolla}}},\ }\href {\doibase 10.1103/PhysRevD.78.084033} {\bibfield
  {journal} {\bibinfo  {journal} {Phys. Rev. D}\ }\textbf {\bibinfo {volume}
  {78}},\ \bibinfo {pages} {084033} (\bibinfo {year} {2008})}\BibitemShut
  {NoStop}%
\bibitem [{\citenamefont {{Anderson}}\ \emph {et~al.}(2008)\citenamefont
  {{Anderson}}, \citenamefont {{Hirschmann}}, \citenamefont {{Lehner}},
  \citenamefont {{Liebling}}, \citenamefont {{Motl}}, \citenamefont
  {{Neilsen}}, \citenamefont {{Palenzuela}},\ and\ \citenamefont
  {{Tohline}}}]{Anderson2007}%
  \BibitemOpen
  \bibfield  {author} {\bibinfo {author} {\bibfnamefont {M.}~\bibnamefont
  {{Anderson}}}, \bibinfo {author} {\bibfnamefont {E.~W.}\ \bibnamefont
  {{Hirschmann}}}, \bibinfo {author} {\bibfnamefont {L.}~\bibnamefont
  {{Lehner}}}, \bibinfo {author} {\bibfnamefont {S.~L.}\ \bibnamefont
  {{Liebling}}}, \bibinfo {author} {\bibfnamefont {P.~M.}\ \bibnamefont
  {{Motl}}}, \bibinfo {author} {\bibfnamefont {D.}~\bibnamefont {{Neilsen}}},
  \bibinfo {author} {\bibfnamefont {C.}~\bibnamefont {{Palenzuela}}}, \ and\
  \bibinfo {author} {\bibfnamefont {J.~E.}\ \bibnamefont {{Tohline}}},\ }\href
  {\doibase 10.1103/PhysRevD.77.024006} {\bibfield  {journal} {\bibinfo
  {journal} {Phys. Rev. D}\ }\textbf {\bibinfo {volume} {77}},\ \bibinfo {eid}
  {024006} (\bibinfo {year} {2008})}\BibitemShut {NoStop}%
\bibitem [{\citenamefont {Liu}\ \emph {et~al.}(2008)\citenamefont {Liu},
  \citenamefont {Shapiro}, \citenamefont {Etienne},\ and\ \citenamefont
  {Taniguchi}}]{Liu:2008xy}%
  \BibitemOpen
  \bibfield  {author} {\bibinfo {author} {\bibfnamefont {Y.~T.}\ \bibnamefont
  {Liu}}, \bibinfo {author} {\bibfnamefont {S.~L.}\ \bibnamefont {Shapiro}},
  \bibinfo {author} {\bibfnamefont {Z.~B.}\ \bibnamefont {Etienne}}, \ and\
  \bibinfo {author} {\bibfnamefont {K.}~\bibnamefont {Taniguchi}},\ }\href
  {\doibase 10.1103/PhysRevD.78.024012} {\bibfield  {journal} {\bibinfo
  {journal} {Phys. Rev. D}\ }\textbf {\bibinfo {volume} {78}},\ \bibinfo
  {pages} {024012} (\bibinfo {year} {2008})}\BibitemShut {NoStop}%
\bibitem [{\citenamefont {{Bernuzzi}}\ \emph
  {et~al.}(2012{\natexlab{a}})\citenamefont {{Bernuzzi}}, \citenamefont
  {{Thierfelder}},\ and\ \citenamefont {{Br{\"u}gmann}}}]{Bernuzzi2011}%
  \BibitemOpen
  \bibfield  {author} {\bibinfo {author} {\bibfnamefont {S.}~\bibnamefont
  {{Bernuzzi}}}, \bibinfo {author} {\bibfnamefont {M.}~\bibnamefont
  {{Thierfelder}}}, \ and\ \bibinfo {author} {\bibfnamefont {B.}~\bibnamefont
  {{Br{\"u}gmann}}},\ }\href {\doibase 10.1103/PhysRevD.85.104030} {\bibfield
  {journal} {\bibinfo  {journal} {Phys. Rev. D}\ }\textbf {\bibinfo {volume}
  {85}},\ \bibinfo {eid} {104030} (\bibinfo {year}
  {2012}{\natexlab{a}})}\BibitemShut {NoStop}%
\bibitem [{\citenamefont {{Berger}}(2014)}]{Berger2013b}%
  \BibitemOpen
  \bibfield  {author} {\bibinfo {author} {\bibfnamefont {E.}~\bibnamefont
  {{Berger}}},\ }\href {\doibase 10.1146/annurev-astro-081913-035926}
  {\bibfield  {journal} {\bibinfo  {journal} {Annual Review of Astron. and
  Astrophys.}\ }\textbf {\bibinfo {volume} {52}},\ \bibinfo {pages} {43}
  (\bibinfo {year} {2014})}\BibitemShut {NoStop}%
\bibitem [{\citenamefont {{Palenzuela}}\ \emph {et~al.}(2013)\citenamefont
  {{Palenzuela}}, \citenamefont {{Lehner}}, \citenamefont {{Ponce}},
  \citenamefont {{Liebling}}, \citenamefont {{Anderson}}, \citenamefont
  {{Neilsen}},\ and\ \citenamefont {{Motl}}}]{Palenzuela2013a}%
  \BibitemOpen
  \bibfield  {author} {\bibinfo {author} {\bibfnamefont {C.}~\bibnamefont
  {{Palenzuela}}}, \bibinfo {author} {\bibfnamefont {L.}~\bibnamefont
  {{Lehner}}}, \bibinfo {author} {\bibfnamefont {M.}~\bibnamefont {{Ponce}}},
  \bibinfo {author} {\bibfnamefont {S.~L.}\ \bibnamefont {{Liebling}}},
  \bibinfo {author} {\bibfnamefont {M.}~\bibnamefont {{Anderson}}}, \bibinfo
  {author} {\bibfnamefont {D.}~\bibnamefont {{Neilsen}}}, \ and\ \bibinfo
  {author} {\bibfnamefont {P.}~\bibnamefont {{Motl}}},\ }\href {\doibase
  10.1103/PhysRevLett.111.061105} {\bibfield  {journal} {\bibinfo  {journal}
  {Phys. Rev. Lett.}\ }\textbf {\bibinfo {volume} {111}},\ \bibinfo {eid}
  {061105} (\bibinfo {year} {2013})}\BibitemShut {NoStop}%
\bibitem [{\citenamefont {{Siegel}}\ \emph {et~al.}(2013)\citenamefont
  {{Siegel}}, \citenamefont {{Ciolfi}}, \citenamefont {{Harte}},\ and\
  \citenamefont {{Rezzolla}}}]{Siegel2013}%
  \BibitemOpen
  \bibfield  {author} {\bibinfo {author} {\bibfnamefont {D.~M.}\ \bibnamefont
  {{Siegel}}}, \bibinfo {author} {\bibfnamefont {R.}~\bibnamefont {{Ciolfi}}},
  \bibinfo {author} {\bibfnamefont {A.~I.}\ \bibnamefont {{Harte}}}, \ and\
  \bibinfo {author} {\bibfnamefont {L.}~\bibnamefont {{Rezzolla}}},\ }\href
  {\doibase 10.1103/PhysRevD.87.121302} {\bibfield  {journal} {\bibinfo
  {journal} {Phys. Rev. D R}\ }\textbf {\bibinfo {volume} {87}},\ \bibinfo
  {eid} {121302} (\bibinfo {year} {2013})}\BibitemShut {NoStop}%
\bibitem [{\citenamefont {{Kiuchi}}\ \emph {et~al.}(2014)\citenamefont
  {{Kiuchi}}, \citenamefont {{Kyutoku}}, \citenamefont {{Sekiguchi}},
  \citenamefont {{Shibata}},\ and\ \citenamefont {{Wada}}}]{Kiuchi2014}%
  \BibitemOpen
  \bibfield  {author} {\bibinfo {author} {\bibfnamefont {K.}~\bibnamefont
  {{Kiuchi}}}, \bibinfo {author} {\bibfnamefont {K.}~\bibnamefont {{Kyutoku}}},
  \bibinfo {author} {\bibfnamefont {Y.}~\bibnamefont {{Sekiguchi}}}, \bibinfo
  {author} {\bibfnamefont {M.}~\bibnamefont {{Shibata}}}, \ and\ \bibinfo
  {author} {\bibfnamefont {T.}~\bibnamefont {{Wada}}},\ }\href {\doibase
  10.1103/PhysRevD.90.041502} {\bibfield  {journal} {\bibinfo  {journal} {Phys.
  Rev. D}\ }\textbf {\bibinfo {volume} {90}},\ \bibinfo {eid} {041502}
  (\bibinfo {year} {2014})}\BibitemShut {NoStop}%
\bibitem [{\citenamefont {{Rezzolla}}\ \emph {et~al.}(2011)\citenamefont
  {{Rezzolla}}, \citenamefont {{Giacomazzo}}, \citenamefont {{Baiotti}},
  \citenamefont {{Granot}}, \citenamefont {{Kouveliotou}},\ and\ \citenamefont
  {{Aloy}}}]{Rezzolla:2011}%
  \BibitemOpen
  \bibfield  {author} {\bibinfo {author} {\bibfnamefont {L.}~\bibnamefont
  {{Rezzolla}}}, \bibinfo {author} {\bibfnamefont {B.}~\bibnamefont
  {{Giacomazzo}}}, \bibinfo {author} {\bibfnamefont {L.}~\bibnamefont
  {{Baiotti}}}, \bibinfo {author} {\bibfnamefont {J.}~\bibnamefont {{Granot}}},
  \bibinfo {author} {\bibfnamefont {C.}~\bibnamefont {{Kouveliotou}}}, \ and\
  \bibinfo {author} {\bibfnamefont {M.~A.}\ \bibnamefont {{Aloy}}},\ }\href
  {\doibase 10.1088/2041-8205/732/1/L6} {\bibfield  {journal} {\bibinfo
  {journal} {Astrophys. J. Letters}\ }\textbf {\bibinfo {volume} {732}},\
  \bibinfo {eid} {L6} (\bibinfo {year} {2011})}\BibitemShut {NoStop}%
\bibitem [{\citenamefont {{Paschalidis}}\ \emph {et~al.}(2014)\citenamefont
  {{Paschalidis}}, \citenamefont {{Ruiz}},\ and\ \citenamefont
  {{Shapiro}}}]{Paschalidis2014}%
  \BibitemOpen
  \bibfield  {author} {\bibinfo {author} {\bibfnamefont {V.}~\bibnamefont
  {{Paschalidis}}}, \bibinfo {author} {\bibfnamefont {M.}~\bibnamefont
  {{Ruiz}}}, \ and\ \bibinfo {author} {\bibfnamefont {S.~L.}\ \bibnamefont
  {{Shapiro}}},\ }\href@noop {} {\bibfield  {journal} {\bibinfo  {journal}
  {arXiv:1410.7392}\ } (\bibinfo {year} {2014})}\BibitemShut {NoStop}%
\bibitem [{\citenamefont {{Flanagan}}\ and\ \citenamefont
  {{Hinderer}}(2008)}]{Flanagan08}%
  \BibitemOpen
  \bibfield  {author} {\bibinfo {author} {\bibfnamefont {{\'E}.~{\'E}.}\
  \bibnamefont {{Flanagan}}}\ and\ \bibinfo {author} {\bibfnamefont
  {T.}~\bibnamefont {{Hinderer}}},\ }\href {\doibase
  10.1103/PhysRevD.77.021502} {\bibfield  {journal} {\bibinfo  {journal} {Phys.
  Rev. D}\ }\textbf {\bibinfo {volume} {77}},\ \bibinfo {pages} {021502}
  (\bibinfo {year} {2008})}\BibitemShut {NoStop}%
\bibitem [{\citenamefont {Baiotti}\ \emph {et~al.}(2010)\citenamefont
  {Baiotti}, \citenamefont {Damour}, \citenamefont {Giacomazzo}, \citenamefont
  {Nagar},\ and\ \citenamefont {Rezzolla}}]{Baiotti:2010}%
  \BibitemOpen
  \bibfield  {author} {\bibinfo {author} {\bibfnamefont {L.}~\bibnamefont
  {Baiotti}}, \bibinfo {author} {\bibfnamefont {T.}~\bibnamefont {Damour}},
  \bibinfo {author} {\bibfnamefont {B.}~\bibnamefont {Giacomazzo}}, \bibinfo
  {author} {\bibfnamefont {A.}~\bibnamefont {Nagar}}, \ and\ \bibinfo {author}
  {\bibfnamefont {L.}~\bibnamefont {Rezzolla}},\ }\href {\doibase
  10.1103/PhysRevLett.105.261101} {\bibfield  {journal} {\bibinfo  {journal}
  {Phys. Rev. Lett.}\ }\textbf {\bibinfo {volume} {105}},\ \bibinfo {pages}
  {261101} (\bibinfo {year} {2010})}\BibitemShut {NoStop}%
\bibitem [{\citenamefont {{Bernuzzi}}\ \emph
  {et~al.}(2012{\natexlab{b}})\citenamefont {{Bernuzzi}}, \citenamefont
  {{Nagar}}, \citenamefont {{Thierfelder}},\ and\ \citenamefont
  {{Br{\"u}gmann}}}]{Bernuzzi2012}%
  \BibitemOpen
  \bibfield  {author} {\bibinfo {author} {\bibfnamefont {S.}~\bibnamefont
  {{Bernuzzi}}}, \bibinfo {author} {\bibfnamefont {A.}~\bibnamefont {{Nagar}}},
  \bibinfo {author} {\bibfnamefont {M.}~\bibnamefont {{Thierfelder}}}, \ and\
  \bibinfo {author} {\bibfnamefont {B.}~\bibnamefont {{Br{\"u}gmann}}},\ }\href
  {\doibase 10.1103/PhysRevD.86.044030} {\bibfield  {journal} {\bibinfo
  {journal} {Phys. Rev. D}\ }\textbf {\bibinfo {volume} {86}},\ \bibinfo {eid}
  {044030} (\bibinfo {year} {2012}{\natexlab{b}})}\BibitemShut {NoStop}%
\bibitem [{\citenamefont {{Radice}}\ \emph
  {et~al.}(2014{\natexlab{a}})\citenamefont {{Radice}}, \citenamefont
  {{Rezzolla}},\ and\ \citenamefont {{Galeazzi}}}]{Radice2013b}%
  \BibitemOpen
  \bibfield  {author} {\bibinfo {author} {\bibfnamefont {D.}~\bibnamefont
  {{Radice}}}, \bibinfo {author} {\bibfnamefont {L.}~\bibnamefont
  {{Rezzolla}}}, \ and\ \bibinfo {author} {\bibfnamefont {F.}~\bibnamefont
  {{Galeazzi}}},\ }\href {\doibase 10.1093/mnrasl/slt137} {\bibfield  {journal}
  {\bibinfo  {journal} {Mon. Not. R. Astron. Soc. L.}\ }\textbf {\bibinfo
  {volume} {437}},\ \bibinfo {pages} {L46} (\bibinfo {year}
  {2014}{\natexlab{a}})}\BibitemShut {NoStop}%
\bibitem [{\citenamefont {{Radice}}\ \emph
  {et~al.}(2014{\natexlab{b}})\citenamefont {{Radice}}, \citenamefont
  {{Rezzolla}},\ and\ \citenamefont {{Galeazzi}}}]{Radice2013c}%
  \BibitemOpen
  \bibfield  {author} {\bibinfo {author} {\bibfnamefont {D.}~\bibnamefont
  {{Radice}}}, \bibinfo {author} {\bibfnamefont {L.}~\bibnamefont
  {{Rezzolla}}}, \ and\ \bibinfo {author} {\bibfnamefont {F.}~\bibnamefont
  {{Galeazzi}}},\ }\href {\doibase 10.1088/0264-9381/31/7/075012} {\bibfield
  {journal} {\bibinfo  {journal} {Class. Quantum Grav.}\ }\textbf {\bibinfo
  {volume} {31}},\ \bibinfo {eid} {075012} (\bibinfo {year}
  {2014}{\natexlab{b}})}\BibitemShut {NoStop}%
\bibitem [{\citenamefont {{Shibata}}(2005)}]{Shibata05d}%
  \BibitemOpen
  \bibfield  {author} {\bibinfo {author} {\bibfnamefont {M.}~\bibnamefont
  {{Shibata}}},\ }\href {\doibase 10.1103/PhysRevLett.94.201101} {\bibfield
  {journal} {\bibinfo  {journal} {Phys. Rev. Lett.}\ }\textbf {\bibinfo
  {volume} {94}},\ \bibinfo {eid} {201101} (\bibinfo {year}
  {2005})}\BibitemShut {NoStop}%
\bibitem [{\citenamefont {Oechslin}\ and\ \citenamefont
  {Janka}(2007)}]{Oechslin07b}%
  \BibitemOpen
  \bibfield  {author} {\bibinfo {author} {\bibfnamefont {R.}~\bibnamefont
  {Oechslin}}\ and\ \bibinfo {author} {\bibfnamefont {H.~T.}\ \bibnamefont
  {Janka}},\ }\href {\doibase 10.1103/PhysRevLett.99.121102} {\bibfield
  {journal} {\bibinfo  {journal} {Phys. Rev. Lett.}\ }\textbf {\bibinfo
  {volume} {99}},\ \bibinfo {pages} {121102} (\bibinfo {year}
  {2007})}\BibitemShut {NoStop}%
\bibitem [{\citenamefont {{Bauswein}}\ and\ \citenamefont
  {{Janka}}(2012)}]{Bauswein2011}%
  \BibitemOpen
  \bibfield  {author} {\bibinfo {author} {\bibfnamefont {A.}~\bibnamefont
  {{Bauswein}}}\ and\ \bibinfo {author} {\bibfnamefont {H.-T.}\ \bibnamefont
  {{Janka}}},\ }\href {\doibase 10.1103/PhysRevLett.108.011101} {\bibfield
  {journal} {\bibinfo  {journal} {Phys. Rev. Lett.}\ }\textbf {\bibinfo
  {volume} {108}},\ \bibinfo {eid} {011101} (\bibinfo {year}
  {2012})}\BibitemShut {NoStop}%
\bibitem [{\citenamefont {{Bauswein}}\ \emph {et~al.}(2012)\citenamefont
  {{Bauswein}}, \citenamefont {{Janka}}, \citenamefont {{Hebeler}},\ and\
  \citenamefont {{Schwenk}}}]{Bauswein2012}%
  \BibitemOpen
  \bibfield  {author} {\bibinfo {author} {\bibfnamefont {A.}~\bibnamefont
  {{Bauswein}}}, \bibinfo {author} {\bibfnamefont {H.-T.}\ \bibnamefont
  {{Janka}}}, \bibinfo {author} {\bibfnamefont {K.}~\bibnamefont {{Hebeler}}},
  \ and\ \bibinfo {author} {\bibfnamefont {A.}~\bibnamefont {{Schwenk}}},\
  }\href {\doibase 10.1103/PhysRevD.86.063001} {\bibfield  {journal} {\bibinfo
  {journal} {Phys. Rev. D}\ }\textbf {\bibinfo {volume} {86}},\ \bibinfo {eid}
  {063001} (\bibinfo {year} {2012})}\BibitemShut {NoStop}%
\bibitem [{\citenamefont {{Hotokezaka}}\ \emph {et~al.}(2013)\citenamefont
  {{Hotokezaka}}, \citenamefont {{Kiuchi}}, \citenamefont {{Kyutoku}},
  \citenamefont {{Muranushi}}, \citenamefont {{Sekiguchi}}, \citenamefont
  {{Shibata}},\ and\ \citenamefont {{Taniguchi}}}]{Hotokezaka2013c}%
  \BibitemOpen
  \bibfield  {author} {\bibinfo {author} {\bibfnamefont {K.}~\bibnamefont
  {{Hotokezaka}}}, \bibinfo {author} {\bibfnamefont {K.}~\bibnamefont
  {{Kiuchi}}}, \bibinfo {author} {\bibfnamefont {K.}~\bibnamefont {{Kyutoku}}},
  \bibinfo {author} {\bibfnamefont {T.}~\bibnamefont {{Muranushi}}}, \bibinfo
  {author} {\bibfnamefont {Y.-i.}\ \bibnamefont {{Sekiguchi}}}, \bibinfo
  {author} {\bibfnamefont {M.}~\bibnamefont {{Shibata}}}, \ and\ \bibinfo
  {author} {\bibfnamefont {K.}~\bibnamefont {{Taniguchi}}},\ }\href {\doibase
  10.1103/PhysRevD.88.044026} {\bibfield  {journal} {\bibinfo  {journal} {Phys.
  Rev. D}\ }\textbf {\bibinfo {volume} {88}},\ \bibinfo {eid} {044026}
  (\bibinfo {year} {2013})}\BibitemShut {NoStop}%
\bibitem [{\citenamefont {{Stergioulas}}\ \emph {et~al.}(2011)\citenamefont
  {{Stergioulas}}, \citenamefont {{Bauswein}}, \citenamefont {{Zagkouris}},\
  and\ \citenamefont {{Janka}}}]{Stergioulas2011b}%
  \BibitemOpen
  \bibfield  {author} {\bibinfo {author} {\bibfnamefont {N.}~\bibnamefont
  {{Stergioulas}}}, \bibinfo {author} {\bibfnamefont {A.}~\bibnamefont
  {{Bauswein}}}, \bibinfo {author} {\bibfnamefont {K.}~\bibnamefont
  {{Zagkouris}}}, \ and\ \bibinfo {author} {\bibfnamefont {H.-T.}\ \bibnamefont
  {{Janka}}},\ }\href {\doibase 10.1111/j.1365-2966.2011.19493.x} {\bibfield
  {journal} {\bibinfo  {journal} {Mon. Not. R. Astron. Soc.}\ }\textbf
  {\bibinfo {volume} {418}},\ \bibinfo {pages} {427} (\bibinfo {year}
  {2011})}\BibitemShut {NoStop}%
\bibitem [{\citenamefont {{Bauswein}}\ \emph {et~al.}(2014)\citenamefont
  {{Bauswein}}, \citenamefont {{Stergioulas}},\ and\ \citenamefont
  {{Janka}}}]{Bauswein2014}%
  \BibitemOpen
  \bibfield  {author} {\bibinfo {author} {\bibfnamefont {A.}~\bibnamefont
  {{Bauswein}}}, \bibinfo {author} {\bibfnamefont {N.}~\bibnamefont
  {{Stergioulas}}}, \ and\ \bibinfo {author} {\bibfnamefont {H.-T.}\
  \bibnamefont {{Janka}}},\ }\href {\doibase 10.1103/PhysRevD.90.023002}
  {\bibfield  {journal} {\bibinfo  {journal} {Phys. Rev. D}\ }\textbf {\bibinfo
  {volume} {90}},\ \bibinfo {eid} {023002} (\bibinfo {year}
  {2014})}\BibitemShut {NoStop}%
\bibitem [{\citenamefont {{Takami}}\ \emph
  {et~al.}(2014{\natexlab{a}})\citenamefont {{Takami}}, \citenamefont
  {{Rezzolla}},\ and\ \citenamefont {{Baiotti}}}]{Takami:2014}%
  \BibitemOpen
  \bibfield  {author} {\bibinfo {author} {\bibfnamefont {K.}~\bibnamefont
  {{Takami}}}, \bibinfo {author} {\bibfnamefont {L.}~\bibnamefont
  {{Rezzolla}}}, \ and\ \bibinfo {author} {\bibfnamefont {L.}~\bibnamefont
  {{Baiotti}}},\ }\href {\doibase 10.1103/PhysRevLett.113.091104} {\bibfield
  {journal} {\bibinfo  {journal} {Phys. Rev. Lett.}\ }\textbf {\bibinfo
  {volume} {113}},\ \bibinfo {eid} {091104} (\bibinfo {year}
  {2014}{\natexlab{a}})}\BibitemShut {NoStop}%
\bibitem [{\citenamefont {{Read}}\ \emph {et~al.}(2013)\citenamefont {{Read}},
  \citenamefont {{Baiotti}}, \citenamefont {{Creighton}}, \citenamefont
  {{Friedman}}, \citenamefont {{Giacomazzo}}, \citenamefont {{Kyutoku}},
  \citenamefont {{Markakis}}, \citenamefont {{Rezzolla}}, \citenamefont
  {{Shibata}},\ and\ \citenamefont {{Taniguchi}}}]{Read2013}%
  \BibitemOpen
  \bibfield  {author} {\bibinfo {author} {\bibfnamefont {J.~S.}\ \bibnamefont
  {{Read}}}, \bibinfo {author} {\bibfnamefont {L.}~\bibnamefont {{Baiotti}}},
  \bibinfo {author} {\bibfnamefont {J.~D.~E.}\ \bibnamefont {{Creighton}}},
  \bibinfo {author} {\bibfnamefont {J.~L.}\ \bibnamefont {{Friedman}}},
  \bibinfo {author} {\bibfnamefont {B.}~\bibnamefont {{Giacomazzo}}}, \bibinfo
  {author} {\bibfnamefont {K.}~\bibnamefont {{Kyutoku}}}, \bibinfo {author}
  {\bibfnamefont {C.}~\bibnamefont {{Markakis}}}, \bibinfo {author}
  {\bibfnamefont {L.}~\bibnamefont {{Rezzolla}}}, \bibinfo {author}
  {\bibfnamefont {M.}~\bibnamefont {{Shibata}}}, \ and\ \bibinfo {author}
  {\bibfnamefont {K.}~\bibnamefont {{Taniguchi}}},\ }\href {\doibase
  10.1103/PhysRevD.88.044042} {\bibfield  {journal} {\bibinfo  {journal} {Phys.
  Rev. D}\ }\textbf {\bibinfo {volume} {88}},\ \bibinfo {eid} {044042}
  (\bibinfo {year} {2013})}\BibitemShut {NoStop}%
\bibitem [{\citenamefont {{Bernuzzi}}\ \emph
  {et~al.}(2014{\natexlab{a}})\citenamefont {{Bernuzzi}}, \citenamefont
  {{Nagar}}, \citenamefont {{Balmelli}}, \citenamefont {{Dietrich}},\ and\
  \citenamefont {{Ujevic}}}]{Bernuzzi2014}%
  \BibitemOpen
  \bibfield  {author} {\bibinfo {author} {\bibfnamefont {S.}~\bibnamefont
  {{Bernuzzi}}}, \bibinfo {author} {\bibfnamefont {A.}~\bibnamefont {{Nagar}}},
  \bibinfo {author} {\bibfnamefont {S.}~\bibnamefont {{Balmelli}}}, \bibinfo
  {author} {\bibfnamefont {T.}~\bibnamefont {{Dietrich}}}, \ and\ \bibinfo
  {author} {\bibfnamefont {M.}~\bibnamefont {{Ujevic}}},\ }\href {\doibase
  10.1103/PhysRevLett.112.201101} {\bibfield  {journal} {\bibinfo  {journal}
  {Phys. Rev. Lett.}\ }\textbf {\bibinfo {volume} {112}},\ \bibinfo {eid}
  {201101} (\bibinfo {year} {2014}{\natexlab{a}})}\BibitemShut {NoStop}%
\bibitem [{rel()}]{relastrowiki41}%
  \BibitemOpen
  \href@noop {} {}\bibinfo {howpublished} {URL
  \url{http://astro.uni-frankfurt.de/relastrowiki}}\BibitemShut {NoStop}%
\bibitem [{\citenamefont {{Baiotti}}\ \emph {et~al.}(2009)\citenamefont
  {{Baiotti}}, \citenamefont {{Giacomazzo}},\ and\ \citenamefont
  {{Rezzolla}}}]{Baiotti:2009gk}%
  \BibitemOpen
  \bibfield  {author} {\bibinfo {author} {\bibfnamefont {L.}~\bibnamefont
  {{Baiotti}}}, \bibinfo {author} {\bibfnamefont {B.}~\bibnamefont
  {{Giacomazzo}}}, \ and\ \bibinfo {author} {\bibfnamefont {L.}~\bibnamefont
  {{Rezzolla}}},\ }\href {\doibase 10.1088/0264-9381/26/11/114005} {\bibfield
  {journal} {\bibinfo  {journal} {Classical and Quantum Gravity}\ }\textbf
  {\bibinfo {volume} {26}},\ \bibinfo {eid} {114005} (\bibinfo {year}
  {2009})}\BibitemShut {NoStop}%
\bibitem [{\citenamefont {{Baiotti}}\ \emph {et~al.}(2010)\citenamefont
  {{Baiotti}}, \citenamefont {{Shibata}},\ and\ \citenamefont
  {{Yamamoto}}}]{Baiotti:2010ka}%
  \BibitemOpen
  \bibfield  {author} {\bibinfo {author} {\bibfnamefont {L.}~\bibnamefont
  {{Baiotti}}}, \bibinfo {author} {\bibfnamefont {M.}~\bibnamefont
  {{Shibata}}}, \ and\ \bibinfo {author} {\bibfnamefont {T.}~\bibnamefont
  {{Yamamoto}}},\ }\href {\doibase 10.1103/PhysRevD.82.064015} {\bibfield
  {journal} {\bibinfo  {journal} {Phys. Rev. D}\ }\textbf {\bibinfo {volume}
  {82}},\ \bibinfo {eid} {064015} (\bibinfo {year} {2010})}\BibitemShut
  {NoStop}%
\bibitem [{\citenamefont {Brown}\ \emph {et~al.}(2009)\citenamefont {Brown},
  \citenamefont {Diener}, \citenamefont {Sarbach}, \citenamefont {Schnetter},\
  and\ \citenamefont {Tiglio}}]{Brown:2008sb}%
  \BibitemOpen
  \bibfield  {author} {\bibinfo {author} {\bibfnamefont {D.}~\bibnamefont
  {Brown}}, \bibinfo {author} {\bibfnamefont {P.}~\bibnamefont {Diener}},
  \bibinfo {author} {\bibfnamefont {O.}~\bibnamefont {Sarbach}}, \bibinfo
  {author} {\bibfnamefont {E.}~\bibnamefont {Schnetter}}, \ and\ \bibinfo
  {author} {\bibfnamefont {M.}~\bibnamefont {Tiglio}},\ }\href
  {http://dx.doi.org/10.1103/PhysRevD.79.044023} {\bibfield  {journal}
  {\bibinfo  {journal} {Phys. Rev. D}\ }\textbf {\bibinfo {volume} {79}},\
  \bibinfo {pages} {044023} (\bibinfo {year} {2009})}\BibitemShut {NoStop}%
\bibitem [{\citenamefont {L{\"{o}}ffler}\ \emph {et~al.}(2012)\citenamefont
  {L{\"{o}}ffler}, \citenamefont {Faber}, \citenamefont {Bentivegna},
  \citenamefont {Bode}, \citenamefont {Diener}, \citenamefont {Haas},
  \citenamefont {Hinder}, \citenamefont {Mundim}, \citenamefont {Ott},
  \citenamefont {Schnetter}, \citenamefont {Allen}, \citenamefont
  {Campanelli},\ and\ \citenamefont {Laguna}}]{Loffler:2011ay}%
  \BibitemOpen
  \bibfield  {author} {\bibinfo {author} {\bibfnamefont {F.}~\bibnamefont
  {L{\"{o}}ffler}}, \bibinfo {author} {\bibfnamefont {J.}~\bibnamefont
  {Faber}}, \bibinfo {author} {\bibfnamefont {E.}~\bibnamefont {Bentivegna}},
  \bibinfo {author} {\bibfnamefont {T.}~\bibnamefont {Bode}}, \bibinfo {author}
  {\bibfnamefont {P.}~\bibnamefont {Diener}}, \bibinfo {author} {\bibfnamefont
  {R.}~\bibnamefont {Haas}}, \bibinfo {author} {\bibfnamefont {I.}~\bibnamefont
  {Hinder}}, \bibinfo {author} {\bibfnamefont {B.~C.}\ \bibnamefont {Mundim}},
  \bibinfo {author} {\bibfnamefont {C.~D.}\ \bibnamefont {Ott}}, \bibinfo
  {author} {\bibfnamefont {E.}~\bibnamefont {Schnetter}}, \bibinfo {author}
  {\bibfnamefont {G.}~\bibnamefont {Allen}}, \bibinfo {author} {\bibfnamefont
  {M.}~\bibnamefont {Campanelli}}, \ and\ \bibinfo {author} {\bibfnamefont
  {P.}~\bibnamefont {Laguna}},\ }\href {\doibase
  doi:10.1088/0264-9381/29/11/115001} {\bibfield  {journal} {\bibinfo
  {journal} {Class. Quantum Grav.}\ }\textbf {\bibinfo {volume} {29}},\
  \bibinfo {pages} {115001} (\bibinfo {year} {2012})}\BibitemShut {NoStop}%
\bibitem [{\citenamefont {{Nakamura}}\ \emph {et~al.}(1987)\citenamefont
  {{Nakamura}}, \citenamefont {{Oohara}},\ and\ \citenamefont
  {{Kojima}}}]{Nakamura87}%
  \BibitemOpen
  \bibfield  {author} {\bibinfo {author} {\bibfnamefont {T.}~\bibnamefont
  {{Nakamura}}}, \bibinfo {author} {\bibfnamefont {K.}~\bibnamefont
  {{Oohara}}}, \ and\ \bibinfo {author} {\bibfnamefont {Y.}~\bibnamefont
  {{Kojima}}},\ }\href {\doibase 10.1143/PTPS.90.1} {\bibfield  {journal}
  {\bibinfo  {journal} {Progress of Theoretical Physics Supplement}\ }\textbf
  {\bibinfo {volume} {90}},\ \bibinfo {pages} {1} (\bibinfo {year}
  {1987})}\BibitemShut {NoStop}%
\bibitem [{\citenamefont {{Shibata}}\ and\ \citenamefont
  {{Nakamura}}(1995)}]{Shibata95}%
  \BibitemOpen
  \bibfield  {author} {\bibinfo {author} {\bibfnamefont {M.}~\bibnamefont
  {{Shibata}}}\ and\ \bibinfo {author} {\bibfnamefont {T.}~\bibnamefont
  {{Nakamura}}},\ }\href {\doibase 10.1103/PhysRevD.52.5428} {\bibfield
  {journal} {\bibinfo  {journal} {Phys. Rev. D}\ }\textbf {\bibinfo {volume}
  {52}},\ \bibinfo {pages} {5428} (\bibinfo {year} {1995})}\BibitemShut
  {NoStop}%
\bibitem [{\citenamefont {{Baumgarte}}\ and\ \citenamefont
  {{Shapiro}}(1999)}]{Baumgarte99}%
  \BibitemOpen
  \bibfield  {author} {\bibinfo {author} {\bibfnamefont {T.~W.}\ \bibnamefont
  {{Baumgarte}}}\ and\ \bibinfo {author} {\bibfnamefont {S.~L.}\ \bibnamefont
  {{Shapiro}}},\ }\href {\doibase 10.1103/PhysRevD.59.024007} {\bibfield
  {journal} {\bibinfo  {journal} {Phys. Rev. D}\ }\textbf {\bibinfo {volume}
  {59}},\ \bibinfo {eid} {024007} (\bibinfo {year} {1999})}\BibitemShut
  {NoStop}%
\bibitem [{\citenamefont {{Alcubierre}}\ \emph {et~al.}(2003)\citenamefont
  {{Alcubierre}}, \citenamefont {{Br{\"u}gmann}}, \citenamefont {{Diener}},
  \citenamefont {{Koppitz}}, \citenamefont {{Pollney}}, \citenamefont
  {{Seidel}},\ and\ \citenamefont {{Takahashi}}}]{Alcubierre02a}%
  \BibitemOpen
  \bibfield  {author} {\bibinfo {author} {\bibfnamefont {M.}~\bibnamefont
  {{Alcubierre}}}, \bibinfo {author} {\bibfnamefont {B.}~\bibnamefont
  {{Br{\"u}gmann}}}, \bibinfo {author} {\bibfnamefont {P.}~\bibnamefont
  {{Diener}}}, \bibinfo {author} {\bibfnamefont {M.}~\bibnamefont {{Koppitz}}},
  \bibinfo {author} {\bibfnamefont {D.}~\bibnamefont {{Pollney}}}, \bibinfo
  {author} {\bibfnamefont {E.}~\bibnamefont {{Seidel}}}, \ and\ \bibinfo
  {author} {\bibfnamefont {R.}~\bibnamefont {{Takahashi}}},\ }\href {\doibase
  10.1103/PhysRevD.67.084023} {\bibfield  {journal} {\bibinfo  {journal} {Phys.
  Rev. D}\ }\textbf {\bibinfo {volume} {67}},\ \bibinfo {eid} {084023}
  (\bibinfo {year} {2003})}\BibitemShut {NoStop}%
\bibitem [{\citenamefont {{Pollney}}\ \emph {et~al.}(2007)\citenamefont
  {{Pollney}}, \citenamefont {{Reisswig}}, \citenamefont {{Rezzolla}},
  \citenamefont {{Szil{\'a}gyi}}, \citenamefont {{Ansorg}}, \citenamefont
  {{Deris}}, \citenamefont {{Diener}}, \citenamefont {{Dorband}}, \citenamefont
  {{Koppitz}}, \citenamefont {{Nagar}},\ and\ \citenamefont
  {{Schnetter}}}]{Pollney:2007ss}%
  \BibitemOpen
  \bibfield  {author} {\bibinfo {author} {\bibfnamefont {D.}~\bibnamefont
  {{Pollney}}}, \bibinfo {author} {\bibfnamefont {C.}~\bibnamefont
  {{Reisswig}}}, \bibinfo {author} {\bibfnamefont {L.}~\bibnamefont
  {{Rezzolla}}}, \bibinfo {author} {\bibfnamefont {B.}~\bibnamefont
  {{Szil{\'a}gyi}}}, \bibinfo {author} {\bibfnamefont {M.}~\bibnamefont
  {{Ansorg}}}, \bibinfo {author} {\bibfnamefont {B.}~\bibnamefont {{Deris}}},
  \bibinfo {author} {\bibfnamefont {P.}~\bibnamefont {{Diener}}}, \bibinfo
  {author} {\bibfnamefont {E.~N.}\ \bibnamefont {{Dorband}}}, \bibinfo {author}
  {\bibfnamefont {M.}~\bibnamefont {{Koppitz}}}, \bibinfo {author}
  {\bibfnamefont {A.}~\bibnamefont {{Nagar}}}, \ and\ \bibinfo {author}
  {\bibfnamefont {E.}~\bibnamefont {{Schnetter}}},\ }\href {\doibase
  10.1103/PhysRevD.76.124002} {\bibfield  {journal} {\bibinfo  {journal} {Phys.
  Rev. D}\ }\textbf {\bibinfo {volume} {76}},\ \bibinfo {eid} {124002}
  (\bibinfo {year} {2007})}\BibitemShut {NoStop}%
\bibitem [{\citenamefont {{Baiotti}}\ \emph {et~al.}(2005)\citenamefont
  {{Baiotti}}, \citenamefont {{Hawke}}, \citenamefont {{Montero}},
  \citenamefont {{L{\"o}ffler}}, \citenamefont {{Rezzolla}}, \citenamefont
  {{Stergioulas}}, \citenamefont {{Font}},\ and\ \citenamefont
  {{Seidel}}}]{Baiotti04}%
  \BibitemOpen
  \bibfield  {author} {\bibinfo {author} {\bibfnamefont {L.}~\bibnamefont
  {{Baiotti}}}, \bibinfo {author} {\bibfnamefont {I.}~\bibnamefont {{Hawke}}},
  \bibinfo {author} {\bibfnamefont {P.~J.}\ \bibnamefont {{Montero}}}, \bibinfo
  {author} {\bibfnamefont {F.}~\bibnamefont {{L{\"o}ffler}}}, \bibinfo {author}
  {\bibfnamefont {L.}~\bibnamefont {{Rezzolla}}}, \bibinfo {author}
  {\bibfnamefont {N.}~\bibnamefont {{Stergioulas}}}, \bibinfo {author}
  {\bibfnamefont {J.~A.}\ \bibnamefont {{Font}}}, \ and\ \bibinfo {author}
  {\bibfnamefont {E.}~\bibnamefont {{Seidel}}},\ }\href {\doibase
  10.1103/PhysRevD.71.024035} {\bibfield  {journal} {\bibinfo  {journal} {Phys.
  Rev. D}\ }\textbf {\bibinfo {volume} {71}},\ \bibinfo {eid} {024035}
  (\bibinfo {year} {2005})}\BibitemShut {NoStop}%
\bibitem [{\citenamefont {{Rezzolla}}\ \emph {et~al.}(2010)\citenamefont
  {{Rezzolla}}, \citenamefont {{Baiotti}}, \citenamefont {{Giacomazzo}},
  \citenamefont {{Link}},\ and\ \citenamefont {{Font}}}]{Rezzolla:2010}%
  \BibitemOpen
  \bibfield  {author} {\bibinfo {author} {\bibfnamefont {L.}~\bibnamefont
  {{Rezzolla}}}, \bibinfo {author} {\bibfnamefont {L.}~\bibnamefont
  {{Baiotti}}}, \bibinfo {author} {\bibfnamefont {B.}~\bibnamefont
  {{Giacomazzo}}}, \bibinfo {author} {\bibfnamefont {D.}~\bibnamefont
  {{Link}}}, \ and\ \bibinfo {author} {\bibfnamefont {J.~A.}\ \bibnamefont
  {{Font}}},\ }\href {\doibase 10.1088/0264-9381/27/11/114105} {\bibfield
  {journal} {\bibinfo  {journal} {Class. Quantum Grav.}\ }\textbf {\bibinfo
  {volume} {27}},\ \bibinfo {pages} {114105} (\bibinfo {year}
  {2010})}\BibitemShut {NoStop}%
\bibitem [{\citenamefont {Tsatsin}\ and\ \citenamefont
  {Marronetti}(2013)}]{Tsatsin2013}%
  \BibitemOpen
  \bibfield  {author} {\bibinfo {author} {\bibfnamefont {P.}~\bibnamefont
  {Tsatsin}}\ and\ \bibinfo {author} {\bibfnamefont {P.}~\bibnamefont
  {Marronetti}},\ }\href {\doibase 10.1103/PhysRevD.88.064060} {\bibfield
  {journal} {\bibinfo  {journal} {Phys. Rev. D}\ }\textbf {\bibinfo {volume}
  {88}},\ \bibinfo {pages} {064060} (\bibinfo {year} {2013})}\BibitemShut
  {NoStop}%
\bibitem [{\citenamefont {{Donat}}\ and\ \citenamefont
  {{Marquina}}(1996)}]{Donat96}%
  \BibitemOpen
  \bibfield  {author} {\bibinfo {author} {\bibfnamefont {R.}~\bibnamefont
  {{Donat}}}\ and\ \bibinfo {author} {\bibfnamefont {A.}~\bibnamefont
  {{Marquina}}},\ }\href {\doibase 10.1006/jcph.1996.0078} {\bibfield
  {journal} {\bibinfo  {journal} {Journal of Computational Physics}\ }\textbf
  {\bibinfo {volume} {125}},\ \bibinfo {pages} {42} (\bibinfo {year}
  {1996})}\BibitemShut {NoStop}%
\bibitem [{\citenamefont {Harten}\ \emph {et~al.}(1983)\citenamefont {Harten},
  \citenamefont {Lax},\ and\ \citenamefont {van Leer}}]{Harten83}%
  \BibitemOpen
  \bibfield  {author} {\bibinfo {author} {\bibfnamefont {A.}~\bibnamefont
  {Harten}}, \bibinfo {author} {\bibfnamefont {P.~D.}\ \bibnamefont {Lax}}, \
  and\ \bibinfo {author} {\bibfnamefont {B.}~\bibnamefont {van Leer}},\ }\href
  {\doibase 10.1137/1025002} {\bibfield  {journal} {\bibinfo  {journal} {SIAM
  Rev.}\ }\textbf {\bibinfo {volume} {25}},\ \bibinfo {pages} {35} (\bibinfo
  {year} {1983})}\BibitemShut {NoStop}%
\bibitem [{\citenamefont {Colella}\ and\ \citenamefont
  {Woodward}(1984)}]{Colella84}%
  \BibitemOpen
  \bibfield  {author} {\bibinfo {author} {\bibfnamefont {P.}~\bibnamefont
  {Colella}}\ and\ \bibinfo {author} {\bibfnamefont {P.~R.}\ \bibnamefont
  {Woodward}},\ }\href {\doibase DOI: 10.1016/0021-9991(84)90143-8} {\bibfield
  {journal} {\bibinfo  {journal} {Journal of Computational Physics}\ }\textbf
  {\bibinfo {volume} {54}},\ \bibinfo {pages} {174} (\bibinfo {year}
  {1984})}\BibitemShut {NoStop}%
\bibitem [{\citenamefont {{Rezzolla}}\ and\ \citenamefont
  {{Zanotti}}(2013)}]{Rezzolla_book:2013}%
  \BibitemOpen
  \bibfield  {author} {\bibinfo {author} {\bibfnamefont {L.}~\bibnamefont
  {{Rezzolla}}}\ and\ \bibinfo {author} {\bibfnamefont {O.}~\bibnamefont
  {{Zanotti}}},\ }\href@noop {} {\emph {\bibinfo {title} {Relativistic
  Hydrodynamics}}}\ (\bibinfo  {publisher} {Oxford University Press},\ \bibinfo
  {address} {Oxford, UK},\ \bibinfo {year} {2013})\BibitemShut {NoStop}%
\bibitem [{\citenamefont {{Kastaun}}\ \emph {et~al.}(2013)\citenamefont
  {{Kastaun}}, \citenamefont {{Galeazzi}}, \citenamefont {{Alic}},
  \citenamefont {{Rezzolla}},\ and\ \citenamefont {{Font}}}]{Kastaun2013}%
  \BibitemOpen
  \bibfield  {author} {\bibinfo {author} {\bibfnamefont {W.}~\bibnamefont
  {{Kastaun}}}, \bibinfo {author} {\bibfnamefont {F.}~\bibnamefont
  {{Galeazzi}}}, \bibinfo {author} {\bibfnamefont {D.}~\bibnamefont {{Alic}}},
  \bibinfo {author} {\bibfnamefont {L.}~\bibnamefont {{Rezzolla}}}, \ and\
  \bibinfo {author} {\bibfnamefont {J.~A.}\ \bibnamefont {{Font}}},\ }\href
  {\doibase 10.1103/PhysRevD.88.021501} {\bibfield  {journal} {\bibinfo
  {journal} {Phys. Rev. D}\ }\textbf {\bibinfo {volume} {88}},\ \bibinfo {eid}
  {021501} (\bibinfo {year} {2013})}\BibitemShut {NoStop}%
\bibitem [{\citenamefont {{Schnetter}}\ \emph {et~al.}(2004)\citenamefont
  {{Schnetter}}, \citenamefont {{Hawley}},\ and\ \citenamefont
  {{Hawke}}}]{Schnetter-etal-03b}%
  \BibitemOpen
  \bibfield  {author} {\bibinfo {author} {\bibfnamefont {E.}~\bibnamefont
  {{Schnetter}}}, \bibinfo {author} {\bibfnamefont {S.~H.}\ \bibnamefont
  {{Hawley}}}, \ and\ \bibinfo {author} {\bibfnamefont {I.}~\bibnamefont
  {{Hawke}}},\ }\href {\doibase 10.1088/0264-9381/21/6/014} {\bibfield
  {journal} {\bibinfo  {journal} {Class. Quantum Grav.}\ }\textbf {\bibinfo
  {volume} {21}},\ \bibinfo {pages} {1465} (\bibinfo {year}
  {2004})}\BibitemShut {NoStop}%
\bibitem [{\citenamefont {{Akmal}}\ \emph {et~al.}(1998)\citenamefont
  {{Akmal}}, \citenamefont {{Pandharipande}},\ and\ \citenamefont
  {{Ravenhall}}}]{Akmal1998a}%
  \BibitemOpen
  \bibfield  {author} {\bibinfo {author} {\bibfnamefont {A.}~\bibnamefont
  {{Akmal}}}, \bibinfo {author} {\bibfnamefont {V.~R.}\ \bibnamefont
  {{Pandharipande}}}, \ and\ \bibinfo {author} {\bibfnamefont {D.~G.}\
  \bibnamefont {{Ravenhall}}},\ }\href {\doibase 10.1103/PhysRevC.58.1804}
  {\bibfield  {journal} {\bibinfo  {journal} {Phys. Rev. C}\ }\textbf {\bibinfo
  {volume} {58}},\ \bibinfo {pages} {1804} (\bibinfo {year}
  {1998})}\BibitemShut {NoStop}%
\bibitem [{\citenamefont {{Alford}}\ \emph {et~al.}(2005)\citenamefont
  {{Alford}}, \citenamefont {{Braby}}, \citenamefont {{Paris}},\ and\
  \citenamefont {{Reddy}}}]{Alford2005}%
  \BibitemOpen
  \bibfield  {author} {\bibinfo {author} {\bibfnamefont {M.}~\bibnamefont
  {{Alford}}}, \bibinfo {author} {\bibfnamefont {M.}~\bibnamefont {{Braby}}},
  \bibinfo {author} {\bibfnamefont {M.}~\bibnamefont {{Paris}}}, \ and\
  \bibinfo {author} {\bibfnamefont {S.}~\bibnamefont {{Reddy}}},\ }\href
  {\doibase 10.1086/430902} {\bibfield  {journal} {\bibinfo  {journal}
  {Astrophys. J.}\ }\textbf {\bibinfo {volume} {629}},\ \bibinfo {pages} {969}
  (\bibinfo {year} {2005})}\BibitemShut {NoStop}%
\bibitem [{\citenamefont {{Douchin}}\ and\ \citenamefont
  {{Haensel}}(2001)}]{Douchin01}%
  \BibitemOpen
  \bibfield  {author} {\bibinfo {author} {\bibfnamefont {F.}~\bibnamefont
  {{Douchin}}}\ and\ \bibinfo {author} {\bibfnamefont {P.}~\bibnamefont
  {{Haensel}}},\ }\href {\doibase 10.1051/0004-6361:20011402} {\bibfield
  {journal} {\bibinfo  {journal} {Astron. Astrophys.}\ }\textbf {\bibinfo
  {volume} {380}},\ \bibinfo {pages} {151} (\bibinfo {year}
  {2001})}\BibitemShut {NoStop}%
\bibitem [{\citenamefont {Glendenning}\ and\ \citenamefont
  {Moszkowski}(1991)}]{GlendenningMoszkowski91}%
  \BibitemOpen
  \bibfield  {author} {\bibinfo {author} {\bibfnamefont {N.~K.}\ \bibnamefont
  {Glendenning}}\ and\ \bibinfo {author} {\bibfnamefont {S.~A.}\ \bibnamefont
  {Moszkowski}},\ }\href {\doibase 10.1103/PhysRevLett.67.2414} {\bibfield
  {journal} {\bibinfo  {journal} {Phys. Rev. Lett.}\ }\textbf {\bibinfo
  {volume} {67}},\ \bibinfo {pages} {2414} (\bibinfo {year}
  {1991})}\BibitemShut {NoStop}%
\bibitem [{\citenamefont {{Glendenning}}(1985)}]{Glendenning1985}%
  \BibitemOpen
  \bibfield  {author} {\bibinfo {author} {\bibfnamefont {N.~K.}\ \bibnamefont
  {{Glendenning}}},\ }\href {\doibase 10.1086/163253} {\bibfield  {journal}
  {\bibinfo  {journal} {Astrophys. J.}\ }\textbf {\bibinfo {volume} {293}},\
  \bibinfo {pages} {470} (\bibinfo {year} {1985})}\BibitemShut {NoStop}%
\bibitem [{\citenamefont {{Antoniadis}}\ \emph {et~al.}(2013)\citenamefont
  {{Antoniadis}}, \citenamefont {{Freire}}, \citenamefont {{Wex}},
  \citenamefont {{Tauris}}, \citenamefont {{Lynch}}, \citenamefont {{van
  Kerkwijk}}, \citenamefont {{Kramer}}, \citenamefont {{Bassa}}, \citenamefont
  {{Dhillon}}, \citenamefont {{Driebe}}, \citenamefont {{Hessels}},
  \citenamefont {{Kaspi}}, \citenamefont {{Kondratiev}}, \citenamefont
  {{Langer}}, \citenamefont {{Marsh}}, \citenamefont {{McLaughlin}},
  \citenamefont {{Pennucci}}, \citenamefont {{Ransom}}, \citenamefont
  {{Stairs}}, \citenamefont {{van Leeuwen}}, \citenamefont {{Verbiest}},\ and\
  \citenamefont {{Whelan}}}]{Antoniadis2013}%
  \BibitemOpen
  \bibfield  {author} {\bibinfo {author} {\bibfnamefont {J.}~\bibnamefont
  {{Antoniadis}}}, \bibinfo {author} {\bibfnamefont {P.~C.~C.}\ \bibnamefont
  {{Freire}}}, \bibinfo {author} {\bibfnamefont {N.}~\bibnamefont {{Wex}}},
  \bibinfo {author} {\bibfnamefont {T.~M.}\ \bibnamefont {{Tauris}}}, \bibinfo
  {author} {\bibfnamefont {R.~S.}\ \bibnamefont {{Lynch}}}, \bibinfo {author}
  {\bibfnamefont {M.~H.}\ \bibnamefont {{van Kerkwijk}}}, \bibinfo {author}
  {\bibfnamefont {M.}~\bibnamefont {{Kramer}}}, \bibinfo {author}
  {\bibfnamefont {C.}~\bibnamefont {{Bassa}}}, \bibinfo {author} {\bibfnamefont
  {V.~S.}\ \bibnamefont {{Dhillon}}}, \bibinfo {author} {\bibfnamefont
  {T.}~\bibnamefont {{Driebe}}}, \bibinfo {author} {\bibfnamefont {J.~W.~T.}\
  \bibnamefont {{Hessels}}}, \bibinfo {author} {\bibfnamefont {V.~M.}\
  \bibnamefont {{Kaspi}}}, \bibinfo {author} {\bibfnamefont {V.~I.}\
  \bibnamefont {{Kondratiev}}}, \bibinfo {author} {\bibfnamefont
  {N.}~\bibnamefont {{Langer}}}, \bibinfo {author} {\bibfnamefont {T.~R.}\
  \bibnamefont {{Marsh}}}, \bibinfo {author} {\bibfnamefont {M.~A.}\
  \bibnamefont {{McLaughlin}}}, \bibinfo {author} {\bibfnamefont {T.~T.}\
  \bibnamefont {{Pennucci}}}, \bibinfo {author} {\bibfnamefont {S.~M.}\
  \bibnamefont {{Ransom}}}, \bibinfo {author} {\bibfnamefont {I.~H.}\
  \bibnamefont {{Stairs}}}, \bibinfo {author} {\bibfnamefont {J.}~\bibnamefont
  {{van Leeuwen}}}, \bibinfo {author} {\bibfnamefont {J.~P.~W.}\ \bibnamefont
  {{Verbiest}}}, \ and\ \bibinfo {author} {\bibfnamefont {D.~G.}\ \bibnamefont
  {{Whelan}}},\ }\href {\doibase 10.1126/science.1233232} {\bibfield  {journal}
  {\bibinfo  {journal} {Science}\ }\textbf {\bibinfo {volume} {340}},\ \bibinfo
  {pages} {1233232} (\bibinfo {year} {2013})}\BibitemShut {NoStop}%
\bibitem [{\citenamefont {{Read}}\ \emph
  {et~al.}(2009{\natexlab{a}})\citenamefont {{Read}}, \citenamefont {{Lackey}},
  \citenamefont {{Owen}},\ and\ \citenamefont {{Friedman}}}]{Read:2009a}%
  \BibitemOpen
  \bibfield  {author} {\bibinfo {author} {\bibfnamefont {J.~S.}\ \bibnamefont
  {{Read}}}, \bibinfo {author} {\bibfnamefont {B.~D.}\ \bibnamefont
  {{Lackey}}}, \bibinfo {author} {\bibfnamefont {B.~J.}\ \bibnamefont
  {{Owen}}}, \ and\ \bibinfo {author} {\bibfnamefont {J.~L.}\ \bibnamefont
  {{Friedman}}},\ }\href {\doibase 10.1103/PhysRevD.79.124032} {\bibfield
  {journal} {\bibinfo  {journal} {Phys. Rev. D}\ }\textbf {\bibinfo {volume}
  {79}},\ \bibinfo {pages} {124032} (\bibinfo {year}
  {2009}{\natexlab{a}})}\BibitemShut {NoStop}%
\bibitem [{lor()}]{lorene41}%
  \BibitemOpen
  \href@noop {} {}\bibinfo {howpublished} {URL
  \url{http://www.lorene.obspm.fr}}\BibitemShut {NoStop}%
\bibitem [{\citenamefont {Janka}\ \emph {et~al.}(1993)\citenamefont {Janka},
  \citenamefont {Zwerger},\ and\ \citenamefont {M{\"o}nchmeyer}}]{Janka93}%
  \BibitemOpen
  \bibfield  {author} {\bibinfo {author} {\bibfnamefont {H.-T.}\ \bibnamefont
  {Janka}}, \bibinfo {author} {\bibfnamefont {T.}~\bibnamefont {Zwerger}}, \
  and\ \bibinfo {author} {\bibfnamefont {R.}~\bibnamefont {M{\"o}nchmeyer}},\
  }\href@noop {} {\bibfield  {journal} {\bibinfo  {journal} {Astron.
  Astrophys.}\ }\textbf {\bibinfo {volume} {268}},\ \bibinfo {pages} {360}
  (\bibinfo {year} {1993})}\BibitemShut {NoStop}%
\bibitem [{\citenamefont {Damour}\ \emph {et~al.}(2012)\citenamefont {Damour},
  \citenamefont {Nagar},\ and\ \citenamefont {Villain}}]{Damour:2012}%
  \BibitemOpen
  \bibfield  {author} {\bibinfo {author} {\bibfnamefont {T.}~\bibnamefont
  {Damour}}, \bibinfo {author} {\bibfnamefont {A.}~\bibnamefont {Nagar}}, \
  and\ \bibinfo {author} {\bibfnamefont {L.}~\bibnamefont {Villain}},\ }\href
  {\doibase 10.1103/PhysRevD.85.123007} {\bibfield  {journal} {\bibinfo
  {journal} {Phys. Rev. D}\ }\textbf {\bibinfo {volume} {85}},\ \bibinfo
  {pages} {123007} (\bibinfo {year} {2012})}\BibitemShut {NoStop}%
\bibitem [{\citenamefont {{Alic}}\ \emph {et~al.}(2012)\citenamefont {{Alic}},
  \citenamefont {{Bona-Casas}}, \citenamefont {{Bona}}, \citenamefont
  {{Rezzolla}},\ and\ \citenamefont {{Palenzuela}}}]{Alic:2011a}%
  \BibitemOpen
  \bibfield  {author} {\bibinfo {author} {\bibfnamefont {D.}~\bibnamefont
  {{Alic}}}, \bibinfo {author} {\bibfnamefont {C.}~\bibnamefont
  {{Bona-Casas}}}, \bibinfo {author} {\bibfnamefont {C.}~\bibnamefont
  {{Bona}}}, \bibinfo {author} {\bibfnamefont {L.}~\bibnamefont {{Rezzolla}}},
  \ and\ \bibinfo {author} {\bibfnamefont {C.}~\bibnamefont {{Palenzuela}}},\
  }\href {\doibase 10.1103/PhysRevD.85.064040} {\bibfield  {journal} {\bibinfo
  {journal} {Phys. Rev. D}\ }\textbf {\bibinfo {volume} {85}},\ \bibinfo {eid}
  {064040} (\bibinfo {year} {2012})}\BibitemShut {NoStop}%
\bibitem [{\citenamefont {{Alic}}\ \emph {et~al.}(2013)\citenamefont {{Alic}},
  \citenamefont {{Kastaun}},\ and\ \citenamefont {{Rezzolla}}}]{Alic2013}%
  \BibitemOpen
  \bibfield  {author} {\bibinfo {author} {\bibfnamefont {D.}~\bibnamefont
  {{Alic}}}, \bibinfo {author} {\bibfnamefont {W.}~\bibnamefont {{Kastaun}}}, \
  and\ \bibinfo {author} {\bibfnamefont {L.}~\bibnamefont {{Rezzolla}}},\
  }\href {\doibase 10.1103/PhysRevD.88.064049} {\bibfield  {journal} {\bibinfo
  {journal} {Phys. Rev. D}\ }\textbf {\bibinfo {volume} {88}},\ \bibinfo {eid}
  {064049} (\bibinfo {year} {2013})}\BibitemShut {NoStop}%
\bibitem [{\citenamefont {{Gourgoulhon}}\ \emph {et~al.}(2001)\citenamefont
  {{Gourgoulhon}}, \citenamefont {{Grandcl{\'e}ment}}, \citenamefont
  {{Taniguchi}}, \citenamefont {{Marck}},\ and\ \citenamefont
  {{Bonazzola}}}]{Gourgoulhon-etal-2000:2ns-initial-data}%
  \BibitemOpen
  \bibfield  {author} {\bibinfo {author} {\bibfnamefont {E.}~\bibnamefont
  {{Gourgoulhon}}}, \bibinfo {author} {\bibfnamefont {P.}~\bibnamefont
  {{Grandcl{\'e}ment}}}, \bibinfo {author} {\bibfnamefont {K.}~\bibnamefont
  {{Taniguchi}}}, \bibinfo {author} {\bibfnamefont {J.-A.}\ \bibnamefont
  {{Marck}}}, \ and\ \bibinfo {author} {\bibfnamefont {S.}~\bibnamefont
  {{Bonazzola}}},\ }\href {\doibase 10.1103/PhysRevD.63.064029} {\bibfield
  {journal} {\bibinfo  {journal} {Phys. Rev. D}\ }\textbf {\bibinfo {volume}
  {63}},\ \bibinfo {eid} {064029} (\bibinfo {year} {2001})}\BibitemShut
  {NoStop}%
\bibitem [{\citenamefont {Goldberg}\ \emph {et~al.}(1967)\citenamefont
  {Goldberg}, \citenamefont {MacFarlane}, \citenamefont {Newman}, \citenamefont
  {Rohrlich},\ and\ \citenamefont {Sudarshan}}]{Goldberg:1967}%
  \BibitemOpen
  \bibfield  {author} {\bibinfo {author} {\bibfnamefont {J.~N.}\ \bibnamefont
  {Goldberg}}, \bibinfo {author} {\bibfnamefont {A.~J.}\ \bibnamefont
  {MacFarlane}}, \bibinfo {author} {\bibfnamefont {E.~T.}\ \bibnamefont
  {Newman}}, \bibinfo {author} {\bibfnamefont {F.}~\bibnamefont {Rohrlich}}, \
  and\ \bibinfo {author} {\bibfnamefont {E.~C.~G.}\ \bibnamefont {Sudarshan}},\
  }\href {\doibase 10.1063/1.1705135} {\bibfield  {journal} {\bibinfo
  {journal} {J. Math. Phys.}\ }\textbf {\bibinfo {volume} {8}},\ \bibinfo
  {pages} {2155} (\bibinfo {year} {1967})}\BibitemShut {NoStop}%
\bibitem [{\citenamefont {{Takami}}\ \emph
  {et~al.}(2014{\natexlab{b}})\citenamefont {{Takami}}, \citenamefont
  {{Rezzolla}},\ and\ \citenamefont {{Baiotti}}}]{Takami2014}%
  \BibitemOpen
  \bibfield  {author} {\bibinfo {author} {\bibfnamefont {K.}~\bibnamefont
  {{Takami}}}, \bibinfo {author} {\bibfnamefont {L.}~\bibnamefont
  {{Rezzolla}}}, \ and\ \bibinfo {author} {\bibfnamefont {L.}~\bibnamefont
  {{Baiotti}}},\ }\href {\doibase 10.1103/PhysRevLett.113.091104} {\bibfield
  {journal} {\bibinfo  {journal} {Phys. Rev. Lett.}\ }\textbf {\bibinfo
  {volume} {113}},\ \bibinfo {eid} {091104} (\bibinfo {year}
  {2014}{\natexlab{b}})}\BibitemShut {NoStop}%
\bibitem [{url()}]{url:adLIGO_Sh_curve}%
  \BibitemOpen
  \href {https://dcc.ligo.org/LIGO-T0900288/public} {}\bibinfo {note} {Advanced
  LIGO anticipated sensitivity curves, LIGO Document No.
  T0900288-v3}\BibitemShut {NoStop}%
\bibitem [{\citenamefont {Punturo}\ \emph
  {et~al.}(2010{\natexlab{a}})\citenamefont {Punturo} \emph
  {et~al.}}]{Punturo:2010}%
  \BibitemOpen
  \bibfield  {author} {\bibinfo {author} {\bibfnamefont {M.}~\bibnamefont
  {Punturo}} \emph {et~al.},\ }\href {\doibase 10.1088/0264-9381/27/8/084007}
  {\bibfield  {journal} {\bibinfo  {journal} {Classical Quantum Gravity}\
  }\textbf {\bibinfo {volume} {27}},\ \bibinfo {pages} {084007} (\bibinfo
  {year} {2010}{\natexlab{a}})}\BibitemShut {NoStop}%
\bibitem [{\citenamefont {Punturo}\ \emph
  {et~al.}(2010{\natexlab{b}})\citenamefont {Punturo} \emph
  {et~al.}}]{Punturo2010b}%
  \BibitemOpen
  \bibfield  {author} {\bibinfo {author} {\bibfnamefont {M.}~\bibnamefont
  {Punturo}} \emph {et~al.},\ }\href {\doibase 10.1088/0264-9381/27/19/194002}
  {\bibfield  {journal} {\bibinfo  {journal} {Class. Quantum Grav.}\ }\textbf
  {\bibinfo {volume} {27}},\ \bibinfo {pages} {194002} (\bibinfo {year}
  {2010}{\natexlab{b}})}\BibitemShut {NoStop}%
\bibitem [{\citenamefont {Messenger}\ \emph {et~al.}(2014)\citenamefont
  {Messenger}, \citenamefont {Takami}, \citenamefont {Gossan}, \citenamefont
  {Rezzolla},\ and\ \citenamefont {Sathyaprakash}}]{Messenger2013}%
  \BibitemOpen
  \bibfield  {author} {\bibinfo {author} {\bibfnamefont {C.}~\bibnamefont
  {Messenger}}, \bibinfo {author} {\bibfnamefont {K.}~\bibnamefont {Takami}},
  \bibinfo {author} {\bibfnamefont {S.}~\bibnamefont {Gossan}}, \bibinfo
  {author} {\bibfnamefont {L.}~\bibnamefont {Rezzolla}}, \ and\ \bibinfo
  {author} {\bibfnamefont {B.~S.}\ \bibnamefont {Sathyaprakash}},\ }\href
  {\doibase 10.1103/PhysRevX.4.041004} {\bibfield  {journal} {\bibinfo
  {journal} {Phys. Rev. X}\ }\textbf {\bibinfo {volume} {4}},\ \bibinfo {pages}
  {041004} (\bibinfo {year} {2014})}\BibitemShut {NoStop}%
\bibitem [{\citenamefont {{Harris}}(1978)}]{Harris78}%
  \BibitemOpen
  \bibfield  {author} {\bibinfo {author} {\bibfnamefont {F.~J.}\ \bibnamefont
  {{Harris}}},\ }\href {\doibase 10.1109/PROC.1978.10837} {\bibfield  {journal}
  {\bibinfo  {journal} {Proceedings of the IEEE}\ }\textbf {\bibinfo {volume}
  {66}},\ \bibinfo {pages} {51} (\bibinfo {year} {1978})}\BibitemShut {NoStop}%
\bibitem [{\citenamefont {{McKechan}}\ \emph {et~al.}(2010)\citenamefont
  {{McKechan}}, \citenamefont {{Robinson}},\ and\ \citenamefont
  {{Sathyaprakash}}}]{McKechan2010}%
  \BibitemOpen
  \bibfield  {author} {\bibinfo {author} {\bibfnamefont {D.~J.~A.}\
  \bibnamefont {{McKechan}}}, \bibinfo {author} {\bibfnamefont
  {C.}~\bibnamefont {{Robinson}}}, \ and\ \bibinfo {author} {\bibfnamefont
  {B.~S.}\ \bibnamefont {{Sathyaprakash}}},\ }\href {\doibase
  10.1088/0264-9381/27/8/084020} {\bibfield  {journal} {\bibinfo  {journal}
  {Classical and Quantum Gravity}\ }\textbf {\bibinfo {volume} {27}},\ \bibinfo
  {eid} {084020} (\bibinfo {year} {2010})}\BibitemShut {NoStop}%
\bibitem [{\citenamefont {{Butterworth}}(1930)}]{Butterworth1930}%
  \BibitemOpen
  \bibfield  {author} {\bibinfo {author} {\bibfnamefont {S.}~\bibnamefont
  {{Butterworth}}},\ }\href@noop {} {\bibfield  {journal} {\bibinfo  {journal}
  {Experimental Wireless and the Wireless Engineer}\ }\textbf {\bibinfo
  {volume} {7}},\ \bibinfo {pages} {536–541} (\bibinfo {year}
  {1930})}\BibitemShut {NoStop}%
\bibitem [{\citenamefont {{Shen}}\ \emph {et~al.}(1998)\citenamefont {{Shen}},
  \citenamefont {{Toki}}, \citenamefont {{Oyamatsu}},\ and\ \citenamefont
  {{Sumiyoshi}}}]{shen98}%
  \BibitemOpen
  \bibfield  {author} {\bibinfo {author} {\bibfnamefont {H.}~\bibnamefont
  {{Shen}}}, \bibinfo {author} {\bibfnamefont {H.}~\bibnamefont {{Toki}}},
  \bibinfo {author} {\bibfnamefont {K.}~\bibnamefont {{Oyamatsu}}}, \ and\
  \bibinfo {author} {\bibfnamefont {K.}~\bibnamefont {{Sumiyoshi}}},\ }\href
  {http://user.numazu-ct.ac.jp/~sumi/eos} {\bibfield  {journal} {\bibinfo
  {journal} {Nucl. Phys. A}\ }\textbf {\bibinfo {volume} {637}},\ \bibinfo
  {pages} {435} (\bibinfo {year} {1998})}\BibitemShut {NoStop}%
\bibitem [{\citenamefont {{Bauswein}}\ \emph {et~al.}(2010)\citenamefont
  {{Bauswein}}, \citenamefont {{Janka}},\ and\ \citenamefont
  {{Oechslin}}}]{Bauswein:2010dn}%
  \BibitemOpen
  \bibfield  {author} {\bibinfo {author} {\bibfnamefont {A.}~\bibnamefont
  {{Bauswein}}}, \bibinfo {author} {\bibfnamefont {H.}~\bibnamefont {{Janka}}},
  \ and\ \bibinfo {author} {\bibfnamefont {R.}~\bibnamefont {{Oechslin}}},\
  }\href {\doibase 10.1103/PhysRevD.82.084043} {\bibfield  {journal} {\bibinfo
  {journal} {Phys. Rev. D}\ }\textbf {\bibinfo {volume} {82}},\ \bibinfo
  {pages} {084043} (\bibinfo {year} {2010})}\BibitemShut {NoStop}%
\bibitem [{\citenamefont {Landau}\ and\ \citenamefont
  {Lifshitz}(1976)}]{Landau-Lifshitz1}%
  \BibitemOpen
  \bibfield  {author} {\bibinfo {author} {\bibfnamefont {L.~D.}\ \bibnamefont
  {Landau}}\ and\ \bibinfo {author} {\bibfnamefont {E.~M.}\ \bibnamefont
  {Lifshitz}},\ }\href@noop {} {\emph {\bibinfo {title} {Mechanics, Volume
  1}}}\ (\bibinfo  {publisher} {Pergamon Press},\ \bibinfo {address} {Oxford},\
  \bibinfo {year} {1976})\BibitemShut {NoStop}%
\bibitem [{\citenamefont {{Zanotti}}\ \emph {et~al.}(2005)\citenamefont
  {{Zanotti}}, \citenamefont {{Font}}, \citenamefont {{Rezzolla}},\ and\
  \citenamefont {{Montero}}}]{Zanotti05}%
  \BibitemOpen
  \bibfield  {author} {\bibinfo {author} {\bibfnamefont {O.}~\bibnamefont
  {{Zanotti}}}, \bibinfo {author} {\bibfnamefont {J.~A.}\ \bibnamefont
  {{Font}}}, \bibinfo {author} {\bibfnamefont {L.}~\bibnamefont {{Rezzolla}}},
  \ and\ \bibinfo {author} {\bibfnamefont {P.~J.}\ \bibnamefont {{Montero}}},\
  }\href {\doibase 10.1111/j.1365-2966.2004.08567.x} {\bibfield  {journal}
  {\bibinfo  {journal} {Mon. Not. R. Astron. Soc.}\ }\textbf {\bibinfo {volume}
  {356}},\ \bibinfo {pages} {1371} (\bibinfo {year} {2005})}\BibitemShut
  {NoStop}%
\bibitem [{\citenamefont {{Clark}}\ \emph {et~al.}(2014)\citenamefont
  {{Clark}}, \citenamefont {{Bauswein}}, \citenamefont {{Cadonati}},
  \citenamefont {{Janka}}, \citenamefont {{Pankow}},\ and\ \citenamefont
  {{Stergioulas}}}]{Clark2014}%
  \BibitemOpen
  \bibfield  {author} {\bibinfo {author} {\bibfnamefont {J.}~\bibnamefont
  {{Clark}}}, \bibinfo {author} {\bibfnamefont {A.}~\bibnamefont {{Bauswein}}},
  \bibinfo {author} {\bibfnamefont {L.}~\bibnamefont {{Cadonati}}}, \bibinfo
  {author} {\bibfnamefont {H.-T.}\ \bibnamefont {{Janka}}}, \bibinfo {author}
  {\bibfnamefont {C.}~\bibnamefont {{Pankow}}}, \ and\ \bibinfo {author}
  {\bibfnamefont {N.}~\bibnamefont {{Stergioulas}}},\ }\href {\doibase
  10.1103/PhysRevD.90.062004} {\bibfield  {journal} {\bibinfo  {journal} {Phys.
  Rev. D}\ }\textbf {\bibinfo {volume} {90}},\ \bibinfo {eid} {062004}
  (\bibinfo {year} {2014})}\BibitemShut {NoStop}%
\bibitem [{\citenamefont {{Yagi}}\ and\ \citenamefont
  {{Yunes}}(2013{\natexlab{a}})}]{Yagi2013a}%
  \BibitemOpen
  \bibfield  {author} {\bibinfo {author} {\bibfnamefont {K.}~\bibnamefont
  {{Yagi}}}\ and\ \bibinfo {author} {\bibfnamefont {N.}~\bibnamefont
  {{Yunes}}},\ }\href@noop {} {\bibfield  {journal} {\bibinfo  {journal}
  {Science}\ }\textbf {\bibinfo {volume} {341}},\ \bibinfo {pages} {365}
  (\bibinfo {year} {2013}{\natexlab{a}})}\BibitemShut {NoStop}%
\bibitem [{\citenamefont {{Yagi}}\ and\ \citenamefont
  {{Yunes}}(2013{\natexlab{b}})}]{Yagi2013b}%
  \BibitemOpen
  \bibfield  {author} {\bibinfo {author} {\bibfnamefont {K.}~\bibnamefont
  {{Yagi}}}\ and\ \bibinfo {author} {\bibfnamefont {N.}~\bibnamefont
  {{Yunes}}},\ }\href {\doibase 10.1103/PhysRevD.88.023009} {\bibfield
  {journal} {\bibinfo  {journal} {Phys. Rev. D}\ }\textbf {\bibinfo {volume}
  {88}},\ \bibinfo {eid} {023009} (\bibinfo {year}
  {2013}{\natexlab{b}})}\BibitemShut {NoStop}%
\bibitem [{\citenamefont {{Urbanec}}\ \emph {et~al.}(2013)\citenamefont
  {{Urbanec}}, \citenamefont {{Miller}},\ and\ \citenamefont
  {{Stuchl{\'{\i}}k}}}]{Urbanec2013}%
  \BibitemOpen
  \bibfield  {author} {\bibinfo {author} {\bibfnamefont {M.}~\bibnamefont
  {{Urbanec}}}, \bibinfo {author} {\bibfnamefont {J.~C.}\ \bibnamefont
  {{Miller}}}, \ and\ \bibinfo {author} {\bibfnamefont {Z.}~\bibnamefont
  {{Stuchl{\'{\i}}k}}},\ }\href {\doibase 10.1093/mnras/stt858} {\bibfield
  {journal} {\bibinfo  {journal} {Mon. Not. R. Astron. Soc.}\ }\textbf
  {\bibinfo {volume} {433}},\ \bibinfo {pages} {1903} (\bibinfo {year}
  {2013})}\BibitemShut {NoStop}%
\bibitem [{\citenamefont {{Doneva}}\ \emph
  {et~al.}(2014{\natexlab{a}})\citenamefont {{Doneva}}, \citenamefont
  {{Yazadjiev}}, \citenamefont {{Stergioulas}},\ and\ \citenamefont
  {{Kokkotas}}}]{Doneva2014a}%
  \BibitemOpen
  \bibfield  {author} {\bibinfo {author} {\bibfnamefont {D.~D.}\ \bibnamefont
  {{Doneva}}}, \bibinfo {author} {\bibfnamefont {S.~S.}\ \bibnamefont
  {{Yazadjiev}}}, \bibinfo {author} {\bibfnamefont {N.}~\bibnamefont
  {{Stergioulas}}}, \ and\ \bibinfo {author} {\bibfnamefont {K.~D.}\
  \bibnamefont {{Kokkotas}}},\ }\href {\doibase 10.1088/2041-8205/781/1/L6}
  {\bibfield  {journal} {\bibinfo  {journal} {Astrophys. J. Letters}\ }\textbf
  {\bibinfo {volume} {781}},\ \bibinfo {eid} {L6} (\bibinfo {year}
  {2014}{\natexlab{a}})}\BibitemShut {NoStop}%
\bibitem [{\citenamefont {{Haskell}}\ \emph {et~al.}(2014)\citenamefont
  {{Haskell}}, \citenamefont {{Ciolfi}}, \citenamefont {{Pannarale}},\ and\
  \citenamefont {{Rezzolla}}}]{Haskell2014}%
  \BibitemOpen
  \bibfield  {author} {\bibinfo {author} {\bibfnamefont {B.}~\bibnamefont
  {{Haskell}}}, \bibinfo {author} {\bibfnamefont {R.}~\bibnamefont {{Ciolfi}}},
  \bibinfo {author} {\bibfnamefont {F.}~\bibnamefont {{Pannarale}}}, \ and\
  \bibinfo {author} {\bibfnamefont {L.}~\bibnamefont {{Rezzolla}}},\ }\href
  {\doibase 10.1093/mnrasl/slt161} {\bibfield  {journal} {\bibinfo  {journal}
  {Mon. Not. R. Astron. Soc. Letters}\ }\textbf {\bibinfo {volume} {438}},\
  \bibinfo {pages} {L71} (\bibinfo {year} {2014})}\BibitemShut {NoStop}%
\bibitem [{\citenamefont {{Pappas}}\ and\ \citenamefont
  {{Apostolatos}}(2014)}]{Pappas2014}%
  \BibitemOpen
  \bibfield  {author} {\bibinfo {author} {\bibfnamefont {G.}~\bibnamefont
  {{Pappas}}}\ and\ \bibinfo {author} {\bibfnamefont {T.~A.}\ \bibnamefont
  {{Apostolatos}}},\ }\href {\doibase 10.1103/PhysRevLett.112.121101}
  {\bibfield  {journal} {\bibinfo  {journal} {Phys. Rev. Lett.}\ }\textbf
  {\bibinfo {volume} {112}},\ \bibinfo {eid} {121101} (\bibinfo {year}
  {2014})}\BibitemShut {NoStop}%
\bibitem [{\citenamefont {{Chakrabarti}}\ \emph {et~al.}(2014)\citenamefont
  {{Chakrabarti}}, \citenamefont {{Delsate}}, \citenamefont {{G{\"u}rlebeck}},\
  and\ \citenamefont {{Steinhoff}}}]{Chakrabarti2014}%
  \BibitemOpen
  \bibfield  {author} {\bibinfo {author} {\bibfnamefont {S.}~\bibnamefont
  {{Chakrabarti}}}, \bibinfo {author} {\bibfnamefont {T.}~\bibnamefont
  {{Delsate}}}, \bibinfo {author} {\bibfnamefont {N.}~\bibnamefont
  {{G{\"u}rlebeck}}}, \ and\ \bibinfo {author} {\bibfnamefont {J.}~\bibnamefont
  {{Steinhoff}}},\ }\href {\doibase 10.1103/PhysRevLett.112.201102} {\bibfield
  {journal} {\bibinfo  {journal} {Phys. Rev. Lett.}\ }\textbf {\bibinfo
  {volume} {112}},\ \bibinfo {eid} {201102} (\bibinfo {year}
  {2014})}\BibitemShut {NoStop}%
\bibitem [{\citenamefont {{Stein}}\ \emph {et~al.}(2014)\citenamefont
  {{Stein}}, \citenamefont {{Yagi}},\ and\ \citenamefont
  {{Yunes}}}]{Stein2014}%
  \BibitemOpen
  \bibfield  {author} {\bibinfo {author} {\bibfnamefont {L.~C.}\ \bibnamefont
  {{Stein}}}, \bibinfo {author} {\bibfnamefont {K.}~\bibnamefont {{Yagi}}}, \
  and\ \bibinfo {author} {\bibfnamefont {N.}~\bibnamefont {{Yunes}}},\ }\href
  {\doibase 10.1088/0004-637X/788/1/15} {\bibfield  {journal} {\bibinfo
  {journal} {Astrophys. J.}\ }\textbf {\bibinfo {volume} {788}},\ \bibinfo
  {eid} {15} (\bibinfo {year} {2014})}\BibitemShut {NoStop}%
\bibitem [{\citenamefont {{Maselli}}\ \emph {et~al.}(2013)\citenamefont
  {{Maselli}}, \citenamefont {{Cardoso}}, \citenamefont {{Ferrari}},
  \citenamefont {{Gualtieri}},\ and\ \citenamefont {{Pani}}}]{Maselli2013}%
  \BibitemOpen
  \bibfield  {author} {\bibinfo {author} {\bibfnamefont {A.}~\bibnamefont
  {{Maselli}}}, \bibinfo {author} {\bibfnamefont {V.}~\bibnamefont
  {{Cardoso}}}, \bibinfo {author} {\bibfnamefont {V.}~\bibnamefont
  {{Ferrari}}}, \bibinfo {author} {\bibfnamefont {L.}~\bibnamefont
  {{Gualtieri}}}, \ and\ \bibinfo {author} {\bibfnamefont {P.}~\bibnamefont
  {{Pani}}},\ }\href {\doibase 10.1103/PhysRevD.88.023007} {\bibfield
  {journal} {\bibinfo  {journal} {Phys. Rev. D}\ }\textbf {\bibinfo {volume}
  {88}},\ \bibinfo {eid} {023007} (\bibinfo {year} {2013})}\BibitemShut
  {NoStop}%
\bibitem [{\citenamefont {{Chatziioannou}}\ \emph {et~al.}(2014)\citenamefont
  {{Chatziioannou}}, \citenamefont {{Yagi}},\ and\ \citenamefont
  {{Yunes}}}]{Chatziioannou2014}%
  \BibitemOpen
  \bibfield  {author} {\bibinfo {author} {\bibfnamefont {K.}~\bibnamefont
  {{Chatziioannou}}}, \bibinfo {author} {\bibfnamefont {K.}~\bibnamefont
  {{Yagi}}}, \ and\ \bibinfo {author} {\bibfnamefont {N.}~\bibnamefont
  {{Yunes}}},\ }\href {\doibase 10.1103/PhysRevD.90.064030} {\bibfield
  {journal} {\bibinfo  {journal} {Phys. Rev. D}\ }\textbf {\bibinfo {volume}
  {90}},\ \bibinfo {eid} {064030} (\bibinfo {year} {2014})}\BibitemShut
  {NoStop}%
\bibitem [{\citenamefont {{Sham}}\ \emph {et~al.}(2014)\citenamefont {{Sham}},
  \citenamefont {{Lin}},\ and\ \citenamefont {{Leung}}}]{Sham2014a}%
  \BibitemOpen
  \bibfield  {author} {\bibinfo {author} {\bibfnamefont {Y.-H.}\ \bibnamefont
  {{Sham}}}, \bibinfo {author} {\bibfnamefont {L.-M.}\ \bibnamefont {{Lin}}}, \
  and\ \bibinfo {author} {\bibfnamefont {P.~T.}\ \bibnamefont {{Leung}}},\
  }\href {\doibase 10.1088/0004-637X/781/2/66} {\bibfield  {journal} {\bibinfo
  {journal} {Astrophys. J.}\ }\textbf {\bibinfo {volume} {781}},\ \bibinfo
  {eid} {66} (\bibinfo {year} {2014})}\BibitemShut {NoStop}%
\bibitem [{\citenamefont {{Pani}}\ and\ \citenamefont
  {{Berti}}(2014)}]{Pani2014}%
  \BibitemOpen
  \bibfield  {author} {\bibinfo {author} {\bibfnamefont {P.}~\bibnamefont
  {{Pani}}}\ and\ \bibinfo {author} {\bibfnamefont {E.}~\bibnamefont
  {{Berti}}},\ }\href {\doibase 10.1103/PhysRevD.90.024025} {\bibfield
  {journal} {\bibinfo  {journal} {Phys. Rev. D}\ }\textbf {\bibinfo {volume}
  {90}},\ \bibinfo {eid} {024025} (\bibinfo {year} {2014})}\BibitemShut
  {NoStop}%
\bibitem [{\citenamefont {{Doneva}}\ \emph
  {et~al.}(2014{\natexlab{b}})\citenamefont {{Doneva}}, \citenamefont
  {{Yazadjiev}}, \citenamefont {{Staykov}},\ and\ \citenamefont
  {{Kokkotas}}}]{Doneva2014b}%
  \BibitemOpen
  \bibfield  {author} {\bibinfo {author} {\bibfnamefont {D.~D.}\ \bibnamefont
  {{Doneva}}}, \bibinfo {author} {\bibfnamefont {S.~S.}\ \bibnamefont
  {{Yazadjiev}}}, \bibinfo {author} {\bibfnamefont {K.~V.}\ \bibnamefont
  {{Staykov}}}, \ and\ \bibinfo {author} {\bibfnamefont {K.~D.}\ \bibnamefont
  {{Kokkotas}}},\ }\href {\doibase 10.1103/PhysRevD.90.104021} {\bibfield
  {journal} {\bibinfo  {journal} {Phys. Rev. D}\ }\textbf {\bibinfo {volume}
  {90}},\ \bibinfo {eid} {104021} (\bibinfo {year}
  {2014}{\natexlab{b}})}\BibitemShut {NoStop}%
\bibitem [{\citenamefont {{Yagi}}\ \emph {et~al.}(2014)\citenamefont {{Yagi}},
  \citenamefont {{Kyutoku}}, \citenamefont {{Pappas}}, \citenamefont
  {{Yunes}},\ and\ \citenamefont {{Apostolatos}}}]{Yagi2014}%
  \BibitemOpen
  \bibfield  {author} {\bibinfo {author} {\bibfnamefont {K.}~\bibnamefont
  {{Yagi}}}, \bibinfo {author} {\bibfnamefont {K.}~\bibnamefont {{Kyutoku}}},
  \bibinfo {author} {\bibfnamefont {G.}~\bibnamefont {{Pappas}}}, \bibinfo
  {author} {\bibfnamefont {N.}~\bibnamefont {{Yunes}}}, \ and\ \bibinfo
  {author} {\bibfnamefont {T.~A.}\ \bibnamefont {{Apostolatos}}},\ }\href
  {\doibase 10.1103/PhysRevD.89.124013} {\bibfield  {journal} {\bibinfo
  {journal} {Phys. Rev. D}\ }\textbf {\bibinfo {volume} {89}},\ \bibinfo {eid}
  {124013} (\bibinfo {year} {2014})}\BibitemShut {NoStop}%
\bibitem [{\citenamefont {{Sham}}\ \emph {et~al.}(2015)\citenamefont {{Sham}},
  \citenamefont {{Chan}}, \citenamefont {{Lin}},\ and\ \citenamefont
  {{Leung}}}]{Sham2014b}%
  \BibitemOpen
  \bibfield  {author} {\bibinfo {author} {\bibfnamefont {Y.-H.}\ \bibnamefont
  {{Sham}}}, \bibinfo {author} {\bibfnamefont {T.~K.}\ \bibnamefont {{Chan}}},
  \bibinfo {author} {\bibfnamefont {L.-M.}\ \bibnamefont {{Lin}}}, \ and\
  \bibinfo {author} {\bibfnamefont {P.~T.}\ \bibnamefont {{Leung}}},\ }\href
  {\doibase 10.1088/0004-637X/798/2/121} {\bibfield  {journal} {\bibinfo
  {journal} {The Astrophysical Journal}\ }\textbf {\bibinfo {volume} {798}},\
  \bibinfo {eid} {121} (\bibinfo {year} {2015})}\BibitemShut {NoStop}%
\bibitem [{\citenamefont {Messenger}\ \emph {et~al.}(2015)\citenamefont
  {Messenger}, \citenamefont {Rezzolla},\ and\ \citenamefont
  {Takami}}]{Messenger2015}%
  \BibitemOpen
  \bibfield  {author} {\bibinfo {author} {\bibfnamefont {C.}~\bibnamefont
  {Messenger}}, \bibinfo {author} {\bibfnamefont {L.}~\bibnamefont {Rezzolla}},
  \ and\ \bibinfo {author} {\bibfnamefont {K.}~\bibnamefont {Takami}},\
  }\href@noop {} {\bibfield  {journal} {\bibinfo  {journal} {in preparation}\ }
  (\bibinfo {year} {2015})}\BibitemShut {NoStop}%
\bibitem [{\citenamefont {{Bernuzzi}}\ \emph
  {et~al.}(2014{\natexlab{b}})\citenamefont {{Bernuzzi}}, \citenamefont
  {{Dietrich}}, \citenamefont {{Tichy}},\ and\ \citenamefont
  {{Br{\"u}gmann}}}]{Bernuzzi2013}%
  \BibitemOpen
  \bibfield  {author} {\bibinfo {author} {\bibfnamefont {S.}~\bibnamefont
  {{Bernuzzi}}}, \bibinfo {author} {\bibfnamefont {T.}~\bibnamefont
  {{Dietrich}}}, \bibinfo {author} {\bibfnamefont {W.}~\bibnamefont {{Tichy}}},
  \ and\ \bibinfo {author} {\bibfnamefont {B.}~\bibnamefont {{Br{\"u}gmann}}},\
  }\href {\doibase 10.1103/PhysRevD.89.104021} {\bibfield  {journal} {\bibinfo
  {journal} {Phys. Rev. D}\ }\textbf {\bibinfo {volume} {89}},\ \bibinfo {eid}
  {104021} (\bibinfo {year} {2014}{\natexlab{b}})}\BibitemShut {NoStop}%
\bibitem [{\citenamefont {Giacomazzo}\ \emph {et~al.}(2011)\citenamefont
  {Giacomazzo}, \citenamefont {Rezzolla},\ and\ \citenamefont
  {Baiotti}}]{Giacomazzo:2010}%
  \BibitemOpen
  \bibfield  {author} {\bibinfo {author} {\bibfnamefont {B.}~\bibnamefont
  {Giacomazzo}}, \bibinfo {author} {\bibfnamefont {L.}~\bibnamefont
  {Rezzolla}}, \ and\ \bibinfo {author} {\bibfnamefont {L.}~\bibnamefont
  {Baiotti}},\ }\href {\doibase 10.1103/PhysRevD.83.044014} {\bibfield
  {journal} {\bibinfo  {journal} {Phys. Rev. D}\ }\textbf {\bibinfo {volume}
  {83}},\ \bibinfo {pages} {044014} (\bibinfo {year} {2011})}\BibitemShut
  {NoStop}%
\bibitem [{\citenamefont {{Read}}\ \emph
  {et~al.}(2009{\natexlab{b}})\citenamefont {{Read}}, \citenamefont
  {{Markakis}}, \citenamefont {{Shibata}}, \citenamefont {{Ury{\= u}}},
  \citenamefont {{Creighton}},\ and\ \citenamefont {{Friedman}}}]{Read:2009b}%
  \BibitemOpen
  \bibfield  {author} {\bibinfo {author} {\bibfnamefont {J.~S.}\ \bibnamefont
  {{Read}}}, \bibinfo {author} {\bibfnamefont {C.}~\bibnamefont {{Markakis}}},
  \bibinfo {author} {\bibfnamefont {M.}~\bibnamefont {{Shibata}}}, \bibinfo
  {author} {\bibfnamefont {K.}~\bibnamefont {{Ury{\= u}}}}, \bibinfo {author}
  {\bibfnamefont {J.~D.~E.}\ \bibnamefont {{Creighton}}}, \ and\ \bibinfo
  {author} {\bibfnamefont {J.~L.}\ \bibnamefont {{Friedman}}},\ }\href
  {\doibase 10.1103/PhysRevD.79.124033} {\bibfield  {journal} {\bibinfo
  {journal} {Phys. Rev. D}\ }\textbf {\bibinfo {volume} {79}},\ \bibinfo {eid}
  {124033} (\bibinfo {year} {2009}{\natexlab{b}})}\BibitemShut {NoStop}%
\end{thebibliography}
